\documentclass[onecolumn,draftcls,10pt]{IEEEtran}
\usepackage{cite}

\ifCLASSINFOpdf
\else
   \usepackage[dvips]{graphicx}
\fi
\usepackage[cmex10]{amsmath}
\usepackage{amssymb,amsthm}
\usepackage{multirow,bigdelim}
\newtheorem{assumption}{Assumption}

\newtheorem{proposition}{Proposition}
\newtheorem{example}{Example}

\newtheorem{lemma}{Lemma}
\newtheorem{corollary}{Corollary}
\newcommand{\argsup}{\mathop{\mathrm{argsup}}\limits} 
\newcommand{\arginf}{\mathop{\mathrm{arginf}}\limits}

\begin{document}
%
\title{On an Achievable Rate of Large Rayleigh Block-Fading MIMO Channels 
with No CSI}
%
%
%

\author{Keigo~Takeuchi,~\IEEEmembership{Member,~IEEE,}
        Ralf~R.~M\"uller,~\IEEEmembership{Senior Member,~IEEE,}
        Mikko~Vehkaper\"a,~\IEEEmembership{Member,~IEEE,}
 and~Toshiyuki~Tanaka,~\IEEEmembership{Member,~IEEE,}
\thanks{Manuscript received August, 2011. 
The work of K.~Takeuchi was in part supported by the Grant-in-Aid for 
Young Scientists~(B) (No.~23760329) from JSPS, Japan. 
The work of M.~Vehkaper\"a was supported by the Norwegian Research Council 
under grant 171133/V30. 
The work of T.~Tanaka was in part supported by the Grant-in-Aid for 
Scientific Research on Priority Areas (No. 18079010) from MEXT, Japan. 
The material in this paper was presented in part at 2009 IEEE 
International Symposium on Information Theory, Seoul, Korea, July 2009, 
and at 2010 International Symposium on Information Theory and its 
Applications \& 2010 International Symposium on Spread Spectrum Techniques 
and Applications, Taichung, Taiwan, Oct.\ 2010.}
\thanks{K.~Takeuchi is with the Department of Communication Engineering 
and Informatics, the University of Electro-Communications, Tokyo 182-8585, 
Japan (e-mail: ktakeuchi@uec.ac.jp).}
\thanks{R.~R.~M\"uller is with the Department of 
Electronics and Telecommunications, the Norwegian University of Science and 
Technology (NTNU), NO--7491 Trondheim, Norway (e-mail: ralf@iet.ntnu.no).}
\thanks{M.~Vehkaper\"a is with the School of Electrical Engineering, Royal 
Institute of Technology (KTH), SE--100 44 Stockholm, 
Sweden (e-mail: mikkov@kth.se).}
\thanks{T.~Tanaka is with the Department of Systems Science, 
Graduate School of Informatics, Kyoto University, Kyoto, 
606-8501, Japan (e-mail: tt@i.kyoto-u.ac.jp).}
}

%
%

\markboth{IEEE transactions on information theory,~Vol.~, No.~, 2011}%
{Takeuchi \MakeLowercase{\textit{et al.}}: }
%

\IEEEpubid{0000--0000/00\$00.00~\copyright~2010 IEEE}


\maketitle

\begin{abstract}
Training-based transmission over Rayleigh block-fading multiple-input 
multiple-output (MIMO) channels is investigated. As a training method  
a combination of a pilot-assisted scheme and a biased signaling scheme is 
considered. The achievable rates of successive decoding (SD) receivers based 
on the linear minimum mean-squared error (LMMSE) channel estimation are 
analyzed in the large-system limit, by using the replica method 
under the assumption of replica symmetry. 
It is shown that negligible pilot information is best in terms of the 
achievable rates of the SD receivers in the large-system limit. 
The obtained analytical formulas of the achievable rates can improve the 
existing lower bound on the capacity of the MIMO channel with no channel state 
information (CSI), derived by Hassibi and Hochwald, for all signal-to-noise 
ratios (SNRs). The comparison between the obtained bound and a high SNR 
approximation of the channel capacity, derived by Zheng and Tse, implies that 
the high SNR approximation is unreliable unless quite high SNR is considered. 
Energy efficiency in the low SNR regime is also investigated in terms of 
the power per information bit required for reliable communication. 
The required minimum power is shown to be achieved at a positive rate 
for the SD receiver with no CSI, whereas it is achieved in the zero-rate limit 
for the case of perfect CSI available at the receiver. 
Moreover, numerical simulations imply that the presented large-system analysis 
can provide a good approximation for not so large systems. 
The results in this paper imply that SD schemes can provide a 
significant performance gain in the low-to-moderate SNR regimes, compared 
to conventional receivers based on one-shot channel estimation. 
\end{abstract}


%
\IEEEpeerreviewmaketitle

\section{Introduction}
\subsection{Motivation} 
\IEEEPARstart{M}{ultiple}-input multiple-output (MIMO) transmission is a 
promising scheme for increasing the spectral efficiency of wireless 
communication systems, and has been applied to several modern standards, 
such as wireless LAN (IEEE 802.11n) and Mobile WiMAX (IEEE 802.16e). 
However, the ultimate achievable rate of MIMO systems is not fully understood. 
Thus, it is an important issue in information theory to elucidate the 
channel capacity of MIMO systems. 

The capacity of MIMO channels with perfect channel state information (CSI) 
at the receiver was analyzed in the early pioneering 
works~\cite{Foschini98,Telatar99}. Telatar~\cite{Telatar99} proved that 
independent and identically distributed (i.i.d.) Gaussian signaling is 
optimal for i.i.d.\ Rayleigh fading MIMO channels with perfect CSI at the 
receiver. See e.g. \cite{Tulino05} for the case of more sophisticated fading 
models. The assumption of perfect CSI available at the receiver is a 
reasonable assumption if the coherence time is sufficiently long compared to 
the number of transmit antennas. However, this assumption becomes unrealistic 
for mobile communications with short coherence time or a large number of 
transmit antennas. Thus, it is worth considering the assumption of CSI 
available neither to the transmitter nor to the receiver, while the receiver 
is assumed to know the statistical model of the channel perfectly. 
In this paper, this assumption is simply referred to as the no CSI assumption.  

Marzetta and Hochwald~\cite{Marzetta99} considered i.i.d.\ Rayleigh 
block-fading MIMO channels with no CSI, and characterized a class of 
capacity-achieving signaling schemes. In block-fading channels, the 
channel is fixed during one fading block and independently changes at the 
beginning of the next fading block. The assumption of block-fading simplifies 
analyzing the capacity, although it might be an idealized 
assumption.\footnote{
The assumption of block-fading is valid for time-division multiple-access 
(TDMA) schemes. 
} See \cite{Lapidoth03,Moser09} for the capacity of time-varying MIMO 
channels with no CSI. In this paper, we consider block-fading MIMO channels 
with no CSI. 

The capacity-achieving inputs are not i.i.d.\ over space or time for 
block-fading MIMO channels with no CSI~\cite{Marzetta99}. These dependencies 
over space and time make it difficult to calculate the capacity. 
In order to circumvent this difficulty, three kinds of strategies have been  
considered in the literature. A first strategy is to obtain a closed form 
for a lower bound on the capacity by considering unitary space-time 
modulation~\cite{Hochwald00}, although the insight provided by the closed 
form is not very clear. It is possible to calculate a lower bound of the 
capacity numerically for all signal-to-noise ratios 
(SNRs)~\cite{Hassibi02,Moustakas06}, while this task is not necessarily 
easy in terms of computational complexity. 

A second strategy is to consider the high or low SNR limits. 
This strategy can provide an analytical formula of the capacity in return 
for giving up the capacity result in the moderate SNR regime. 
High SNR approximations of the capacity were derived in \cite{Zheng02,Yang13}. 
The approximations tolerate an error of $o(1)$ in the high SNR limit.  
The analytical formula by Zheng and Tse~\cite{Zheng02} provides a useful 
geometric insight, i.e., the capacity of MIMO channels with no CSI has an 
interpretation as sphere packing in the Grassmann manifold, while the capacity 
for the case of perfect CSI available at the receiver has an interpretation 
in terms of sphere packing in the Euclidean space.   
 
The power per information bit $E_{\mathrm{b}}$ required for reliable 
communication is a key performance measure in the low SNR regime. 
Verd\'u~\cite{Verdu02} proved that the SNR per information bit 
$E_{\mathrm{b}}/N_{0}$, with $N_{0}$ denoting noise power, required for MIMO 
channels with no CSI achieves the minimum 
$NE_{\mathrm{b}}/N_{0}=\ln 2\approx -1.59$~dB 
in the low SNR limit, with $N$ denoting the number of receive antennas. 
This result provides a fundamental limit in terms of energy efficiency. 
See \cite{Zheng07,Ray07} for more detailed analysis. 

The last strategy is to analyze the achievable rate of a training-based 
system, which obviously provides a lower bound on the capacity. 
Since accurate channel estimates are assumed to be obtained by training, 
i.i.d.\ signaling over space and time is commonly used for training-based 
systems. This signaling contains practical modulation schemes, such as 
quadrature phase shift keying (QPSK) or quadrature amplitude modulation (QAM). 
Results based on this strategy are less explored than those based on 
the other two strategies. An advantage of the training-based strategy is 
that it is possible to obtain an {\em analytical} bound that can be easily 
evaluated for all SNRs. Hassibi and Hochwald~\cite{Hassibi03} derived an 
analytical lower bound on the achievable rate of a pilot-assisted system, 
called the Hassibi-Hochwald (HH) bound in this paper. Another advantage is 
that it can provide a useful guideline for designing practical 
training-based MIMO systems. In fact, it was shown 
in \cite{Hassibi03} that the optimal number of pilot symbols is equal to 
the number of transmit antennas in terms of their lower bound.  
A weakness is that lower bounds derived by the last strategy might be looser 
than those derived by the first strategy. 
In this paper, we focus on the last strategy and improve 
the existing lower bound of the capacity based on training. 

Hassibi and Hochwald~\cite{Hassibi03} used a method for lower-bounding the 
achievable rate of a pilot-assisted system, developed by M\'edard 
in \cite{Medard00}. As shown in \cite{Takeuchi101}, using this method requires 
the assumption of one-shot channel estimation, under which the decoder regards 
the channel estimates provided by the channel estimator as the true ones. In 
other words, the decoded data symbols are not re-utilized for refining the 
channel estimates. Thus, a lower bound based on training-based systems should 
improve by refining the channel estimates with the decoded data symbols. 

We follow a successive decoding (SD) strategy considered in 
\cite{Li07,Padmanabhan08,Takeuchi101}, in which 
the data symbols decoded in the preceding stages are utilized for refining 
the channel estimates. In the initial channel estimation, the channel 
estimator utilizes pilot signals transmitted by using a fraction of resources.  
As a training-based scheme suitable for the SD strategy, we consider a 
combination of the conventional pilot-assisted scheme 
and a bias-based scheme~\cite{Takeuchi092}. In the bias-based scheme, 
a probabilistic bias of transmitted symbols is used for the initial 
channel estimation, while time-division multiplexed pilot symbols are 
utilized in the pilot-assisted scheme. The bias-based scheme was numerically 
shown to outperform pilot-assisted schemes for practical iterative 
receivers~\cite{Takeuchi111,Takeuchi112,Takeuchi12}. 
The goal of this paper is to derive an analytical bound based on the SD 
strategy with a combination of the pilot-assisted scheme and the bias-based 
scheme.  

\subsection{Contributions \& Methodology} 
The main contribution of this paper is to derive lower bounds on the 
achievable rates of SD receivers in the large-system limit, 
where the number of transmit antennas, the number of receive antennas, 
and coherence time tend to infinity while their ratios are kept constant. 
The lower bounds can be evaluated easily, whereas Padmanabhan et 
al.~\cite{Padmanabhan08} calculated bounds on the corresponding achievable 
rate by numerical simulations. Numerical simulations in this paper show that 
the large-system results can provide a good approximation for not so large 
systems. The derived lower bounds are used to optimize the overhead for 
training. It is shown that negligibly small overhead for training is best in 
terms of the lower bounds. The optimized bound outperforms the HH 
bound~\cite{Hassibi03} for all SNRs. Furthermore, the comparison between our 
bound and the high-SNR approximation of the capacity~\cite{Zheng02,Yang13} 
implies that the high-SNR approximation is valid only for quite high SNR. 

The derivation of the proposed bounds consists of two steps: 
First, the optimal channel estimator is replaced by the linear minimum 
mean-squared error (LMMSE) channel estimator. Since the optimal channel 
estimator is nonlinear in general, the distribution of the channel estimates 
becomes non-Gaussian. This non-Gaussianity makes it difficult to calculate 
the achievable rates of the SD receivers with the optimal channel estimator. 
In order to circumvent this difficulty, we consider a lower bound based on 
LMMSE channel estimation. 


Next, we take the large-system limit to obtain analytical results. 
The large-system limit has been extensively considered in the analysis of 
code-division multiple-access (CDMA) and 
MIMO systems with perfect CSI at the receiver, by using random matrix 
theory~\cite{Tse99,Verdu99,Evans00,Shamai01} and the replica 
method~\cite{Tanaka02,Moustakas03,Mueller04,Guo05,Takeda06,Wen07,Takeuchi082}. 
The advantage of taking the large-system limit is that several performance 
measures, such as mutual information and signal-interference-plus-noise 
ratio (SINR), are expected to be self-averaging, i.e., they converge 
in probability to deterministic values in the large-system limit. 
This self-averaging property allows us to obtain analytical results. 
The large-system limit in previous works for the perfect CSI case may be 
regarded as the limit in which the numbers of transmit and receive antennas 
tend to infinity at the same rate {\em after} taking the long coherence-time 
limit, since the receiver can obtain accurate channel estimates in these 
limits. In order to consider the no CSI case, on the other hand, the 
coherence time {\em and} the numbers of transmit and receive antennas tend to 
infinity at the same rate in this paper. Note that the coherence time must 
also tend to infinity to obtain meaningful results when the number of antennas 
tends to infinity, since there is no point in using transmit antennas more 
than the coherence time~\cite{Marzetta99}.   

We use the replica method to evaluate the achievable rates in the 
large-system limit. The replica method was originally developed in 
statistical physics~\cite{Sherrington75}. See 
\cite{Nishimori01,Mezard87,Fischer91} for the details of 
the replica method. Recently, it has been recognized 
that the replica method is useful for analyzing nonlinear receivers~\cite{Tanaka02,Moustakas03,Mueller04,Guo05,Takeda06,Wen07,Takeuchi082}. 
A weakness of the replica method is that it is based on several non-rigorous 
assumptions in the present time. See \cite{Guerra032,Talagrand06} for a recent 
remarkable progress with respect to the replica method. 

\subsection{Notation} 
For a complex number $z\in\mathbb{C}$, throughout this paper, 
$\mathrm{j}$, $\Re[z]$, $\Im[z]$, 
and $z^{*}$ denote the imaginary unit, the real and imaginary parts of $z$, 
and the complex conjugate of $z$, respectively. 
For a complex matrix $\boldsymbol{A}$, $\boldsymbol{A}^{\mathrm{T}}$, 
$\boldsymbol{A}^{\mathrm{H}}$, $\mathrm{Tr}(\boldsymbol{A})$, and 
$\det\boldsymbol{A}$ represent the transpose, the conjugate transpose, 
the trace, and the determinant of $\boldsymbol{A}$, respectively. 
The vector $\boldsymbol{1}_{n}$ denotes the $n$-dimensional vector whose 
elements are all one. The $n\times n$ identity matrix is denoted by 
$\boldsymbol{I}_{n}$.  
The operator $\otimes$ denotes the Kronecker product of two matrices. 
The matrix $\mathrm{diag}(a_{1},\ldots,a_{n})$ represents 
the diagonal matrix with $a_{i}$ as the $i$th diagonal element. 
$\mathcal{M}_{n}^{+}$ denotes the set of all positive definite $n\times n$ 
Hermitian matrices.  
$\log x$, $\ln x$, 
$\delta(\cdot)$, and $\delta_{a,b}$ denote $\log_{2}x$, $\log_{\mathrm{e}}x$, 
the Dirac delta function, and the Kronecker delta, respectively. 
For random variables 
$X$, $Y$, and $Z$, $I(X;Y|Z)$ denotes the conditional mutual information 
between $X$ and $Y$ given $Z$ with the logarithm to base~$2$. For 
a complex random vector $\boldsymbol{x}$ and a random variable $Y$, 
$\mathrm{cov}[\boldsymbol{x}|Y]$ represents the covariance matrix of 
$\boldsymbol{x}$ given $Y$. 
$\mathcal{CN}(\boldsymbol{m},\boldsymbol{\Sigma})$ 
denotes a proper complex Gaussian distribution with mean~$\boldsymbol{m}$ 
and a covariance matrix~$\boldsymbol{\Sigma}$~\cite{Neeser93}. 
For covariance matrices $\boldsymbol{\Sigma}$ and 
$\tilde{\boldsymbol{\Sigma}}$, 
$D_{a}(\boldsymbol{\Sigma}\|\tilde{\boldsymbol{\Sigma}})$ represents the 
Kullback-Leibler divergence with the logarithm to base~$a$ between 
$\mathcal{CN}(\boldsymbol{0},\boldsymbol{\Sigma})$ and 
$\mathcal{CN}(\boldsymbol{0},\tilde{\boldsymbol{\Sigma}})$. 

As notational convenience for subsets of the natural numbers $\mathbb{N}$, 
we use $[a,b) = \{i\in\mathbb{N}: a\leq i < b\}$ for integers $a$ and 
$b$ ($>a$). The other sets $[a,b]$, $(a,b)$, and so on are defined in the same 
manner. The set $\mathcal{J}\backslash\{j\}=\{j'\in\mathcal{J}: j'\neq j\}$ 
denotes the set obtained by eliminating the element $j$ from a set 
of indices~$\mathcal{J}$. When $\mathcal{J}$ equals the set of all indices, 
$\mathcal{J}\backslash\{j\}$ is simply written as $\backslash j$. 
 
For a set of indices $\mathcal{J}=\{j_{1},\ldots,j_{n}\}$ and 
scalars $\{v_{j}:j\in\mathcal{J}\}$, $\boldsymbol{v}_{\mathcal{J}}$ denotes 
the column vector $\boldsymbol{v}=(v_{j_{1}},\ldots,v_{j_{n}})^{\mathrm{T}}$, 
while $\vec{\boldsymbol{v}}_{\mathcal{J}}$ does the row vector 
$\vec{\boldsymbol{v}}_{\mathcal{J}}=(v_{j_{1}},\ldots,v_{j_{n}})$.  
For example, $\boldsymbol{v}_{[a,b)}=(v_{a},\ldots,v_{b-1})^{\mathrm{T}}$  
and $\boldsymbol{v}_{\mathcal{J}\backslash\{j_{2}\}}=(v_{j_{1}},v_{j_{3}},\ldots,
v_{j_{n}})^{\mathrm{T}}$. Note that 
$\boldsymbol{v}_{\mathcal{J}\backslash\{j\}}$ is written as 
$\boldsymbol{v}_{\backslash j}$ when $\mathcal{J}$ is the set of all indices, 
since $\mathcal{J}\backslash\{j\}$ is abbreviated as $\backslash j$.  
For column vectors $\{\boldsymbol{a}_{j}:j\in\mathcal{J}\}$, similarly, 
$\boldsymbol{A}_{\mathcal{J}}$ denotes the matrix 
$\boldsymbol{A}_{\mathcal{J}}=(\boldsymbol{a}_{j_{1}},\ldots,
\boldsymbol{a}_{j_{n}})$. For example, 
$\boldsymbol{A}_{[a,b)}=(\boldsymbol{a}_{a},\ldots,\boldsymbol{a}_{b-1})$ 
and $\boldsymbol{A}_{\mathcal{J}\backslash\{j_{2}\}}
=(\boldsymbol{a}_{j_{1}},\boldsymbol{a}_{j_{3}},\ldots,
\boldsymbol{a}_{j_{n}})$. 
We use symbols with tildes and hats to represent random variables for 
postulated (or virtual) channels and estimates of random variables, 
respectively. Underlined symbols are used to represent random variables for 
decoupled channels. 

The remainder of this paper is organized as follows: A Rayleigh block-fading 
MIMO channel is introduced in Section~\ref{sec_channel_model}. 
The achievable rates of SD receivers based on LMMSE channel estimation are  
formulated in Section~\ref{sec_receivers}. The main results of this 
paper are presented in Section~\ref{sec_main_results}. The obtained analytical 
bounds are compared to existing results in Section~\ref{sec_comparison}. We 
conclude this paper in Section~\ref{sec_conclusion}. The derivation of the 
main results is summarized in appendices.

\section{Channel Model} \label{sec_channel_model}
\subsection{MIMO Channel} 
A narrowband MIMO system with $M$ transmit antennas and $N$ receive 
antennas is considered. 
We assume block-fading with coherence time $T_{\mathrm{c}}$, i.e., 
the channel matrix $\boldsymbol{H}\in\mathbb{C}^{N\times M}$ is kept 
constant during one fading block consisting of $T_{\mathrm{c}}$ symbol periods, 
and at the beginning of the next fading block the channel matrix is 
independently sampled from a distribution. The received vector 
$\boldsymbol{y}_{t}\in\mathbb{C}^{N}$ in the $t$th symbol period within a 
fading block is given by 
\begin{equation} \label{MIMO} 
\boldsymbol{y}_{t} = \frac{1}{\sqrt{M}}\boldsymbol{H}\boldsymbol{x}_{t} 
+ \boldsymbol{n}_{t}, 
\quad t=1,\ldots,T_{\mathrm{c}}, 
\end{equation}
where $\boldsymbol{x}_{t}=(x_{1,t},\ldots,x_{M,t})^{\mathrm{T}}$ and 
$\boldsymbol{n}_{t}\sim\mathcal{CN}(\boldsymbol{0},N_{0}\boldsymbol{I}_{N})$ 
denote the transmitted vector in the $t$th symbol period and an 
additive white Gaussian noise (AWGN) vector with a covariance matrix 
$N_{0}\boldsymbol{I}_{N}$, respectively. 
The MIMO channel~(\ref{MIMO}) can be represented in matrix form as 
\begin{equation} \label{matrix_MIMO}
\boldsymbol{Y} = \frac{1}{\sqrt{M}}\boldsymbol{H}\boldsymbol{X} + 
\boldsymbol{N}, 
\end{equation}
with $\boldsymbol{Y}=(\boldsymbol{y}_{1},\ldots,
\boldsymbol{y}_{T_{\mathrm{c}}})$, 
$\boldsymbol{X}=(\boldsymbol{x}_{1},\ldots,\boldsymbol{x}_{T_{\mathrm{c}}})$, 
and $\boldsymbol{N}=(\boldsymbol{n}_{1},\ldots,
\boldsymbol{n}_{T_{\mathrm{c}}})$. 

For the simplicity of analysis, we assume i.i.d.\ Rayleigh fading MIMO 
channels, i.e., the channel matrix $\boldsymbol{H}$ has mutually independent 
entries, and each entry $h_{n,m}=(\boldsymbol{H})_{n,m}$ is 
drawn from the circularly symmetric complex Gaussian (CSCG) distribution 
$\mathcal{CN}(0,1)$ with unit variance. Note that the assumption of i.i.d.\ 
Rayleigh fading might be an idealized assumption since there can be 
correlations between the elements of the channel matrix in practice. 

We impose a power constraint 
\begin{equation} \label{weak_power_constraint} 
\frac{1}{MT_{\mathrm{c}}}
\sum_{m=1}^{M}\sum_{t=1}^{T_{\mathrm{c}}}\mathbb{E}\left[
 |x_{m,t}|^{2}
\right]\leq P, 
\end{equation}
for $P>0$. Marzetta and Hochwald~\cite{Marzetta99} proved that the capacity 
does not decrease even if the power constraint is strengthened to a power 
constraint on each transmitted symbol, 
\begin{equation} \label{power_constraint}
\mathbb{E}\left[
 |x_{m,t}|^{2}
\right] \leq P. 
\end{equation} 
The former power constraint~(\ref{weak_power_constraint}) allows us to use 
power allocation over space and time, whereas 
the latter power constraint~(\ref{power_constraint}) does not. In this paper, 
we only consider the latter power constraint~(\ref{power_constraint}), which 
simplifies the analysis. 

\subsection{Training-Based Transmission}
We assume that neither the transmitter nor the receiver has CSI. 
More precisely, only the statistical properties of 
the MIMO channel~(\ref{MIMO}) are assumed to be known to the receiver. 
The previous works~\cite{Zheng02,Hassibi03} showed that pilot-assisted 
channel estimation can achieve the capacity in the leading order of 
SNR in the high SNR regime, i.e., the full spatial multiplexing gain, while 
the obtained lower bounds are loose in the low-to-moderate SNR regime. 
Channel estimation based on pilot information is also considered in this 
paper. The main difference between the previous works and this paper appears 
in the receiver structure. 
We consider joint channel and data estimation based on SD, whereas in the 
previous works data symbols decoded successfully were not utilized for 
refining channel estimates. 

One fading block is decomposed into the training phase 
$\mathcal{T}_{T_{\mathrm{tr}}}
=\{1,\ldots,T_{\mathrm{tr}}\}$ and the communication 
phase $\mathcal{C}_{T_{\mathrm{tr}}+1}=
\{T_{\mathrm{tr}}+1,\ldots,T_{\mathrm{c}}\}$, which consist 
of the first $T_{\mathrm{tr}}$ symbol periods and of the remaining 
$(T_{\mathrm{c}}-T_{\mathrm{tr}})$ symbol periods, respectively. 
The transmitter sends pilot symbol vectors in the training phase, and 
transmits data symbol vectors in the communication phase. Therefore, the 
transmitted vector $\boldsymbol{x}_{t}$ is assumed to be known to the receiver 
for $t\in\mathcal{T}_{T_{\mathrm{tr}}}$. 
For simplicity, we assume that the pilot symbol matrix  
$\boldsymbol{X}_{\mathcal{T}_{T_{\mathrm{tr}}}}=
(\boldsymbol{x}_{1},\ldots,\boldsymbol{x}_{\mathrm{Tr}})
\in\mathbb{C}^{M\times T_{\mathrm{tr}}}$ has 
zero-mean i.i.d.\ entries with i.i.d.\ real and imaginary parts.  
Furthermore, we assume that each pilot symbol satisfies 
$\mathbb{E}[|x_{m,t}|^{2}]=P$ for $t\in\mathcal{T}_{T_{\mathrm{tr}}}$,  
since the accuracy of channel estimation should improve as the power of 
pilot symbols increases. The transmission of i.i.d.\ data symbols can  
achieve the capacity of the MIMO channel~(\ref{MIMO}) with perfect CSI at 
the receiver. If accurate channel estimates are obtained by joint channel 
and data estimation, thus, i.i.d.\ signaling should be a reasonable option  
for training-based transmissions.  
We assume that the data symbols 
$\{x_{m,t}:t\in\mathcal{C}_{T_{\mathrm{tr}}+1}\}$ are 
i.i.d.\ random variables with i.i.d.\ real and imaginary parts 
for all $m$ and $t\in\mathcal{C}_{T_{\mathrm{tr}}+1}$. 
Note that zero-mean is not assumed for the data symbols. 
Under this assumption, the achievable rate is monotonically increasing 
with the power of each data symbol. We hereinafter let 
$\mathbb{E}[|x_{m,t}|^{2}]=P$.  
 
In this paper, we consider a biased signaling scheme, in which the mean 
$\mathbb{E}[x_{m,t}]=\theta_{m,t}$ of the data symbol for 
$t\in\mathcal{C}_{T_{\mathrm{tr}}+1}$ 
is biased while the long-term average $(T_{\mathrm{c}}-T_{\mathrm{tr}})^{-1}
\sum_{t=T_{\mathrm{tr}}+1}^{T_{\mathrm{c}}}\theta_{m,t}$ 
tends to zero as $T_{\mathrm{c}}\rightarrow\infty$. In order to apply the 
replica method, we assume that $\{\Re[\theta_{m,t}],
\Im[\theta_{m,t}]:\hbox{for all $m$, $t$}\}$ are independently drawn 
from a zero-mean hyperprior probability density function (pdf)\footnote{
When $\theta_{m,t}$ is discrete, $p(\theta)$ denotes a probability 
mass function (pmf). 
} $p(\theta)$ with variance $\sigma_{\theta}^{2}/2$. 
The transmitter informs the receiver in advance about the bias matrix  
$\boldsymbol{\Theta}=(\boldsymbol{O},\boldsymbol{\theta}_{T_{\mathrm{tr}}+1},
\ldots,\boldsymbol{\theta}_{T_{\mathrm{c}}})
\in\mathbb{C}^{M\times T_{\mathrm{c}}}$, 
with $\boldsymbol{\theta}_{t}=(\theta_{1,t},\ldots,\theta_{M,t})^{\mathrm{T}}$. 
In other words, $\boldsymbol{\Theta}$ is assumed to be known to the receiver.  
The biased signaling can reduce the overhead for training~\cite{Takeuchi092} 
compared to the conventional pilot-assisted schemes. 
We present two examples of biased signaling: biased QPSK and 
biased Gaussian signaling. See \cite{Takeuchi111,Takeuchi112,Takeuchi12} for  
implementations of biased QPSK. 

\begin{example}[Biased QPSK]
For $\Re[x_{m,t}], \Im[x_{m,t}]\in\{\pm\sqrt{P/2}\}$, 
the prior pmf of $x_{m,t}$ for biased QPSK is given by  
\begin{equation}
p(x_{m,t}|\theta_{m,t}) 
= \frac{1+\Re[\theta_{m,t}]/\Re[x_{m,t}]}{2}
\frac{1+\Im[\theta_{m,t}]/\Im[x_{m,t}]}{2}.  
\end{equation}
The non-negativity of probability restricts the support of the hyperprior pdf 
$p(\theta)$ to $\Re[\theta_{m,t}]\in[-\sqrt{P/2},\sqrt{P/2}]$ and 
$\Im[\theta_{m,t}]\in[-\sqrt{P/2},\sqrt{P/2}]$. It is straightforward 
to check that $\mathbb{E}[x_{m,t}|\theta_{m,t}]=\theta_{m,t}$ and 
$\mathbb{E}[|x_{m,t}|^{2}|\theta_{m,t}]=P$. 
\end{example}
\begin{example}[Biased Gaussian Signaling]
The prior pdf of $x_{m,t}\in\mathbb{C}$ for biased Gaussian signaling is 
given by 
\begin{equation}
p(x_{m,t}|\theta_{m,t}) = \frac{1}{\pi(P-|\theta_{m,t}|^{2})}
\mathrm{e}^{-\frac{|x_{m,t}-\theta_{m,t}|^{2}}{P-|\theta_{m,t}|^{2}}}. 
\end{equation}
Note that the support of the hyperprior pdf $p(\theta)$ is restricted to 
$|\theta_{m,t}|<\sqrt{P}$, due to the positivity of variance.  
\end{example}

The main results presented in this paper hold for a general prior of $x_{m,t}$ 
with finite moments. The performance for the biased Gaussian signaling 
corresponds to a performance bound for multilevel modulation with trellis 
shaping~\cite{Forney92,Takeuchi102}. 

\section{Receivers} \label{sec_receivers} 
\subsection{Successive Decoding} 
\begin{figure}[t]
\begin{center}
\includegraphics[width=0.5\hsize]{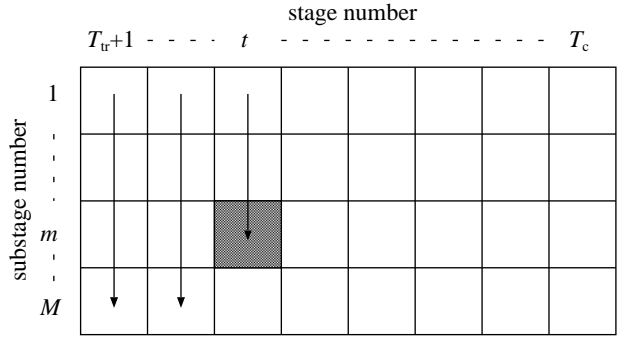}
\caption{
Successive decoding. 
}
\label{fig1} 
\end{center}
\end{figure}

We consider an SD receiver~\cite{Li07,Padmanabhan08} (See Fig.~\ref{fig1}). 
The data symbol vectors $\{\boldsymbol{x}_{t}\}$ are decoded in the order 
$t=T_{\mathrm{tr}}+1,\ldots,T_{\mathrm{c}}$. In stage~$t$, the matrix 
$\boldsymbol{X}_{(T_{\mathrm{tr}},t)}=
(\boldsymbol{x}_{T_{\mathrm{tr}}+1},\ldots,\boldsymbol{x}_{t-1})
\in\mathbb{C}^{M\times(t-T_{\mathrm{tr}}-1)}$ contains 
the data symbol vectors decoded in the preceding stages. 
Stage~$t$ consists of $M$ substages, in which the elements  
$\{x_{m,t}\}$ are decoded in the order $m=1,\ldots,M$. 
In substage~$m$ within stage~$t$, the vector 
$\boldsymbol{x}_{[1,m),t}=(x_{1,t},\ldots,x_{m-1,t})^{\mathrm{T}}
\in\mathbb{C}^{m-1}$ consists of the data symbols decoded in the preceding 
substages. The channel estimator utilizes the data symbols 
$\boldsymbol{X}_{(T_{\mathrm{tr}},t)}$ decoded in the 
preceding stages, along with the received matrix 
$\boldsymbol{Y}_{\backslash t}\in\mathbb{C}^{N\times(T_{\mathrm{c}}-1)}$ 
and the pilot information 
$\{\boldsymbol{X}_{\mathcal{T}_{T_{\mathrm{tr}}}}, 
\boldsymbol{\Theta}_{(t,T_{\mathrm{c}}]}\}$, 
in which the matrices $\boldsymbol{Y}_{\backslash t}$ and 
$\boldsymbol{\Theta}_{(t,T_{\mathrm{c}}]}\in\mathbb{C}^{M\times
(T_{\mathrm{c}}-t)}$ are obtained by 
eliminating the $t$th column vector from the received matrix 
$\boldsymbol{Y}$ and the first $t$ column vectors from the 
bias matrix $\boldsymbol{\Theta}$, respectively. 
We write the information used for channel estimation in stage~$t$
as $\mathcal{I}_{t}=\{\bar{\boldsymbol{X}}_{\backslash t},
\boldsymbol{Y}_{\backslash t}\}$, with 
$\bar{\boldsymbol{X}}_{\backslash t}=(\boldsymbol{X}_{\mathcal{T}_{T_{\mathrm{tr}}}},
\boldsymbol{X}_{(T_{\mathrm{tr}},t)},\boldsymbol{\Theta}_{(t,T_{\mathrm{c}}]})
\in\mathbb{C}^{M\times(T_{\mathrm{c}}-1)}$. In substage~$m$ within stage~$t$, 
the detector with successive interference cancellation (SIC) uses the data 
symbols $\boldsymbol{x}_{[1,m),t}$ decoded 
in the preceding substages and the bias vector 
$\boldsymbol{\theta}_{[m,M],t}=(\theta_{m,t},\ldots,
\theta_{M,t})^{\mathrm{T}}$ to subtract inter-stream 
interference from the received vector $\boldsymbol{y}_{t}$, and then 
perform multiuser detection (MUD). 

Let us define the constrained capacity of the MIMO system based on the pilot 
information $\{\boldsymbol{X}_{\mathcal{T}_{T_{\mathrm{tr}}}},\boldsymbol{\Theta}\}$ 
as the conditional mutual information per symbol period between all 
data symbol vectors $\{\boldsymbol{x}_{t}:t\in\mathcal{C}_{T_{\mathrm{tr}}+1}\}$ 
and the received matrix $\boldsymbol{Y}$ conditioned on the pilot symbol 
matrix $\boldsymbol{X}_{\mathcal{T}_{T_{\mathrm{tr}}}}$ and the bias matrix 
$\boldsymbol{\Theta}$~\cite{Cover06} 
\begin{equation} \label{capacity} 
C = \frac{1}{T_{\mathrm{c}}}
I(\{\boldsymbol{x}_{t}:t\in\mathcal{C}_{T_{\mathrm{tr}}+1}\};\boldsymbol{Y}|
\boldsymbol{X}_{\mathcal{T}_{T_{\mathrm{tr}}}},\boldsymbol{\Theta}). 
\end{equation}
It is straightforward to confirm that the optimal SD receiver can 
achieve the constrained capacity~(\ref{capacity}).  
Applying the chain rule for mutual information to (\ref{capacity}) 
repeatedly~\cite{Cover06}, we obtain   
\begin{IEEEeqnarray}{rl}
C =& \frac{1}{T_{\mathrm{c}}}\sum_{t=T_{\mathrm{tr}}+1}^{T_{\mathrm{c}}}
I(\boldsymbol{x}_{t};\boldsymbol{Y}| \boldsymbol{X}_{\mathcal{T}_{T_{\mathrm{tr}}}},
\boldsymbol{\Theta},\boldsymbol{X}_{(T_{\mathrm{tr}},t)}) \nonumber \\ 
=& \frac{1}{T_{\mathrm{c}}}\sum_{t=T_{\mathrm{tr}}+1}^{T_{\mathrm{c}}}
I(\boldsymbol{x}_{t};\boldsymbol{y}_{t}| \mathcal{I}_{t},
\boldsymbol{\Theta}_{(T_{\mathrm{tr}},t)}, 
\boldsymbol{\theta}_{t}) \nonumber \\
=& \frac{1}{T_{\mathrm{c}}}\sum_{t=T_{\mathrm{tr}}+1}^{T_{\mathrm{c}}}
\sum_{m=1}^{M}
I(x_{m,t};\boldsymbol{y}_{t}| \mathcal{I}_{t},
\boldsymbol{x}_{[1,m),t},
\boldsymbol{\theta}_{[m,M],t}), \label{capacity_SD} 
\end{IEEEeqnarray}
with $\boldsymbol{\Theta}_{(T_{\mathrm{tr}},t)}=(
\boldsymbol{\theta}_{T_{\mathrm{tr}}+1},\ldots,\boldsymbol{\theta}_{t-1})$. 
In the derivation of the second 
equality, we have used the fact that $\boldsymbol{x}_{t}$ and 
$\boldsymbol{Y}_{\backslash t}$ are independent of each other, 
due to the i.i.d.\ assumption of 
the data symbols. In the last expression, we have omitted conditioning 
with respect to $\boldsymbol{\Theta}_{(T_{\mathrm{tr}},t)}$ and 
$\boldsymbol{\theta}_{[1,m),t}=(\theta_{1,t},\ldots,
\theta_{m-1,t})^{\mathrm{T}}$, which are not utilized by the receiver in 
substage~$m$ within stage~$t$ since they are the 
parameters of the known data symbols $\boldsymbol{X}_{(T_{\mathrm{tr}},t)}$ and 
$\boldsymbol{x}_{[1,m),t}$. 
For notational simplicity, this omission is applied throughout this paper. 
Note that $\boldsymbol{\Theta}_{(T_{\mathrm{tr}},t)}$ and 
$\boldsymbol{\theta}_{[1,m),t}$ affect the achievable rate~(\ref{capacity_SD}). 
Expression~(\ref{capacity_SD}) implies that the SD scheme 
results in no loss of information if the detector with SIC can achieve the 
mutual information $I(x_{m,t};\boldsymbol{y}_{t}| \mathcal{I}_{t},
\boldsymbol{x}_{[1,m),t},\boldsymbol{\theta}_{[m,M],t})$ in substage~$m$ 
within stage~$t$. 

It is difficult to evaluate the mutual information 
$I(x_{m,t};\boldsymbol{y}_{t}| \mathcal{I}_{t},  
\boldsymbol{x}_{[1,m),t},
\boldsymbol{\theta}_{[m,M],t})$ exactly. Instead, we derive a lower 
bound based on LMMSE channel estimation. 
We first introduce the optimal 
channel estimator and then define the LMMSE channel estimator. 

\subsection{Channel Estimators} \label{sec_lower_bound}
\subsubsection{Optimal Channel Estimator} 
We focus on stage~$t$ in this section. 
The optimal channel estimator uses the information $\mathcal{I}_{t}$ to 
estimate the channel matrix $\boldsymbol{H}$, and sends   
the joint posterior pdf    
\begin{equation} \label{posterior_H} 
p(\boldsymbol{H}|\mathcal{I}_{t}) = \frac{
 p(\boldsymbol{Y}_{\backslash t} | \boldsymbol{H},
 \bar{\boldsymbol{X}}_{\backslash t})p(\boldsymbol{H})
}
{
 \int p(\boldsymbol{Y}_{\backslash t} | \boldsymbol{H},
 \bar{\boldsymbol{X}}_{\backslash t})
 p(\boldsymbol{H})d\boldsymbol{H} 
}, 
\end{equation}
to the detector with SIC. In (\ref{posterior_H}), 
the pdf $p(\boldsymbol{Y}_{\backslash t} | \boldsymbol{H},
\bar{\boldsymbol{X}}_{\backslash t})$ is decomposed into the product of pdfs 
$p(\boldsymbol{Y}_{\backslash t} | \boldsymbol{H},
\bar{\boldsymbol{X}}_{\backslash t})=  
\prod_{t'=1}^{t-1}p(\boldsymbol{y}_{t'} | \boldsymbol{H},\boldsymbol{x}_{t'})
\prod_{t'=t+1}^{T_{\mathrm{c}}}
p(\boldsymbol{y}_{t'} | \boldsymbol{H},\boldsymbol{\theta}_{t'})$,  
given by 
\begin{equation} \label{marginal_channel} 
p(\boldsymbol{y}_{t'} | \boldsymbol{H},\boldsymbol{\theta}_{t'}) = 
\int p(\boldsymbol{y}_{t'} | \boldsymbol{H},\boldsymbol{x}_{t'})
p(\boldsymbol{x}_{t'}|\boldsymbol{\theta}_{t'})d\boldsymbol{x}_{t'}, 
\end{equation}
where $p(\boldsymbol{y}_{t'} | \boldsymbol{H},\boldsymbol{x}_{t'})$ represents 
the MIMO channel~(\ref{MIMO}). 
Note that the joint posterior pdf~(\ref{posterior_H}) is decomposed into the 
product $\prod_{n=1}^{N}p(\vec{\boldsymbol{h}}_{n}|\mathcal{I}_{t})$ of 
the marginal posterior pdfs, with 
$\vec{\boldsymbol{h}}_{n}\in\mathbb{C}^{1\times M}$ denoting the $n$th row 
vector of $\boldsymbol{H}$, due to the assumption of i.i.d.\ fading. 

The optimal channel estimator is nonlinear in general, which 
makes it difficult to analyze detectors with SIC, 
while it is possible to evaluate the performance of the optimal channel 
estimator. In order to circumvent this difficulty, we reduce the optimal 
channel estimator to an LMMSE channel estimator by considering a virtual 
MIMO channel. 

\subsubsection{LMMSE Channel Estimator} \label{sec_LMMSE_channel_estimator}  
We use M\'edard's method~\cite{Medard00} to replace the MIMO 
channels~(\ref{MIMO}) for $t'=t+1,\ldots,T_{\mathrm{c}}$ by 
virtual MIMO channels  
\begin{equation} \label{virtual_MIMO} 
\tilde{\boldsymbol{y}}_{t'} 
= \frac{1}{\sqrt{M}}\boldsymbol{H}\boldsymbol{\theta}_{t'}
+ \boldsymbol{w}_{t'} + \boldsymbol{n}_{t'}, 
\end{equation}
where $\boldsymbol{w}_{t'}\in\mathbb{C}^{N}$ denotes a CSCG random vector 
with the covariance matrix $(P-\sigma_{t'}^{2})\boldsymbol{I}_{N}$, with 
$\sigma_{t'}^{2}= M^{-1}\sum_{m=1}^{M}|\theta_{m,t'}|^{2}$. 
The virtual MIMO channel~(\ref{virtual_MIMO}) is obtained by extracting 
the term $M^{-1/2}\boldsymbol{H}(\boldsymbol{x}_{t'}
-\boldsymbol{\theta}_{t'})$ from the first term of the right-hand side 
(RHS) in the original MIMO channel~(\ref{MIMO}) and then replacing it by  
the AWGN term $\boldsymbol{w}_{t'}$ with the covariance matrix 
$\mathrm{cov}[M^{-1/2}\boldsymbol{H}(\boldsymbol{x}_{t'}
-\boldsymbol{\theta}_{t'})|\boldsymbol{\theta}_{t'}]$. 
This replacement implies that information about the channel matrix included 
in $\boldsymbol{H}(\boldsymbol{x}_{t'}-\boldsymbol{\theta}_{t'})$ is 
discarded. Thus, channel estimation based on the virtual MIMO 
channel~(\ref{virtual_MIMO}) should be inferior to that based on 
the original MIMO channel~(\ref{MIMO}). In other words, 
the mutual information $I(x_{m,t};\boldsymbol{y}_{t}| 
\mathcal{I}_{t}, \boldsymbol{x}_{[1,m),t},
\boldsymbol{\theta}_{[m,M],t})$ should be bounded from below by 
\begin{equation} \label{lower_bound_mutual_inf} 
I(x_{m,t};\boldsymbol{y}_{t}| 
\mathcal{I}_{t}, \boldsymbol{x}_{[1,m),t},
\boldsymbol{\theta}_{[m,M],t}) \geq 
I(x_{m,t};\boldsymbol{y}_{t}| 
\tilde{\mathcal{I}}_{t}, \boldsymbol{x}_{[1,m),t},
\boldsymbol{\theta}_{[m,M],t}), 
\end{equation}
which denotes the constrained capacity  of the MIMO channel~(\ref{MIMO}) in 
symbol period~$t$ with side information 
$\tilde{\mathcal{I}}_{t}=
\{\bar{\boldsymbol{X}}_{\backslash t},\tilde{\boldsymbol{Y}}_{\backslash t}\}$, 
$\boldsymbol{x}_{[1,m),t}$, and 
$\boldsymbol{\theta}_{[m,M],t}$, 
in which $\tilde{\boldsymbol{Y}}_{\backslash t}=
(\boldsymbol{y}_{1},\ldots,\boldsymbol{y}_{t-1},\tilde{\boldsymbol{y}}_{t+1},
\ldots,\tilde{\boldsymbol{y}}_{T_{\mathrm{c}}})$ contains the received 
vectors of the virtual MIMO channel~(\ref{virtual_MIMO}) in the 
last $(T_{\mathrm{c}}-t)$ elements while the first $(t-1)$ elements of 
$\tilde{\boldsymbol{Y}}_{\backslash t}$ are the same as those of the original 
one $\boldsymbol{Y}_{\backslash t}$. From (\ref{MIMO}) and 
(\ref{virtual_MIMO}), the matrix $\tilde{\boldsymbol{Y}}_{\backslash t}$ 
is explicitly given by 
\begin{equation} \label{MIMO_channel_estimation}
\tilde{\boldsymbol{Y}}_{\backslash t} = 
\frac{1}{\sqrt{M}}\boldsymbol{H}\bar{\boldsymbol{X}}_{\backslash t} 
+ (\boldsymbol{O}, \boldsymbol{W}_{(t,T_{\mathrm{c}}]}) 
+ \boldsymbol{N}_{\backslash t}, 
\end{equation}
with $\boldsymbol{W}_{(t,T_{\mathrm{c}}]}=(\boldsymbol{w}_{t+1},\ldots, 
\boldsymbol{w}_{T_{\mathrm{c}}})\in\mathbb{C}^{N\times (T_{\mathrm{c}}-t)}$. 
In (\ref{MIMO_channel_estimation}), the matrix 
$\boldsymbol{N}_{\backslash t}\in\mathbb{C}^{N\times(T_{\mathrm{c}}-1)}$ 
is obtained by eliminating the $t$th column vector from the noise 
matrix $\boldsymbol{N}$. 

Let us consider channel estimation based on the information 
$\tilde{\mathcal{I}}_{t}$. The optimal channel estimator for this case 
constructs the joint posterior pdf $p(\boldsymbol{H}|\tilde{\mathcal{I}}_{t})$ 
and feeds it to the detector with SIC. 
The joint posterior pdf of $\boldsymbol{H}$ given $\tilde{\boldsymbol{I}}_{t}$ 
is defined by (\ref{posterior_H}) in which 
the pdf~(\ref{marginal_channel}) for $t'=t+1,\ldots,T_{\mathrm{c}}$ is 
replaced by 
\begin{equation}
p(\tilde{\boldsymbol{y}}_{t'} | \boldsymbol{H},
\boldsymbol{\theta}_{t'}) 
= \int p(\tilde{\boldsymbol{y}}_{t'} | \boldsymbol{H},
\boldsymbol{\theta}_{t'},\boldsymbol{w}_{t'})
p(\boldsymbol{w}_{t'} | \boldsymbol{\theta}_{t'})
d\boldsymbol{w}_{t'}, 
\end{equation}
where $ p(\tilde{\boldsymbol{y}}_{t} | \boldsymbol{H},
\boldsymbol{\theta}_{t'},\boldsymbol{w}_{t'})$ represents 
the virtual MIMO channel~(\ref{virtual_MIMO}). 
A straightforward calculation indicates that the joint posterior pdf 
$p(\boldsymbol{H} | \tilde{\mathcal{I}}_{t})$ is a proper complex 
Gaussian pdf with mean $\hat{\boldsymbol{H}}_{t}\in\mathbb{C}^{N\times M}$ 
and covariance $\mathrm{cov}[(\vec{\boldsymbol{h}}_{1},\ldots,
\vec{\boldsymbol{h}}_{N})^{\mathrm{T}}|\tilde{\mathcal{I}}_{t}]=
\boldsymbol{I}_{N}\otimes\boldsymbol{\Xi}_{t}$, 
given by 
\begin{equation} \label{LMMSE_estimate} 
\hat{\boldsymbol{H}}_{t} = \frac{1}{\sqrt{M}}\left[
 \sum_{t'=1}^{t-1}\frac{\boldsymbol{y}_{t'}\boldsymbol{x}_{t'}^{\mathrm{H}}}{N_{0}}
 + \sum_{t'=t+1}^{T_{\mathrm{c}}}\frac{\tilde{\boldsymbol{y}}_{t'}
 \boldsymbol{\theta}_{t'}^{\mathrm{H}}}{P-\sigma_{t'}^{2}+N_{0}}
\right]\boldsymbol{\Xi}_{t}, 
\end{equation}
\begin{equation} \label{error_covariance} 
\boldsymbol{\Xi}_{t} = \left[
 \boldsymbol{I}_{M} + \frac{1}{M}\left( 
  \sum_{t'=1}^{t-1}\frac{\boldsymbol{x}_{t'}\boldsymbol{x}_{t'}^{\mathrm{H}}}{N_{0}}
  + \sum_{t'=t+1}^{T_{\mathrm{c}}}\frac{\boldsymbol{\theta}_{t'}
  \boldsymbol{\theta}_{t'}^{\mathrm{H}}}{P-\sigma_{t'}^{2}+N_{0}}
 \right) 
\right]^{-1}. 
\end{equation}
The posterior mean $\hat{\boldsymbol{H}}_{t}$ coincides with the 
LMMSE estimator of $\boldsymbol{H}$ based on the received 
matrix $\tilde{\boldsymbol{Y}}_{\backslash t}$ and the known information 
$\bar{\boldsymbol{X}}_{\backslash t}$. Furthermore, $\boldsymbol{\Xi}_{t}$ is 
equal to the error covariance matrix of the LMMSE estimator for 
each row vector of $\boldsymbol{H}$. Thus, we refer to the optimal 
channel estimator for the virtual MIMO channel~(\ref{virtual_MIMO}) as 
the LMMSE channel estimator. 

Note that the linear filter given by (\ref{LMMSE_estimate}) provides the 
LMMSE estimates of $\boldsymbol{H}$ for the original MIMO channel~(\ref{MIMO}). 
One should not confuse the LMMSE channel estimator for the virtual MIMO 
channel~(\ref{virtual_MIMO}) with that for the original MIMO 
channel~(\ref{MIMO}). The former, which is considered in this paper, is the 
optimal channel estimator for the virtual MIMO channel~(\ref{virtual_MIMO}), 
while the latter is a suboptimal channel estimator for the original MIMO 
channel~(\ref{MIMO}).

\subsection{Detectors}
\subsubsection{Optimal Detector}  
We focus on substage~$m$ within stage~$t$ and define the optimal detector with 
SIC, which achieves the lower bound~(\ref{lower_bound_mutual_inf}).  
The optimal detector with SIC feeds to the associated decoder the posterior 
pdf\footnote{The marginal posterior pdf~(\ref{posterior_x}) is 
replaced by the posterior pmf of $x_{m,t}$ 
if $x_{m,t}$ is a discrete random variable.} of $x_{m,t}$ based on the 
knowledge about the received vector $\boldsymbol{y}_{t}$, the data symbols 
$\boldsymbol{x}_{[1,m),t}$ decoded in the preceding substages, 
the bias $\boldsymbol{\theta}_{[m,M],t}$ for the unknown data 
symbols $\boldsymbol{x}_{[m,M],t}=(x_{m,t},\ldots,x_{M,t})^{\mathrm{T}}$, 
and the joint posterior pdf $p(\boldsymbol{H}|\tilde{\mathcal{I}}_{t})$ 
provided by the LMMSE channel estimator, given by    
\begin{equation} \label{posterior_x} 
p(x_{m,t} | \boldsymbol{y}_{t}, \tilde{\mathcal{I}}_{t},
\boldsymbol{x}_{[1,m),t},\boldsymbol{\theta}_{[m,M],t})
= \frac{ 
 \int p(\boldsymbol{y}_{t} | \boldsymbol{x}_{t},\tilde{\mathcal{I}}_{t}) 
 p(\boldsymbol{x}_{[m,M],t}|\boldsymbol{\theta}_{[m,M],t})
 d\boldsymbol{x}_{(m,M],t} 
}
{
  \int p(\boldsymbol{y}_{t} | 
  \boldsymbol{x}_{t},\tilde{\mathcal{I}}_{t}) 
  p(\boldsymbol{x}_{[m,M],t}|
  \boldsymbol{\theta}_{[m,M],t})
 d\boldsymbol{x}_{[m,M],t}
}
\end{equation}
with $\boldsymbol{x}_{(m,M],t}=
(x_{m+1,t},\ldots,x_{M,t})^{\mathrm{T}}\in\mathbb{C}^{M-m}$. 
In (\ref{posterior_x}), 
the pdf~$p(\boldsymbol{y}_{t} | \boldsymbol{x}_{t},\tilde{\mathcal{I}}_{t})$ 
is given by 
\begin{equation} \label{conditional_channel} 
p(\boldsymbol{y}_{t} | \boldsymbol{x}_{t},\tilde{\mathcal{I}}_{t}) 
= \int p(\boldsymbol{y}_{t} | \boldsymbol{H},\boldsymbol{x}_{t}) 
p(\boldsymbol{H} | \tilde{\mathcal{I}}_{t})d\boldsymbol{H}, 
\end{equation}
where $p(\boldsymbol{y}_{t} | \boldsymbol{H},\boldsymbol{x}_{t})$ represents 
the MIMO channel~(\ref{MIMO}).  
The use of SIC appears in the pdf~(\ref{conditional_channel}), which is a 
proper complex Gaussian pdf with mean 
$M^{-1/2}\hat{\boldsymbol{H}}_{t}\boldsymbol{x}_{t}$ and 
covariance $(N_{0}+M^{-1}\boldsymbol{x}_{t}^{\mathrm{H}}\boldsymbol{\Xi}_{t}
\boldsymbol{x}_{t})\boldsymbol{I}_{N}$.  
Expression~(\ref{posterior_x}) implies that the optimal detector with SIC 
subtracts the known inter-stream interference $M^{-1/2}\hat{\boldsymbol{H}}_{t} 
({\boldsymbol{x}_{[1,m),t}}^{\mathrm{T}},0,
{\boldsymbol{\theta}_{(m,M],t}}^{\mathrm{T}})^{\mathrm{T}}$  
from the received vector $\boldsymbol{y}_{t}$, with 
$\boldsymbol{\theta}_{(m,M],t}=
(\theta_{m+1,t},\ldots,\theta_{M,t})^{\mathrm{T}}$, and then mitigates residual 
inter-stream interference by performing the optimal nonlinear MUD. 

Let $\tilde{x}_{m,t}\in\mathbb{C}$ denote a random variable following the 
marginal posterior pdf~(\ref{posterior_x}). 
Since the lower bound~(\ref{lower_bound_mutual_inf}) is equal to 
the mutual information $I(x_{m,t};\tilde{x}_{m,t}|\tilde{\mathcal{I}}_{t},
\boldsymbol{x}_{[1,m),t},
\boldsymbol{\theta}_{[m,M],t})$, 
the achievable rate~(\ref{capacity_SD}) of the optimal SD receiver is 
bounded from below by 
\begin{equation} \label{capacity_nonlinear} 
C \geq \frac{1}{T_{\mathrm{c}}}\sum_{t=T_{\mathrm{tr}}+1}^{T_{\mathrm{c}}}
\sum_{m=1}^{M}
I(x_{m,t};\tilde{x}_{m,t}| \tilde{\mathcal{I}}_{t},
\boldsymbol{x}_{[1,m),t},
\boldsymbol{\theta}_{[m,M],t}),  
\end{equation}
which is given via the equivalent channel 
between the data symbol $x_{m,t}$ and the associated decoder  
\begin{IEEEeqnarray}{rl}  
&p(\tilde{x}_{m,t}| x_{m,t}, \tilde{\mathcal{I}}_{t},
\boldsymbol{x}_{[1,m),t},\boldsymbol{\theta}_{[m,M],t}) 
\nonumber \\ 
=& \int p(x_{m,t}=\tilde{x}_{m,t} | \boldsymbol{y}_{t},\tilde{\mathcal{I}}_{t},
\boldsymbol{x}_{[1,m),t},\boldsymbol{\theta}_{[m,M],t})
p(\boldsymbol{y}_{t}|\boldsymbol{x}_{t},\tilde{\mathcal{I}}_{t})
p(\boldsymbol{x}_{(m,M],t}|
\boldsymbol{\theta}_{(m,M],t})
d\boldsymbol{x}_{(m,M],t}d\boldsymbol{y}_{t}.  
\label{equivalent_channel}
\end{IEEEeqnarray}

\subsubsection{LMMSE Detector} 
Since the optimal detector is infeasible in terms of the complexity, 
it is important in practice to obtain a lower bound based on the LMMSE 
detector with SIC. We follow \cite{Takeuchi101} to derive the LMMSE detector 
that feeds to the associated detector an approximate posterior pdf of 
$x_{m,t}$ based on the knowledge about the received vector 
$\boldsymbol{y}_{t}$, the data symbols $\boldsymbol{x}_{[1,m),t}$ decoded 
in the preceding substages, the bias $\boldsymbol{\theta}_{[m,M],t}$ 
for the unknown data symbols 
$\boldsymbol{x}_{[m,M],t}=(x_{m,t},\ldots,x_{M,t})^{\mathrm{T}}$, 
and the joint posterior pdf $p(\boldsymbol{H}|\tilde{\mathcal{I}}_{t})$ 
provided by the LMMSE channel estimator. 

We first divide the RHS of the MIMO channel~(\ref{MIMO}) into two terms: 
\begin{equation} \label{LMMSE_MIMO1} 
\boldsymbol{y}_{t} 
= \frac{1}{\sqrt{M}}\hat{\boldsymbol{H}}_{t}\boldsymbol{x}_{t} 
+ \frac{1}{\sqrt{M}}(\boldsymbol{H} - \hat{\boldsymbol{H}}_{t})
\boldsymbol{x}_{t} + \boldsymbol{n}_{t}, 
\end{equation}
with $\hat{\boldsymbol{H}}_{t}$ denoting the LMMSE 
estimate~(\ref{LMMSE_estimate}) of the channel matrix. Let 
$\hat{\boldsymbol{H}}_{[1,m),t}\in\mathbb{C}^{N\times (m-1)}$ and 
$\hat{\boldsymbol{H}}_{[m,M],t}\in\mathbb{C}^{N\times(M-m+1)}$ denote the matrices 
that consist of the first $(m-1)$ column vectors of the LMMSE 
estimate~(\ref{LMMSE_estimate}) and of the last $(M-m+1)$ column vectors, 
respectively. The first term of the RHS in (\ref{LMMSE_MIMO1}) is further 
decomposed into two terms: 
\begin{equation} \label{LMMSE_MIMO2} 
\boldsymbol{y}_{t} 
= \frac{1}{\sqrt{M}}\hat{\boldsymbol{H}}_{[1,m),t}\boldsymbol{x}_{[1,m),t} 
+ \frac{1}{\sqrt{M}}\hat{\boldsymbol{H}}_{[m,M],t}\boldsymbol{x}_{[m,M],t}
+ \frac{1}{\sqrt{M}}(\boldsymbol{H} - \hat{\boldsymbol{H}}_{t})
\boldsymbol{x}_{t} + \boldsymbol{n}_{t}, 
\end{equation}
where the first term of the RHS is a known quantity. From this expression, 
the LMMSE detector is defined via the postulated MIMO channel 
\begin{equation}
\tilde{\boldsymbol{y}}_{t}^{(\mathrm{L})} 
- \frac{1}{\sqrt{M}}\hat{\boldsymbol{H}}_{[1,m),t}\boldsymbol{x}_{[1,m),t}
= \frac{1}{\sqrt{M}}\hat{\boldsymbol{H}}_{[m,M],t}
\tilde{\boldsymbol{x}}_{[m,M],t}^{(\mathrm{L})} 
+ \tilde{\boldsymbol{w}}_{t}^{(\mathrm{L})} + \boldsymbol{n}_{t}, 
\end{equation}
where $\tilde{\boldsymbol{y}}_{t}^{(\mathrm{L})}\in\mathbb{C}^{N}$ and 
$\tilde{\boldsymbol{x}}_{[m,M],t}^{(\mathrm{L})}
=(\tilde{x}_{m,t}^{(\mathrm{L})},\ldots,\tilde{x}_{M,t}^{(\mathrm{L})})
^{\mathrm{T}}\in\mathbb{C}^{M-m+1}$ denote the received vector and 
the data symbol vector postulated by the LMMSE detector, 
respectively. The postulated data symbol vector  
$\tilde{\boldsymbol{x}}_{[m,M],t}^{(\mathrm{L})}$ is assumed to be a proper 
complex Gaussian random vector with mean 
$\boldsymbol{\theta}_{[m,M],t}$ and covariance 
\begin{equation}
\boldsymbol{\Sigma}_{m,t} 
= P\boldsymbol{I}_{M-m+1} - \mathrm{diag}(|\theta_{m,t}|^{2},\ldots,
|\theta_{M,t}|^{2}), 
\end{equation} 
which are the same as the mean and covariance of the original vector  
$\boldsymbol{x}_{[m,M],t}$, respectively. 
Furthermore, $\tilde{\boldsymbol{w}}_{t}^{(\mathrm{L})}\in\mathbb{C}^{N}$ 
is a CSCG random vector with the covariance matrix 
$\zeta_{m,t}\boldsymbol{I}_{N}$, in which 
\begin{equation}
\zeta_{m,t}  
= \frac{1}{M}\mathrm{Tr}\left(
 \mathbb{E}[\boldsymbol{x}_{t}\boldsymbol{x}_{t}^{\mathrm{H}}
 |\boldsymbol{x}_{[1,m),t}]\boldsymbol{\Xi}_{t}
\right), 
\end{equation}
with 
\begin{equation}
\mathbb{E}[\boldsymbol{x}_{t}\boldsymbol{x}_{t}^{\mathrm{H}}
|\boldsymbol{x}_{[1,m),t}]
= \mathrm{diag}(|x_{1,t}|^{2},\ldots,|x_{m-1,t}|^{2},P,\ldots,P). 
\end{equation}
Note that $\tilde{\boldsymbol{w}}_{t}^{(\mathrm{L})}$ has the same first 
and second moments as those of the third term on the RHS of 
(\ref{LMMSE_MIMO2}). 

The posterior distribution of $\tilde{\boldsymbol{x}}_{[m,M],t}^{(\mathrm{L})}$ 
given $\tilde{\boldsymbol{y}}_{t}^{(\mathrm{L})}$, $\tilde{\mathcal{I}}_{t}$, 
$\boldsymbol{x}_{[1,m),t}$, and $\boldsymbol{\theta}_{[m,M],t}$ is a proper 
complex Gaussian distribution with mean 
$\hat{\boldsymbol{x}}_{[m,M],t}^{(\mathrm{L})}$ and 
covariance $\boldsymbol{\Xi}_{m,t}^{(\mathrm{L})}$, given by 
\begin{equation} \label{LMMSE_detector}
\hat{\boldsymbol{x}}_{[m,M],t}^{(\mathrm{L})} 
= (\boldsymbol{\Xi}_{m,t}^{(\mathrm{L})})^{-1} 
\left\{
 \frac{1}{\sqrt{M}}(N_{0}+\zeta_{m,t})^{-1}
 \hat{\boldsymbol{H}}_{[m,M],t}^{\mathrm{H}}\left(
  \tilde{\boldsymbol{y}}_{t}^{(\mathrm{L})} 
  - \frac{1}{\sqrt{M}}\hat{\boldsymbol{H}}_{[1,m),t}\boldsymbol{x}_{[1,m),t} 
 \right)
 + \boldsymbol{\Sigma}_{m,t}^{-1}\boldsymbol{\theta}_{[m,M],t} 
\right\}, 
\end{equation}
\begin{equation}
\boldsymbol{\Xi}_{m,t}^{(\mathrm{L})} 
= \left(
 \boldsymbol{\Sigma}_{m,t}^{-1} 
 + \frac{1}{M}(N_{0} + \zeta_{m,t})^{-1}
 \hat{\boldsymbol{H}}_{[m,M],t}^{\mathrm{H}}\hat{\boldsymbol{H}}_{[m,M],t}
\right)^{-1}. 
\end{equation} 
Note that the posterior mean~(\ref{LMMSE_detector}) with 
$\tilde{\boldsymbol{y}}_{t}^{(\mathrm{L})}=\boldsymbol{y}_{t}$ is equal to 
the LMMSE estimate of $\boldsymbol{x}_{[m,M],t}$.  
The LMMSE detector with SIC sends the marginal posterior pdf 
$p(\tilde{x}_{m,t}^{(\mathrm{L})}|
\tilde{\boldsymbol{y}}_{t}^{(\mathrm{L})}=\boldsymbol{y}_{t}, 
\tilde{\mathcal{I}}_{t},\boldsymbol{x}_{[1,m),t},
\boldsymbol{\theta}_{[m,M],t})$, corresponding to (\ref{posterior_x}), 
given via the marginalization of the joint posterior pdf 
$p(\tilde{\boldsymbol{x}}_{[m,M],t}^{(\mathrm{L})}|
\tilde{\boldsymbol{y}}_{t}^{(\mathrm{L})}, \tilde{\mathcal{I}}_{t}, 
\boldsymbol{x}_{[1,m),t},\boldsymbol{\theta}_{[m,M],t})$.  
Thus, we have a lower bound of the achievable rate based on the LMMSE 
detector with SIC 
\begin{equation} \label{capacity_LMMSE} 
C \geq \frac{1}{T_{\mathrm{c}}}\sum_{t=T_{\mathrm{tr}}+1}^{T_{\mathrm{c}}}
\sum_{m=1}^{M}
I(x_{m,t};\tilde{x}_{m,t}^{(\mathrm{L})}| \tilde{\mathcal{I}}_{t},
\boldsymbol{x}_{[1,m),t},
\boldsymbol{\theta}_{[m,M],t}),  
\end{equation}
which is given via the equivalent channel between the data symbol $x_{m,t}$ 
and the associated LMMSE detector in the same manner as in 
(\ref{capacity_nonlinear}). 

The goal of this paper is to optimize the length of training phase 
$T_{\mathrm{tr}}$, the prior pdf of the data symbols, and the hyperprior pdf 
of the bias in terms of the two lower bounds~(\ref{capacity_nonlinear})  
and (\ref{capacity_LMMSE}). 

\section{Main Results} \label{sec_main_results} 
\subsection{Large-System Analysis} 
First, the lower bound~(\ref{capacity_nonlinear}) based on the optimal 
detector is evaluated in the large-system limit. 
We focus on substage~$m$ within stage~$t$ in the SD receiver. 
In order to calculate 
$I(x_{m,t};\tilde{x}_{m,t}|\tilde{\mathcal{I}}_{t},
\boldsymbol{x}_{[1,m),t},
\boldsymbol{\theta}_{[m,M],t})$, we have to evaluate the 
distribution of the equivalent channel~(\ref{equivalent_channel}), which 
is a probability distribution on the space of distributions and 
depends on the omitted variables $\boldsymbol{\Theta}_{(T_{\mathrm{tr}},t)}$ 
and $\boldsymbol{\theta}_{[1,m),t}$ implicitly through the posterior 
pdf $p(\boldsymbol{H}|\tilde{\mathcal{I}}_{t})$ and 
$\boldsymbol{x}_{[1,m),t}$. 
This evaluation is quite difficult in general for finite-sized systems. 
A key assumption of circumventing this difficulty is the assumption of 
the large-system limit in which $M$, $N$, $T_{\mathrm{c}}$, $T_{\mathrm{tr}}$, 
$t$, and $m$ tend to infinity while 
their ratios $\alpha=M/N$, $\beta=M/T_{\mathrm{c}}$, 
$\tau_{0}=T_{\mathrm{tr}}/T_{\mathrm{c}}$, 
$\tau=t/T_{\mathrm{c}}$, and $\mu=m/M$ are kept constant. 
The self-averaging property for the equivalent 
channel~(\ref{equivalent_channel}) is expected to hold in the large-system 
limit: The distribution of the equivalent channel~(\ref{equivalent_channel}) 
converges to a Dirac measure on the space of distributions 
in the large-system limit. See \cite{Ledoux01} for a mathematical treatment 
of self-averaging. The assumption of self-averaging implies that the detection 
performance for each data symbol coincides with the corresponding 
average performance in the large-system limit. Under this assumption, 
the replica method allows us to analyze the equivalent 
channel~(\ref{equivalent_channel}) in the large-system limit.    

The self-averaging properties are classified into those for 
extensive quantities and those for non-extensive quantities.  
The former quantities are proportional to the size of systems, while 
the latter quantities are not.  
The self-averaging property for extensive quantities, such as sum capacity 
and the so-called free-energy in statistical physics, has been rigorously 
justified for linear systems~\cite{Verdu99} and general 
systems~\cite{Guerra02,Guerra031}. 
It might be possible to prove the self-averaging property for the 
lower bound~(\ref{capacity_nonlinear}) 
by using the method developed in \cite{Guerra02,Guerra031}. 
However, we need the self-averaging property for {\em each} equivalent 
channel~(\ref{equivalent_channel}), which is non-extensive. 
The self-averaging property for non-extensive quantities is less understood 
except for several simple cases.  

We also need the self-averaging property for each element of the error 
covariance matrix~(\ref{error_covariance}), along with that for the equivalent 
channel~(\ref{equivalent_channel}). Note that the error covariance 
matrix~(\ref{error_covariance}) is a random matrix 
depending on $\bar{\boldsymbol{X}}_{\backslash t}$ explicitly and on 
$\boldsymbol{\Theta}_{(T_{\mathrm{tr}},t)}$ implicitly through the data 
symbol vectors decoded in the preceding stages. 
See \cite{Cottatellucci05} for the self-averaging property of each 
diagonal element for random covariance matrices. 

\begin{assumption} \label{assumption1} 
Each element of the error covariance matrix~(\ref{error_covariance}) for the 
LMMSE channel estimation converges in probability 
to a deterministic value in the large-system limit, i.e., 
\begin{equation} \label{error_covariance_limit} 
(\boldsymbol{\Xi}_{t})_{\tilde{m},\tilde{m}'} \rightarrow \left\{
\begin{array}{cc}
\xi^{2}(\tau) & \hbox{for $\tilde{m}=\tilde{m}'$} \\
\rho(\tau) & \hbox{for $\tilde{m}<\tilde{m}'$} \\ 
\rho^{*}(\tau) & \hbox{for $\tilde{m}>\tilde{m}'$.}
\end{array}
\right.
\end{equation}
\end{assumption}

The error covariance matrix~(\ref{error_covariance}) does not depend on 
the number of receive antennas~$N$ or the current substage~$m$. Thus, 
the limits~(\ref{error_covariance_limit}) do not depend on $\alpha$ or 
$\mu$, while they may depend on $\beta$, $\tau_{0}$, and $\tau$. 
Assumption~\ref{assumption1} has been rigorously proved for the unbiased 
case $\theta_{m,t}=0$ in~\cite{Evans00}. 

\begin{assumption} \label{assumption2}
The equivalent channel~(\ref{equivalent_channel}) is self-averaging with 
respect to $\tilde{\boldsymbol{Y}}_{\backslash t}$ and 
$\boldsymbol{x}_{[1,m),t}$:  
(\ref{equivalent_channel}) converges in law to a 
conditional pdf of $\tilde{x}_{m,t}$ given $x_{m,t}$, 
$\bar{\boldsymbol{X}}_{\backslash t}$, and $\boldsymbol{\theta}_{t}$, 
which does not depend on $\tilde{\boldsymbol{Y}}_{\backslash t}$ or 
$\boldsymbol{x}_{[1,m),t}$, in the large-system limit. 
\end{assumption}

The self-averaging property for equivalent channels has been rigorously 
proved in the case of linear receivers by using random matrix 
theory~\cite{Tse99,Evans00}, while its justification is still open for 
nonlinear receivers. 
The equivalent channel~(\ref{equivalent_channel}) is also expected to be 
self-averaging with respect to the other random variables. However, 
Assumption~\ref{assumption2} is sufficient for using the replica method. 
We postulate Assumptions~\ref{assumption1} and \ref{assumption2} 
since their justification is beyond the scope of this paper. 

Following the replica methodology raises the issue of whether replica 
symmetry (RS) or replica symmetry breaking (RSB) should be assumed. 
The formal definition of the RS assumption will be presented in 
Appendices~\ref{sec_deriv_proposition2} and \ref{sec_deriv_proposition3}. 
Roughly speaking, RSB should be assumed for complicated optimization problems 
in which the object function has multi-valley 
structure~\cite{Mezard87,Nishimori01}. There are many local optima 
for such an object function. On the other hand, the RS assumption is 
applicable to simple problems such that the object function has the unique 
global optimum or a few local optima. Nishimori's rigorous 
result~\cite{Nishimori021} suggests that the individually optimal (IO)  
detection~\cite{Verdu98} considered in this paper should have a simple 
structure corresponding to the RS assumption, while the jointly optimal (JO) 
detection~\cite{Verdu98} should be a complicated problem corresponding 
to the RSB assumption. A recent rigorous study~\cite{Korada11} also supports 
the RS assumption for the IO detection. Thus, the RS assumption is postulated 
in this paper. Note that the self-averaging property for the equivalent 
channel might not hold if the system had a complicated structure corresponding 
to the RSB assumption~\cite{Mezard87,Pastur91}. 

The lower bound~(\ref{capacity_nonlinear}) is 
given as a double integral of the constrained capacity for AWGN  
channels. We first derive the AWGN channels in a heuristic 
manner. The heuristic derivation described below provides an intuitive  
interpretation of the AWGN channels, although the formal 
derivation is based on the replica method.  
In the current stage~$t=\tau T_{\mathrm{c}}$, with $\tau_{0}\leq\tau\leq1$, 
we consider fading channels with time diversity 
for channel estimation, 
\begin{equation} \label{fading_channel_tr}
\underline{y}_{n,t'} = \frac{1}{\sqrt{M}}h_{n,m}x_{m,t'} + 
\underline{w}_{n,t'}, 
\quad \hbox{for all $t'< t$,}  
\end{equation}
\begin{equation} \label{fading_channel_c} 
\underline{y}_{n,t'} = \frac{1}{\sqrt{M}}h_{n,m}\theta_{m,t'} + 
\underline{w}_{n,t'}, 
\quad \hbox{for all $t'> t$,}  
\end{equation}
with $\underline{w}_{n,t'}\sim\mathcal{CN}(0,\sigma_{\mathrm{tr}}^{2})$ 
for $t'<t$ and with 
$\underline{w}_{n,t'}\sim\mathcal{CN}(0,\sigma_{\mathrm{c}}^{2})$ 
for $t'>t$. 
The fading channels are obtained by extracting the first terms in 
(\ref{fading_channel_tr}) and (\ref{fading_channel_c}) from the original 
and virtual MIMO channels~(\ref{MIMO}) and~(\ref{virtual_MIMO}), and then by 
approximating the remaining terms by CSCG random vectors with covariance 
$\sigma_{\mathrm{tr}}^{2}\boldsymbol{I}_{N}$ 
and $\sigma_{\mathrm{c}}^{2}\boldsymbol{I}_{N}$, respectively. 
We apply maximal-ratio combining (MRC) to (\ref{fading_channel_tr}) and 
(\ref{fading_channel_c}),  
\begin{equation} \label{z_tr} 
\underline{z}_{\mathrm{tr}} = \frac{1}{\sqrt{P(\tau T_{\mathrm{c}}-1})}
\sum_{t'=1}^{\tau T_{\mathrm{c}}-1}x_{m,t'}^{*}\underline{y}_{n,t'},
\end{equation}
\begin{equation}
\underline{z}_{\mathrm{c}} 
= \frac{1}{\sqrt{\sigma_{\theta}^{2}(1-\tau)T_{\mathrm{c}}}}
\sum_{t'=\tau T_{\mathrm{c}}+1}^{T_{\mathrm{c}}}\theta_{m,t'}^{*}
\underline{y}_{n,t'}. 
\end{equation}
Taking the large-system limit, due to the weak law of large numbers, 
we obtain   
\begin{equation} \label{AWGN_channel_estimation}
\begin{pmatrix}
\underline{z}_{\mathrm{tr}} \\ 
\underline{z}_{\mathrm{c}}  
\end{pmatrix} 
= \frac{1}{\sqrt{\beta}}
\begin{pmatrix}
\sqrt{\tau P} \\ 
\sqrt{(1-\tau)\sigma_{\theta}^{2}} 
\end{pmatrix}h_{n,m} 
+ \begin{pmatrix}
w_{\mathrm{tr}} \\ 
w_{\mathrm{c}} 
\end{pmatrix},
\end{equation}
where $w_{\mathrm{tr}}$ and $w_{\mathrm{c}}$ are mutually independent CSCG 
random variables with variances 
$\sigma_{\mathrm{tr}}^{2}$ and $\sigma_{\mathrm{c}}^{2}$, respectively.  
The minimum mean-squared error (MMSE) estimate of $h_{n,m}$ for the 
channel~(\ref{AWGN_channel_estimation}) is given by~\cite{Tse05} 
\begin{equation} \label{decoupled_LMMSE_estimate}
\underline{\hat{h}}_{n,m} = 
\frac{1}{\sqrt{\beta}\xi^{2}(\tau)}
\left(
 \frac{\sqrt{\tau P}}{\sigma_{\mathrm{tr}}^{2}}\underline{z}_{\mathrm{tr}} 
 + \frac{\sqrt{(1-\tau)\sigma_{\theta}^{2}}}
 {\sigma_{\mathrm{c}}^{2}}\underline{z}_{\mathrm{c}} 
\right). 
\end{equation}
where the mean-squared error (MSE) $\xi^{2}(\tau)$ for the MMSE 
estimate~(\ref{decoupled_LMMSE_estimate}) is explicitly given by 
\begin{equation} \label{xi}
\xi^{2}(\tau) = \left(
 1 + \frac{\tau P}{\sigma_{\mathrm{tr}}^{2}\beta}
 + \frac{(1-\tau)\sigma_{\theta}^{2}}{\sigma_{\mathrm{c}}^{2}\beta}
\right)^{-1}. 
\end{equation}
It is well known that the MMSE estimate $\underline{\hat{h}}_{n,m}$ and the 
estimation error 
$\Delta \underline{h}_{n,m}=h_{n,m}-\underline{\hat{h}}_{n,m}$ are 
uncorrelated CSCG random variables with 
variances $(1-\xi^{2}(\tau))$ and $\xi^{2}(\tau)$, respectively. 

We next consider fading channels with spatial diversity for data estimation 
in the current substage~$m=\mu M$ for $0\leq\mu\leq1$,  
\begin{equation}
\underline{y}_{n,t} = \frac{1}{\sqrt{M}}\left\{
 \underline{\hat{h}}_{n,m}x_{m,t} + \Delta \underline{h}_{n,m}
 x_{m,t}
\right\} + \underline{w}_{n,t}, 
\quad n=1,\ldots,N,  
\end{equation}
where $\underline{w}_{n,t}\in\mathbb{C}$ denotes a CSCG random variable 
with variance $\sigma^{2}(\tau,\mu)$.  
Applying the MRC to $\{\underline{y}_{n,t}\}$, we obtain 
\begin{equation} 
\underline{z} = \frac{1}{\sqrt{N(1-\xi^{2}(\tau))}}
\sum_{n=1}^{N}\underline{\hat{h}}_{n,m}^{*}
\underline{y}_{n,t}. 
\end{equation}
Taking the large-system limit gives the AWGN channel 
\begin{equation} \label{AWGN} 
\underline{z} = \sqrt{\frac{1-\xi^{2}(\tau)}{\alpha}}x_{m,t} + \underline{w}, 
\end{equation}
with $\underline{w}\sim\mathcal{CN}(0,\sigma^{2}(\tau,\mu))$.   
The MMSE estimate $\underline{\hat{x}}_{m,t}$ of $x_{m,t}$ for the AWGN 
channel~(\ref{AWGN}) is given as the mean 
$\underline{\hat{x}}_{m,t}=\int x_{m,t}
p(x_{m,t}|\underline{z},\theta_{m,t})dx_{m,t}$ with respect to the 
posterior pdf 
\begin{equation} \label{decoupled_posterior} 
p(x_{m,t}|\underline{z},\theta_{m,t}) = \frac{
p(\underline{z}|x_{m,t})p(x_{m,t}|\theta_{m,t})
}
{
\int p(\underline{z}|x_{m,t})p(x_{m,t}|\theta_{m,t})x_{m,t}
}, 
\end{equation}
where $p(\underline{z}|x_{m,t})$ represents the AWGN channel~(\ref{AWGN}). 
The MSE for the MMSE estimate $\underline{\hat{x}}_{m,t}$ given $\theta_{m,t}$ 
is defined as 
\begin{equation} \label{MSE} 
\mathrm{MSE}(\sigma^{2},\theta_{m,t}) = \mathbb{E}\left[
 \left. 
  |x_{m,t} - \underline{\hat{x}}_{m,t}|^{2} 
 \right| \theta_{m,t} 
\right]. 
\end{equation}  

We have not so far specified the variances $\sigma_{\mathrm{tr}}^{2}$,  
$\sigma_{\mathrm{c}}^{2}$, and $\sigma^{2}(\tau,\mu)$. The constrained 
capacity of the AWGN channel~(\ref{AWGN}) corresponds to the mutual 
information in the lower bound~(\ref{capacity_nonlinear}) when 
the three variances are determined as solutions to fixed-point equations.  

\begin{proposition}[Optimal Detector] \label{proposition1} 
Suppose that Assumption~\ref{assumption1}, Assumption~\ref{assumption2}, 
and the RS assumption hold. Then, 
the constrained capacity~(\ref{capacity}) per transmit antenna is bounded 
from below by  
\begin{equation} \label{capacity_lower_bound} 
\frac{C}{M} \geq \int_{\tau\in[\tau_{0},1]}\int_{\mu\in[0,1]}
I(x_{m,t};\underline{z}|\theta_{m,t})d\tau d\mu,  
\end{equation} 
in the large-system limit, in which the mutual information 
$I(x_{m,t};\underline{z}|\theta_{m,t})$ is equal to  
the constrained capacity of the AWGN channel~(\ref{AWGN}). 
In evaluating $I(x_{m,t};\underline{z}|\theta_{m,t})$, 
$\{\sigma_{\mathrm{tr}}^{2},\sigma_{\mathrm{c}}^{2}\}$ is given as the 
solution to the coupled fixed-point equations  
\begin{equation} \label{fixed_point_tr} 
\sigma_{\mathrm{tr}}^{2} = N_{0} + P\xi^{2}(\tau), 
\end{equation}
\begin{equation} \label{fixed_point_c} 
\sigma_{\mathrm{c}}^{2} = N_{0} + (P-\sigma_{\theta}^{2}) + 
\sigma_{\theta}^{2}\xi^{2}(\tau), 
\end{equation} 
where $\xi^{2}(\tau)$ is given by (\ref{xi}). 
Furthermore, $\sigma^{2}(\tau,\mu)$ is given as a solution $\sigma^{2}$ to 
the fixed-point equation 
\begin{equation} \label{fixed_point} 
\sigma^{2} = N_{0} + P\xi^{2}(\tau) + (1-\mu)(1-\xi^{2}(\tau))
\mathbb{E}\left[
 \mathrm{MSE}(\sigma^{2},\theta_{m,t}) 
\right], 
\end{equation} 
where $\mathrm{MSE}(\sigma^{2},\theta_{m,t})$ is given by (\ref{MSE}). 
If the fixed-point equation~(\ref{fixed_point}) has multiple solutions, 
one should choose the solution minimizing the following quantity 
\begin{equation} \label{free_energy} 
(1-\mu)I(x_{m,t};\underline{z}|\theta_{m,t}) + \frac{1}{\alpha}\left[
 D_{2}(N_{0}\|\sigma^{2}) 
 + \frac{\xi^{2}(\tau)}{\sigma^{2}}\log_{2}\mathrm{e} 
\right]. 
\end{equation}
\end{proposition} 
\begin{IEEEproof}[Derivation of Proposition~\ref{proposition1}]
See Appendix~\ref{sec_derivation}. 
\end{IEEEproof}

Note that the last terms in the coupled fixed-point 
equations~(\ref{fixed_point_tr}) and (\ref{fixed_point_c}) depend on 
$\sigma_{\mathrm{tr}}^{2}$ and $\sigma_{\mathrm{c}}^{2}$ through (\ref{xi}). 
Equation~(\ref{fixed_point}) 
for given $\xi^{2}(\tau)$ provides a fixed-point equation with respect to 
$\sigma^{2}$. The second and last terms in the RHS of 
(\ref{fixed_point}) correspond to contributions from channel estimation 
errors and inter-stream interference, respectively. The integrand in 
(\ref{capacity_lower_bound}) depends on the variables $\tau$ and $\mu$ 
through the SNR $P(1-\xi^{2}(\tau))/(\alpha\sigma^{2}(\tau,\mu))$. 

The existence of multiple solutions in (\ref{fixed_point}) relates to 
the so-called phase transition in statistical physics. See 
\cite{Tanaka02,Guo05} for an interpretation in the context of communications. 
Numerical evaluation of (\ref{fixed_point}) for QPSK modulation implies that 
multiple solutions do not appear when $\alpha\leq1$. 
In the high SNR regime there is no point to use transmit antennas more than 
receive antennas or half the coherence time. In fact, Zheng and 
Tse~\cite{Zheng02} proved that the full spatial multiplexing gain of the 
MIMO channel with no CSI is given by $\bar{M}(1-\bar{M}/T_{\mathrm{c}})$, 
with $\bar{M}=\min\{M,N,\lfloor T_{\mathrm{c}}/2\rfloor\}$, which is achieved 
by using $\min\{N,\lfloor T_{\mathrm{c}}/2\rfloor\}$ transmit antennas out of 
$M$ transmit antennas if $M>N$ or $M>\lfloor T_{\mathrm{c}}/2\rfloor$. 
Thus, we consider $\alpha\leq1$ and $\beta\leq1/2$ in the high SNR regime. 

Next, the lower bound~(\ref{capacity_LMMSE}) based on the LMMSE detector 
is evaluated in the large-system limit. The following proposition is obtained 
in the same manner as in the derivation of Proposition~\ref{proposition1}. 
\begin{assumption} \label{assumption3}
The equivalent channel for the LMMSE detector is self-averaging with 
respect to $\tilde{\boldsymbol{Y}}_{\backslash t}$ and 
$\boldsymbol{x}_{[1,m),t}$ in the large-system limit. 
\end{assumption}
\begin{proposition}[LMMSE Detector] \label{proposition_LMMSE} 
Suppose that Assumption~\ref{assumption1}, Assumption~\ref{assumption3}, 
and the RS assumption hold. Then, 
the constrained capacity~(\ref{capacity}) per transmit antenna is bounded 
from below by (\ref{capacity_lower_bound}) in the large-system limit.  
In evaluating (\ref{capacity_lower_bound}), 
$\{\sigma_{\mathrm{tr}}^{2},\sigma_{\mathrm{c}}^{2}\}$ is given as the 
solution to the coupled fixed-point equations~(\ref{fixed_point_tr}) 
and (\ref{fixed_point_c}). On the other hand,   
$\sigma^{2}(\tau,\mu)$ is given as the unique solution $\sigma^{2}$ to 
the fixed-point equation 
\begin{equation} \label{fixed_point_Gauss} 
\sigma^{2} = N_{0} + P\xi^{2}(\tau) + (1-\mu)(1-\xi^{2}(\tau))
\mathbb{E}\left[
 \frac{(P-|\theta_{m,t}|^{2})\alpha\sigma^{2}}
 {(1-\xi^{2}(\tau))(P-|\theta_{m,t}|^{2}) + \alpha\sigma^{2}}
\right]. 
\end{equation} 
\end{proposition} 
\begin{IEEEproof}[Derivation of Proposition~\ref{proposition_LMMSE}]
Proposition~\ref{proposition_LMMSE} is obtained by repeating the derivation of 
Proposition~\ref{proposition1}. Thus, 
we only prove the uniqueness of the solution to the fixed-point 
equation~(\ref{fixed_point_Gauss}). The RHS of 
(\ref{fixed_point_Gauss}) is a concave function of $\sigma^{2}$, which 
intersects with a straight line passing the origin with slope $1$ at two 
points. Since the concave function passes the point 
$(0,N_{0}+\xi^{2}(\tau))$, which is above the origin, one intersection 
must be in $\sigma^{2}<0$. Thus, the fixed-point 
equation~(\ref{fixed_point_Gauss}) has the unique solution in the 
region $\sigma^{2}>0$.   
\end{IEEEproof}
The expectation in the RHS of (\ref{fixed_point_Gauss}) corresponds to 
the MSE for the LMMSE estimator of the data symbol $x_{m,t}$ transmitted 
through the AWGN channel~(\ref{AWGN}), while (\ref{MSE}) in the fixed-point 
equation~(\ref{fixed_point}) is the MSE for the MMSE estimator. 

\subsection{Optimization} 
The next goal is to optimize the lower bound~(\ref{capacity_lower_bound}) 
based on the optimal detector with 
respect to $\tau_{0}$, the hyperprior pdf of $\theta_{m,t}$, and the prior pdf 
of $x_{m,t}$. 
We first notice that the lower bound~(\ref{capacity_lower_bound}) is 
monotonically nonincreasing with respect to $\tau_{0}$ since the integrand is 
non-negative and does not depend on $\tau_{0}$. Thus, the lower 
bound~(\ref{capacity_lower_bound}) is maximized as $\tau_{0}\rightarrow0$.  
Note that the limit $\tau_{0}\rightarrow0$ does not necessarily indicate 
no pilot symbols, since we have taken the limit $\tau_{0}\rightarrow0$ after 
the large-system limit. In other words, the effect of pilot symbols is 
negligible in Proposition~\ref{proposition1} if $T_{\mathrm{tr}}$ is sublinear 
in $T_{\mathrm{c}}$, i.e., $T_{\mathrm{tr}}=o(T_{\mathrm{c}})$. 


Next, we maximize the lower bound~(\ref{capacity_lower_bound}) with respect to 
the hyperprior pdf of $\theta_{m,t}$. For a fixed hyperprior pdf, the SNR 
$P(1-\xi^{2}(\tau))/(\alpha\sigma^{2}(\tau,\mu))$ improves as the variance 
$\sigma_{\theta}^{2}$ grows, since the increase of $\sigma_{\theta}^{2}$ 
results in reductions of the channel estimator error $\xi^{2}(\tau)$ and the 
inter-stream interference given by the last term in the RHS of 
(\ref{fixed_point}). However, increasing $\sigma_{\theta}^{2}$ reduces 
the mutual information $I(x_{m,t};\underline{z}|\theta_{m,t})$ for a fixed SNR, 
due to the reduction of payload. Interestingly, numerical results presented in 
Section~\ref{sec_comparison} show that the lower 
bound~(\ref{capacity_lower_bound}) is maximized as 
$\sigma_{\theta}^{2}\rightarrow0$ for a fixed hyperprior pdf of 
$\theta_{m,t}$. This indicates that the lower 
bound~(\ref{capacity_lower_bound}) is maximized when $\theta_{m,t}=0$ 
with probability one.  
 
The arguments described above indicate that negligible pilot information, 
or more precisely, the limits $\tau_{0},\sigma_{\theta}^{2}\rightarrow0$ are   
best in the large-system limit, while $\tau_{0}=\beta$ is best in 
terms of the HH bound~\cite{Hassibi03}. Thus, we can conclude 
that the SD scheme can reduce the overhead for training significantly. 
It is worth noting that using a capacity-achieving error-correcting code is 
assumed in our analysis. We conjecture that if some practical coding is used 
finite $\tau_{0}$ or $\sigma_{\theta}^{2}$ are required for getting 
accurate channel estimates in the initial stage.  
See \cite{Vehkaperae093} for the case of practical coding. 

Finally, we optimize the lower bound~(\ref{capacity_lower_bound}) with 
respect to the prior pdf of $x_{m,t}$. This optimization problem is 
nonlinear since the prior pdf of $x_{m,t}$ depends on $\sigma^{2}$ through 
the last term in the RHS of the fixed-point 
equation~(\ref{fixed_point}). 
Instead of solving the nonlinear optimization problem exactly, we consider 
the biased Gaussian signaling 
$x_{m,t}\sim\mathcal{CN}(\theta_{m,t},P-|\theta_{m,t}|^{2})$ as a suboptimal 
solution. This choice of the prior pdf should be reasonable since Gaussian 
signaling is optimal if accurate channel estimates can be obtained. In this 
case, Proposition~\ref{proposition1} reduces to the following corollary. 

\begin{corollary} \label{corollary1}
Suppose that Assumption~\ref{assumption1}, Assumption~\ref{assumption2}, 
and the RS assumption hold. 
If $x_{m,t}\sim\mathcal{CN}(\theta_{m,t},P-|\theta_{m,t}|^{2})$, then the 
constrained capacity~(\ref{capacity}) per transmit antenna is bounded 
from below by $c_{\mathrm{g}}$ given by  
\begin{equation} \label{lower_bound_Gauss} 
\frac{C}{M}\geq c_{\mathrm{g}} = \int_{\tau\in[\tau_{0},1]}
\int_{\mu\in[0,1]}
\mathbb{E}\left[
 \log\left(
  1 + \frac{(1-\xi^{2}(\tau))(P-|\theta_{m,t}|^{2})}
  {\alpha\sigma^{2}(\tau,\mu)} 
 \right)
\right]d\tau d\mu,  
\end{equation} 
in the large-system limit. 
In evaluating the integrand in (\ref{lower_bound_Gauss}), 
$\xi^{2}(\tau)$ is given by (\ref{xi}), defined via (\ref{fixed_point_tr}) 
and (\ref{fixed_point_c}). 
Furthermore, $\sigma^{2}(\tau,\mu)$ is given as the unique 
solution $\sigma^{2}$ to the fixed-point equation~(\ref{fixed_point_Gauss}).  
\end{corollary}
\begin{IEEEproof}[Proof of Corollary~\ref{corollary1}]
It is straightforward to confirm that the lower 
bound~(\ref{capacity_lower_bound}) and the fixed-point 
equation~(\ref{fixed_point}) reduce to (\ref{lower_bound_Gauss}) and 
(\ref{fixed_point_Gauss}) under the biased Gaussian signaling, respectively. 
\end{IEEEproof}

We believe that the biased Gaussian signaling maximizes the 
quantity~(\ref{free_energy}), following the argument in \cite{Kitagawa10}.  
If $\xi^{2}(\tau)=0$ and $\sigma^{2}=N_{0}$ were satisfied for all 
$\tau$ and $\mu$, the biased Gaussian signaling would maximize the 
lower bound~(\ref{capacity_lower_bound}). However, $\xi^{2}(\tau)$ is bounded 
from below by a positive value for $\tau<\beta$. This implies the 
suboptimality of the i.i.d.\ Gaussian signaling. 

Proposition~\ref{proposition_LMMSE} and Corollary~\ref{corollary1} imply 
that the large-system performance of the LMMSE detector coincides with that 
of the optimal detector when the i.i.d.\ Gaussian signaling is used. Note that 
this observation is not necessarily trivial, since we have made the Gaussian 
approximation of the third term on the RHS of (\ref{LMMSE_MIMO2}) 
in the derivation of the LMMSE detector. Our results imply that, for 
the i.i.d.\ Gaussian signaling, the performance loss due to the Gaussian 
approximation vanishes in the large-system limit.  

\subsection{High SNR Regime} 
In the high SNR limit $N_{0}\rightarrow0$, the lower 
bound~(\ref{lower_bound_Gauss}) is shown to achieve the full spatial 
multiplexing gain when $T_{\mathrm{tr}}\leq M$. 
\begin{proposition} \label{corollary2}
Suppose that Assumption~\ref{assumption1}, Assumption~\ref{assumption2}, 
and the RS assumption hold. 
For $\alpha\leq1$ and $\tau_{0}\leq \beta\leq1/2$, the lower 
bound~(\ref{lower_bound_Gauss}) with the biased Gaussian signaling 
$x_{m,t}\sim\mathcal{CN}(\theta_{m,t},P-|\theta_{m,t}|^{2})$ achieves the full 
spatial multiplexing gain in the high SNR limit $N_{0}\rightarrow0$, i.e., 
\begin{equation}
\liminf_{N_{0}\rightarrow0}\frac{c_{\mathrm{g}}}{\log(P/N_{0})} = 
1-\beta.  
\end{equation}
\end{proposition} 
\begin{IEEEproof}[Proof of Proposition~\ref{corollary2}] 
We first prove that the solution $\sigma_{\mathrm{tr}}^{2}$ 
to the coupled fixed-point equations~(\ref{fixed_point_tr}) and 
(\ref{fixed_point_c}) converges to zero in the high SNR limit 
for $\tau>\beta$. The proof is by contradiction. 
Suppose that $\sigma_{\mathrm{tr}}^{2}$ is  
strictly positive in the high SNR limit. Dividing both sides of 
(\ref{fixed_point_tr}) by $\sigma_{\mathrm{tr}}^{2}$ and taking the high 
SNR limit, we have 
\begin{equation} \label{fixed_point_tr_tmp1} 
1 = \frac{\beta P\sigma_{\mathrm{c}}^{2}}
{\beta\sigma_{\mathrm{tr}}^{2}\sigma_{\mathrm{c}}^{2} 
 + \tau P\sigma_{\mathrm{c}}^{2} + 
 (1-\tau)\sigma_{\theta}^{2}\sigma_{\mathrm{tr}}^{2}}. 
\end{equation}  
Rearranging (\ref{fixed_point_tr_tmp1}), we obtain
\begin{equation} \label{fixed_point_tr_tmp2}
\sigma_{\mathrm{tr}}^{2} = -\frac{(\tau-\beta)P\sigma_{\mathrm{c}}^{2}}
{\beta\sigma_{\mathrm{c}}^{2} + (1-\tau)\sigma_{\theta}^{2}}. 
\end{equation} 
However, the RHS of (\ref{fixed_point_tr_tmp2}) is negative, due to 
$\tau>\beta$, which 
is a contradiction. Thus, the solution $\sigma_{\mathrm{tr}}^{2}$ must 
converge to zero in the high SNR limit for $\tau>\beta$. This result implies 
that the MSE~(\ref{xi}) also converges to zero in the high SNR limit 
for $\tau>\beta$. 

It is straightforward to show in a similar manner that the 
solution~$\sigma^{2}$ to the fixed-point equation~(\ref{fixed_point_Gauss}) is 
$O(N_{0})$ in the high SNR limit for $\alpha\leq 1$ when the MSE~(\ref{xi}) 
converges to zero. Thus, we have 
\begin{equation}
\liminf_{N_{0}\rightarrow0}\frac{c_{\mathrm{g}}}{\log(P/N_{0})} 
= 1-\beta, 
\end{equation} 
which is equal to the full spatial multiplexing gain for $\alpha\leq1$ and 
$\beta\leq1/2$. 
\end{IEEEproof}

The proof of Proposition~\ref{corollary2} indicates that in the first $M$ 
stages the performance of the SD receiver is limited by channel estimation 
errors, rather than inter-stream interference in MUD. This phenomenon is 
robust in the sense that it occurs regardless of the prior of data symbols. 

\subsection{Low SNR Regime} \label{sec_low_SNR}  
The power per information bit $E_{\mathrm{b}}$ required for reliable 
communication is a key performance measure in the low SNR regime. 
Verd\'u~\cite{Verdu02} proved that the capacity $C_{\mathrm{opt}}$ of the 
MIMO channel~(\ref{MIMO}) wit no CSI is given by 
$C_{\mathrm{opt}}=NP/(N_{0}\ln2) + o(N_{0})$ in the low SNR limit 
$N_{0}\rightarrow\infty$, or $NE_{\mathrm{b}}/N_{0}\geq 
\lim_{N_{0}\rightarrow\infty}NP/(N_{0}C_{\mathrm{opt}})=\ln2\approx-1.59$~dB. 
Since using multiple transmit antennas wastes valuable power in the low SNR 
regime, the number of transmit antennas used should be reduced as $N_{0}$ 
increases. One option is to increase $M^{-1}$ and $N_{0}$ at the same rate. 
Thus, we consider the limit, in which $\alpha, \beta\rightarrow0$ 
and $N_{0}\rightarrow\infty$ while $\beta/\alpha$ and $s=P/(\beta N_{0})$ are 
kept constant. The following proposition provides an upper bound on the 
normalized SNR $NE_{\mathrm{b}}/N_{0}= NP/(N_{0}C)$ required for the optimal 
SD receiver, with $C$ denoting the achievable rate~(\ref{capacity_SD}) of 
the optimal SD receiver.

\begin{proposition} \label{corollary3}
Suppose that the optimal SD receiver achieves a rate $R/M$. Then,  
the normalized SNR $NE_{\mathrm{b}}/N_{0}$ is bounded from above by 
\begin{equation} \label{normalized_SNR} 
N\frac{E_{\mathrm{b}}}{N_{0}} \leq \frac{\beta s}{\alpha R} + o(N_{0}), 
\end{equation}
in the limit where $\alpha, \beta\rightarrow0$ 
and $N_{0}\rightarrow\infty$ while $\beta/\alpha$ and $s=P/(\beta N_{0})$ are 
kept constant. In (\ref{normalized_SNR}), $s$ is implicitly given by  
\begin{equation} \label{lower_bound_low_SNR} 
R = \left[
 1 + \left(
  s + \frac{\beta}{\alpha}s^{2}
 \right)^{-1}
\right]\log\left(
 1 + s + \frac{\beta}{\alpha}s^{2}
\right) - \left(
 1 + \frac{1}{s}
\right)\log(1+s). 
\end{equation}
\end{proposition} 
\begin{IEEEproof}[Proof of Proposition~\ref{corollary3}]
Using the lower bound~(\ref{lower_bound_Gauss}) for Gaussian signaling, we 
obtain an upper bound $NE_{\mathrm{b}}/N_{0}\leq NP/(MN_{0}c_{\mathrm{g}})
=\beta s/(\alpha c_{\mathrm{g}})$. Thus, it is sufficient to prove that   
the maximum of $c_{\mathrm{g}}$ with respect to $\tau_{0}$ and 
$\sigma_{\theta}^{2}$ is given by the RHS of 
(\ref{lower_bound_low_SNR}).  

We evaluate the solutions to the fixed-point 
equations~(\ref{fixed_point_tr}), (\ref{fixed_point_c}), and 
(\ref{fixed_point_Gauss}). It is straightforward to find that  
$\sigma_{\mathrm{tr}}^{2}/N_{0}$, $\sigma_{\mathrm{c}}^{2}/N_{0}$, and 
$\sigma^{2}/N_{0}$ tend to $1$ as $N_{0}\rightarrow\infty$, since 
(\ref{xi}) and (\ref{MSE}) are bounded. This observation implies 
\begin{equation} \label{lower_bound_low_SNR_tmp1} 
c_{\mathrm{g}} \leq \int_{\tau_{0}}^{1}
\log\left[
 1 + \frac{\beta s}{\alpha}\left(
  1 - \frac{\sigma_{\theta}^{2}}{P}
 \right)
 \frac{\tau + (1-\tau)\sigma_{\theta}^{2}/P}
 {s^{-1} + \tau + (1-\tau)\sigma_{\theta}^{2}/P}
\right]d\tau + o(N_{0}), 
\end{equation}
in the limit described in Proposition~\ref{corollary3}. 
In the derivation of (\ref{lower_bound_low_SNR_tmp1}), we have used Jensen's 
inequality. The equality holds only when $|\theta_{m,t}|^{2}$ takes 
$\sigma_{\theta}^{2}$ with probability one. It is easy to confirm that the 
integrand in (\ref{lower_bound_low_SNR_tmp1}) is monotonically decreasing 
with respect to $\sigma_{\theta}^{2}$. Thus, the maximum of $c_{\mathrm{g}}$ 
is achieved at $\tau_{0}=0$ and $\sigma_{\theta}^{2}=0$, and given by 
\begin{equation} \label{lower_bound_low_SNR_tmp2}
\max_{\tau_{0},\sigma_{\theta}^{2}\geq0}c_{\mathrm{g}} 
= \int_{0}^{1}\log\left[
 1 + \frac{\beta}{\alpha}
 \frac{s^{2}\tau}{1 + s\tau}
\right]d\tau + o(N_{0}). 
\end{equation}
Calculating the integration in (\ref{lower_bound_low_SNR_tmp2}), we find that 
the first term in the RHS of (\ref{lower_bound_low_SNR_tmp2}) 
is equal to the RHS of (\ref{lower_bound_low_SNR}).  
\end{IEEEproof}

In the proof of Proposition~\ref{corollary3}, we have proved that 
the lower bound~(\ref{lower_bound_Gauss}) is maximized at $\tau_{0}=0$ and 
$\sigma_{\theta}^{2}=0$ in the low SNR regime. This result implies that 
negligible pilot information is best in terms of the lower 
bound~(\ref{lower_bound_Gauss}) in the low SNR regime.  

It is interesting to note that the achievable 
rate~(\ref{lower_bound_low_SNR}) is approximated by 
$R=\beta s^{2}/(2\alpha\ln2) + O(s^{3})$ as $s\rightarrow0$, which implies 
that $NE_{\mathrm{b}}/N_{0}\leq \sqrt{2\beta\ln2/\alpha R} + o(N_{0},R)$ 
in the low rate regime, i.e., the upper bound~(\ref{normalized_SNR}) diverges 
in $R\rightarrow0$. In other words, the minimum of the upper 
bound~(\ref{normalized_SNR}) is achieved at a strictly positive rate, 
as shown in Section~\ref{sec_comparison}. We remark that a similar result 
was reported in \cite{Gursoy09}. 

The reason why the minimum is achieved at a positive rate is because we have 
spread power over all time slots. It is well known that on-off keying is 
optimal in the low SNR regime, in other words, that spreading power over 
all time slots results in a waste of valuable power.  
If on-off keying was used, the normalized SNR required would reduce 
monotonically  as the achievable rate decreases, as shown in \cite{Gursoy09}. 
However, on-off keying requires high peak-to-average power ratio (PAPR), 
which is unfavorable in practice. 
Thus, the minimum of the upper bound~(\ref{normalized_SNR}) may be 
interpreted as a practical performance bound in terms of energy efficiency.  

\begin{figure}[t]
\includegraphics[width=\hsize]{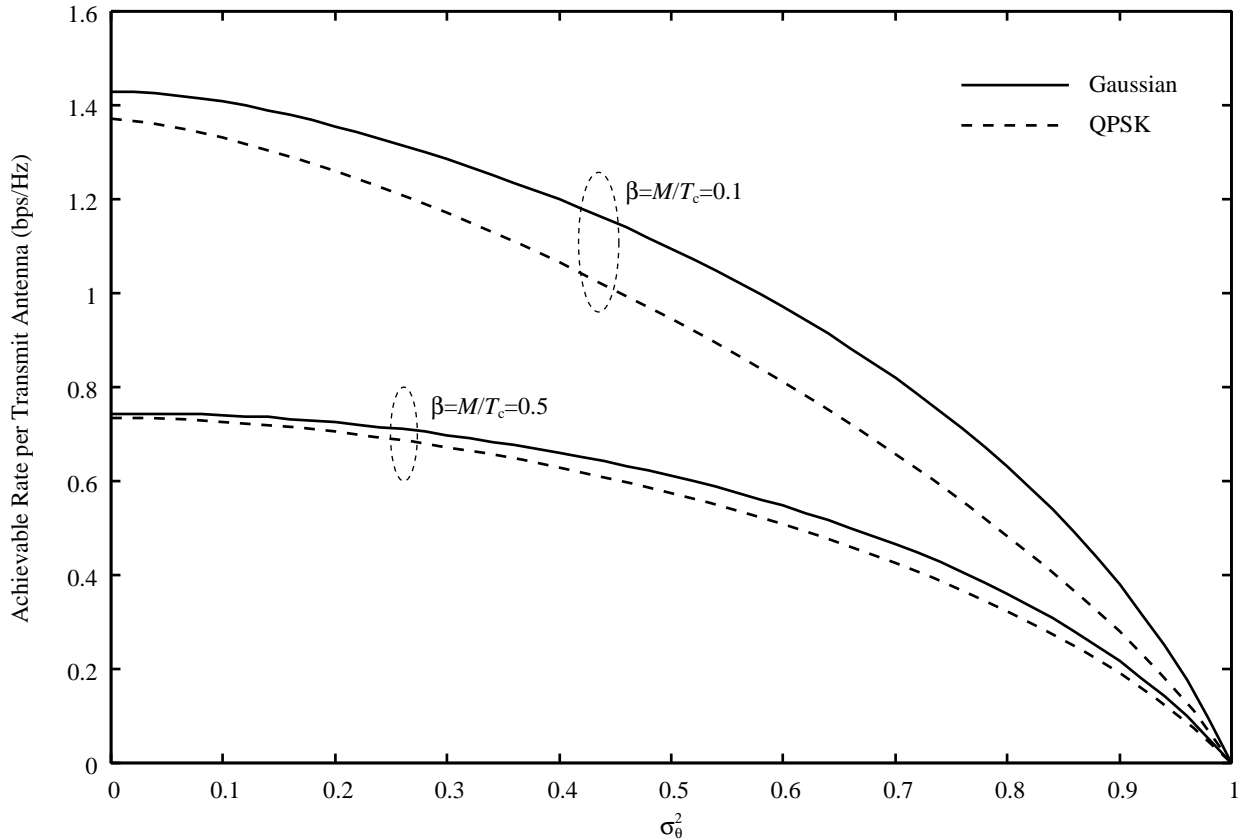}
\caption{
Achievable rate versus the variance of the bias~$\sigma_{\theta}^{2}$ for 
the SNR~$P/N_{0}=6$~dB, $\alpha=M/N=1$, and 
$\tau_{0}=T_{\mathrm{tr}}/T_{\mathrm{c}}=0$. 
}
\label{fig2} 
\end{figure}

\section{Numerical Results} \label{sec_comparison} 
\subsection{Large Systems} 
The lower bound~(\ref{lower_bound_Gauss}) for the biased Gaussian signaling is 
compared to two existing bounds in this section. 
One is the HH bound~\cite{Hassibi03}, 
which corresponds to the achievable rate of receivers based on one-shot 
channel estimation, in which the decoded data symbols are not re-utilized for 
refining the channel estimates. The other one is the high-SNR approximation 
of the capacity~\cite{Zheng02}, which includes a deviation of $o(1)$ from the 
capacity at high SNR. In all numerical results, we chose 
$\tau_{0}=0$ since the lower bound~(\ref{lower_bound_Gauss}) is maximized at 
$\tau_{0}=0$. In order to investigate the optimal choice of 
$\sigma_{\theta}^{2}$, we display the lower bound~(\ref{lower_bound_Gauss}) 
for the biased Gaussian signaling with respect to $\sigma_{\theta}^{2}$ in 
Fig.~\ref{fig2}. The lower bound~(\ref{capacity_lower_bound}) for the biased 
QPSK signaling is also shown in the same figure. We have used 
$p(\theta)=[\delta(\theta-\sigma_{\theta})+\delta(\theta+\sigma_{\theta})]/2$ 
as the distribution of $\theta_{m,t}$, i.e., $\theta_{m,t}$ takes 
$\pm\sigma_{\theta}$ with equal probability. We find that the lower 
bound for the biased Gaussian signaling is larger than that for the biased 
QPSK signaling for all $\sigma_{\theta}^{2}$. Furthermore, both lower bounds 
are monotonically decreasing as $\sigma_{\theta}^{2}$ grows. The latter 
observation implies that the lower bounds are maximized at 
$\sigma_{\theta}^{2}=0$, in other words, negligible pilot information is 
best in terms of the lower bounds. 
Hereinafter, we consider the unbiased case $\sigma_{\theta}^{2}=0$. 

\begin{figure}[t]
\includegraphics[width=\hsize]{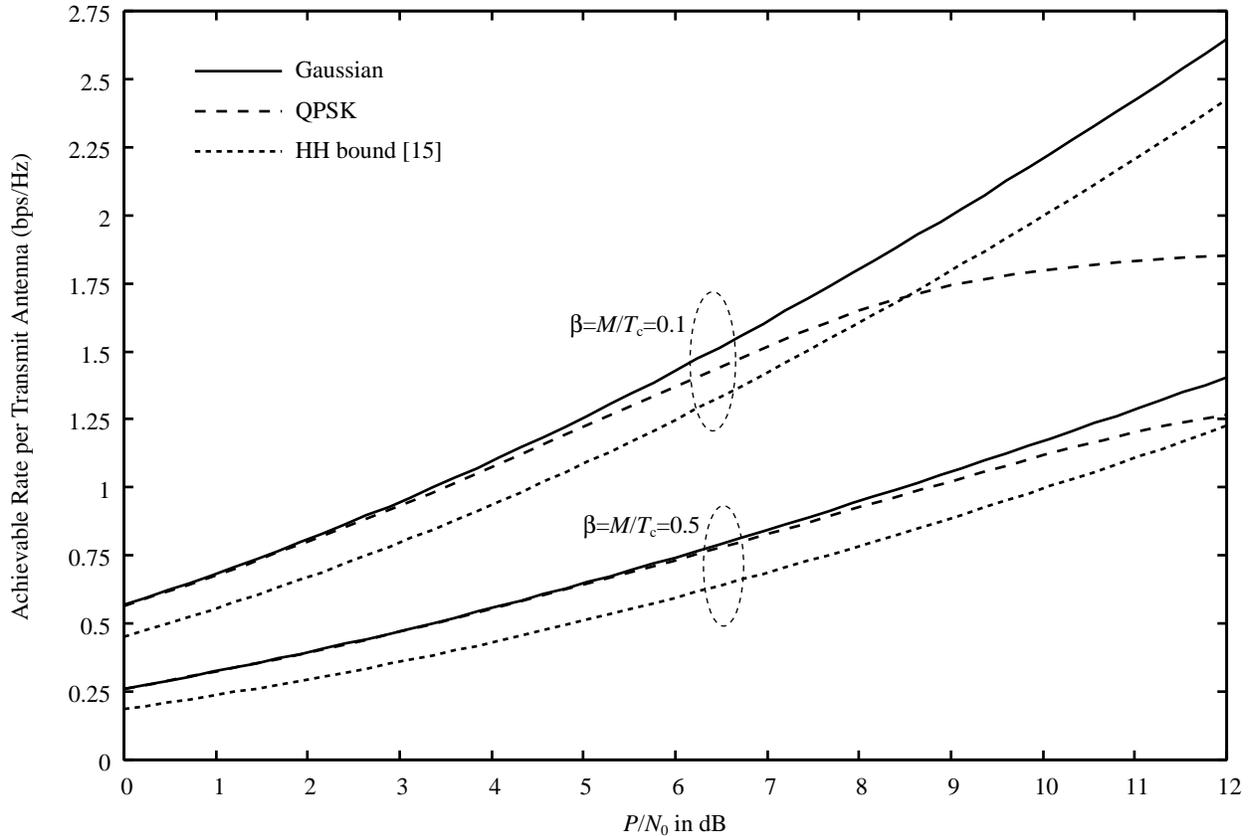}
\caption{
Achievable rate versus SNR in the moderate SNR regime for 
the variance of the bias~$\sigma_{\theta}^{2}=0$, $\alpha=M/N=1$, 
and $\tau_{0}=T_{\mathrm{tr}}/T_{\mathrm{c}}=0$. 
}
\label{fig3} 
\end{figure}

Figure~\ref{fig3} provides a comparison between the lower 
bound~(\ref{lower_bound_Gauss}) for unbiased Gaussian signaling and the HH 
bound in the moderate SNR regime. The lower bound~(\ref{capacity_lower_bound}) 
for QPSK modulation is also displayed. The HH bound is a lower bound on the 
capacity for finite-sized systems~\cite[Theorem~3]{Hassibi03}. We have used a 
large-system formula of the HH bound, which is easily derived in the same 
manner as in \cite{Verdu99}. There is a significant gap of $1$~dB to 
$1.8$~dB between the lower 
bound for unbiased Gaussian signaling and the HH bound for all SNRs. Moreover, 
the HH bound is inferior even to the lower bound for QPSK modulation in the 
case of short coherence time ($\beta=0.5$). These observations imply that 
SD receivers can provide a substantial performance gain in the moderate SNR 
regime, compared to receivers based on one-shot channel estimation, since 
they can reduce overhead for training significantly. 

\begin{figure}[t]
\includegraphics[width=\hsize]{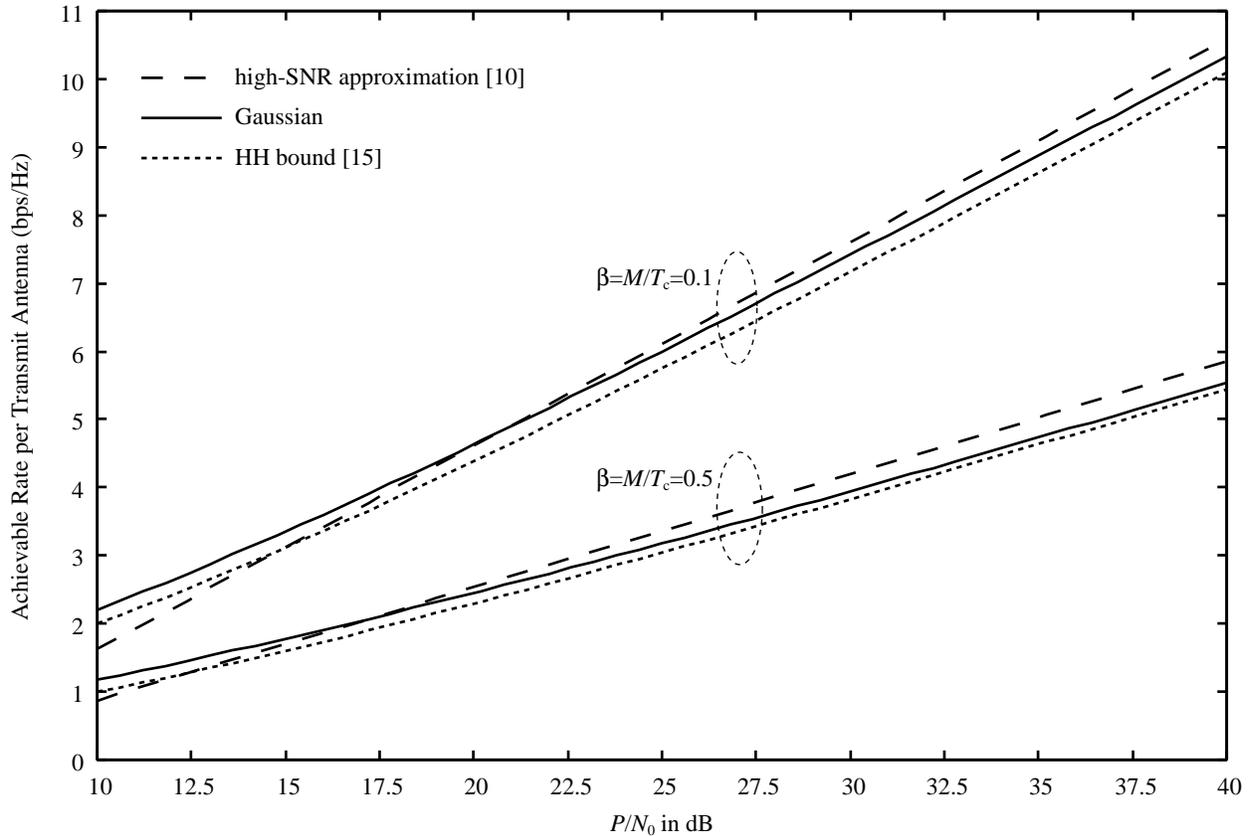}
\caption{
Achievable rate versus SNR in the high SNR regime 
for the variance of the bias~$\sigma_{\theta}^{2}=0$, $\alpha=M/N=1$, 
and $\tau_{0}=T_{\mathrm{tr}}/T_{\mathrm{c}}=0$. 
}
\label{fig4} 
\end{figure}

Next, we compare the lower bound for unbiased Gaussian signaling with the 
high-SNR approximation of the capacity~\cite[Corollary~11]{Zheng02} 
in the high-SNR regime in Fig.~\ref{fig4}. The HH bound is also displayed 
in the same figure. Note that in the high-SNR approximation the large-system 
limit is taken after the high SNR limit. Thus, the comparison makes sense 
under the assumption that the large-system limit and the high SNR limit 
commute for the high-SNR approximation. We find that the high-SNR 
approximation is smaller than the lower bound for unbiased Gaussian signaling 
in the SNR region of below $17.5$~dB for 
$\beta=0.5$ or below $20$~dB for $\beta=0.1$. 
This implies that the high-SNR approximation derived 
by Zheng and Tse~\cite{Zheng02} is valid only for quite high SNR.  
The lower bound for unbiased Gaussian signaling is close to the HH bound, 
rather than the high-SNR approximation, in the quite high SNR regime, which 
indicates the suboptimality of Gaussian signaling in the quite high SNR regime. 

\begin{figure}[t]
\includegraphics[width=\hsize]{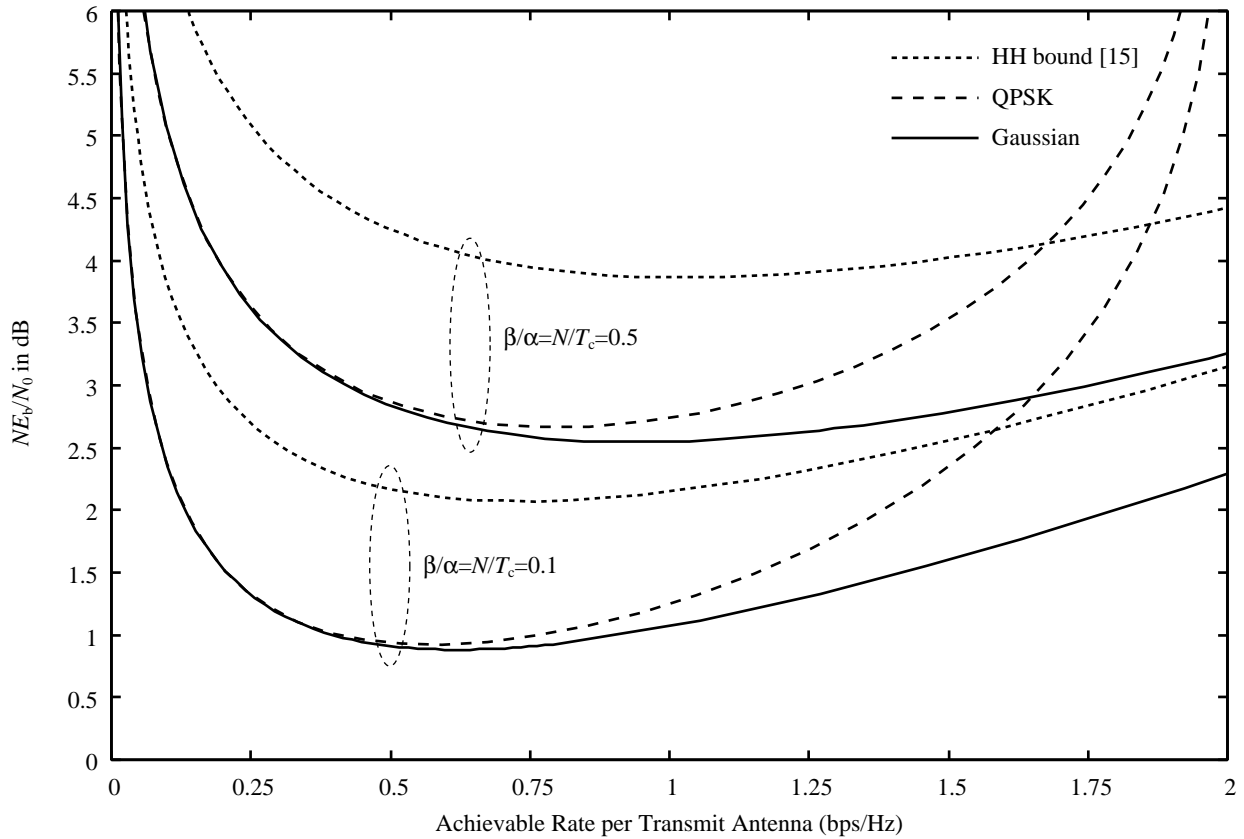}
\caption{
Normalized SNR versus achievable rate in the low SNR regime for 
the variance of the bias~$\sigma_{\theta}^{2}=0$ and 
$\tau_{0}=T_{\mathrm{tr}}/T_{\mathrm{c}}=0$. 
}
\label{fig5} 
\end{figure}

Finally, we consider the low SNR regime and take the limit described in 
Proposition~\ref{corollary3}, in which the power per information bit 
$E_{\mathrm{b}}$ required for reliable communication is a key performance 
measure. Figure~\ref{fig5} displays the upper bound~(\ref{normalized_SNR}) 
of $NE_{\mathrm{b}}/N_{0}$ required for the SD receiver with unbiased Gaussian 
signaling as a function of the achievable rate per transmit antenna. The 
normalized SNRs $NE_{\mathrm{b}}/N_{0}$ are also plotted for the SD receiver 
with QPSK modulation and for the HH bound. We find that 
$NE_{\mathrm{b}}/N_{0}$ has a minimum at a positive achievable rate. 
This observation is due to the suboptimality of i.i.d.\ Gaussian signaling. 
If perfect CSI was available at the receiver, the normalized SNR 
$NE_{\mathrm{b}}/N_{0}$ would be monotonically decreasing with the reduction 
of the achievable rate~\cite{Verdu02}, since i.i.d.\ Gaussian signaling is 
optimal in that case. For the case of no CSI, however, the normalized SNR 
$NE_{\mathrm{b}}/N_{0}$ diverges as the achievable rate tends to zero, since 
i.i.d.\ signaling wastes valuable power in the low SNR regime, as discussed in 
Section~\ref{sec_low_SNR}. Another observation is that there is a large  
gap of $1$~dB to $1.5$~dB between the minimal normalized SNRs for the SD 
receivers and the HH bound, while all bounds are far from the ultimate limit 
$NE_{\mathrm{b}}/N_{0}\approx-1.59$~dB. This result implies that the SD scheme 
can significantly improve the HH bound in the low SNR regime. 

\subsection{Finite-Sized Systems} 
\begin{figure}[t]
\includegraphics[width=\hsize]{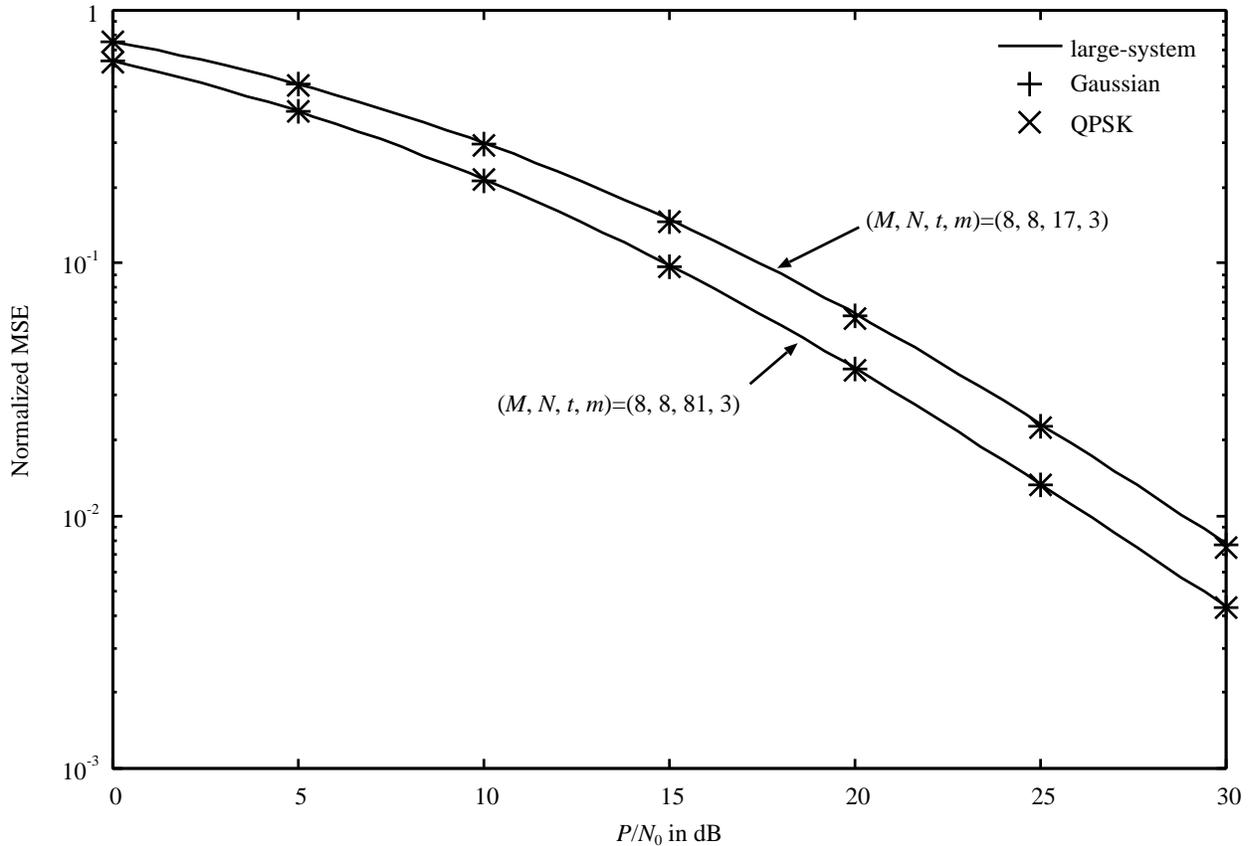}
\caption{
Normalized MSE for the LMMSE detector versus SNR in substage~$m$ 
within stage~$t$ for the SD receiver. $m=3$, $t=17,81$, $M=8$ transmit 
antennas, $N=8$ receive antennas, and the variance of the 
bias~$\sigma_{\theta}^{2}=0$. 
}
\label{fig6} 
\end{figure}

We have so far considered the large-system performance. In order to 
confirm the usefulness of the large-system analysis in practice, 
numerical simulations are presented for finite-sized systems with no CSI. 
Since the optimal detector has high complexity, only the performance of 
the LMMSE detector is investigated. Unbiased QPSK and unbiased Gaussian 
signaling are considered. Note that, for the unbiased case, the performance 
of the LMMSE channel estimator depends on stage~$t$, rather than 
the coherence time $T_{\mathrm{c}}$. Figure~\ref{fig6} shows the normalized MSE 
for the LMMSE estimate~(\ref{LMMSE_detector}) of the data symbols in 
substage~$m$ within stage~$t$ for the SD receiver.  
For comparison, we plot the large-system results based on 
Proposition~\ref{proposition_LMMSE} with $\alpha=M/N$, 
$\beta/\tau=M/(t-1)$, and $\mu=(m-1)/M$. The normalized MSE is given by 
the expectation in the RHS of (\ref{fixed_point_Gauss}) divided by $P$ in 
the large-system limit. A correction of minus one for the 
stage index~$t$ is because the performance in stage~$t$ is approximated by 
that in stage~$t+1$ in the large-system limit. See 
(\ref{z_tr})--(\ref{AWGN_channel_estimation}). A correction for the substage 
index is also due to the same reason. We find that the large-system 
results are in good agreement with those for not so large systems. 

\section{Conclusions} \label{sec_conclusion}  
We have investigated the achievable rates of SD receivers for the Rayleigh 
block-fading MIMO channel with no CSI. 
Analytical formulas for the achievable rates have been derived 
in the large-system limit, by using the replica method. 
It has been shown that negligible pilot information is best in terms of 
the information-theoretical achievable rates. From a theoretical 
point of view, the formulas provide the best lower bound on the capacity 
among existing {\em analytical} lower bounds that can be easily evaluated 
for all SNRs, while it is far from the true capacity in the low or quite high 
SNR regimes. From a practical point of view, the analytical lower bounds 
derived in this paper can be regarded as a fundamental performance limit for 
practical training-based systems with QPSK or multilevel modulation. 
We conclude that the SD receiver can reduce overhead for training 
significantly. Thus, it provides a substantial performance gain, 
compared to receivers based on one-shot channel estimation, especially in 
the low-to-moderate SNR regime. 

One important future work is to investigate spatially correlated MIMO channels 
with no CSI. On the one hand, spatial correlations cause a reduction of 
diversity. On the other hand, they make it possible to estimate the channel 
matrix more accurately than without correlations, since one can utilize the 
knowledge about the correlations for channel estimation. Thus, it should be 
worth investigating impacts of these two effects onto the performance 
for training-based systems. 
It does not seem to be straightforward to extend the results presented in this 
paper to the case of spatially correlated MIMO channels. 

\appendices

\section{Derivation of Proposition~\ref{proposition1}} \label{sec_derivation} 
\subsection{Sketch} \label{sec_sketch} 
Let us consider substage~$m=\mu M$ within stage~$t=\tau T_{\mathrm{c}}$ in 
the SD receiver for $0\leq\tau\leq1$ and $0\leq\mu\leq1$. For some 
$\tilde{m}\in\mathbb{N}$, we decompose the lower 
bound~(\ref{capacity_nonlinear}) into two terms, 
\begin{equation} \label{capacity_tmp} 
\frac{C}{M} \geq \frac{1}{T_{\mathrm{c}}M}
\sum_{t=T_{\mathrm{tr}}+1}^{T_{\mathrm{c}}}\left[ 
 \sum_{m=1}^{\tilde{m}}I(x_{m,t};\tilde{x}_{m,t} | \tilde{\mathcal{I}}_{t},
 \boldsymbol{x}_{[1,m),t},\boldsymbol{\theta}_{[m,M],t}) 
 + \sum_{m=\tilde{m}+1}^{M}I(x_{m,t};\tilde{x}_{m,t} | \tilde{\mathcal{I}}_{t},
 \boldsymbol{x}_{[1,m),t},\boldsymbol{\theta}_{[m,M],t})
\right], 
\end{equation} 
and then take the limit in which $M$, $N$, $T_{\mathrm{c}}$, 
$T_{\mathrm{tr}}$, $t$, $m$, and $\tilde{m}$ tend to infinity while their 
ratios $\alpha=M/N$, $\beta=M/T_{\mathrm{c}}$, 
$\tau_{0}=T_{\mathrm{tr}}/T_{\mathrm{c}}$, $\tau=t/T_{\mathrm{c}}$, 
$\mu=m/M$, and $\mu_{0}=\tilde{m}/M$  are kept constant. 
The second term consists of the mutual information for the decoding problem 
after an extensive number of users have been decoded, while the 
first term contains the mutual information for the problem after a finite 
number of users have been decoded. We will show below that 
$I(x_{m,t};\tilde{x}_{m,t} | \tilde{\mathcal{I}}_{t},
\boldsymbol{x}_{[1,m),t},
\boldsymbol{\theta}_{[m,M],t})$ for $\mu\geq\mu_{0}$ converges to 
the integrand in (\ref{capacity_lower_bound}) in the large-system limit. 
The definition of the Riemann integral implies that 
the sum $(T_{\mathrm{c}}M^{-1})\sum_{t=T_{\mathrm{tr}}+1}^{T_{\mathrm{c}}}
\sum_{m=\tilde{m}+1}^{M}$ in the second term of (\ref{capacity_tmp}) tends to 
$\int_{\tau_{0}}^{1}d\tau\int_{\mu_{0}}^{1}d\mu$. 
Taking the limit $\mu_{0}\rightarrow0$, we arrive at 
Proposition~\ref{proposition1}, since the first 
term of (\ref{capacity_tmp}) tends to zero as $\mu_{0}\rightarrow0$.   

The evaluation of $I(x_{m,t};\tilde{x}_{m,t} | \tilde{\mathcal{I}}_{t},
\boldsymbol{x}_{[1,m),t},
\boldsymbol{\theta}_{[m,M],t})$ consists of two parts: analysis 
of the error covariance matrix~(\ref{error_covariance}) and analysis of 
the equivalent channel~(\ref{equivalent_channel}). We apply the replica method 
to both analyses. 

\subsection{Analysis of Channel Estimator} 
We evaluate the error covariance matrix~(\ref{error_covariance}) for the LMMSE 
channel estimation in the large-system limit. Since the joint 
posterior pdf $p(\boldsymbol{H}|\tilde{\mathcal{I}}_{t})$ is decomposed into 
$\prod_{n=1}^{N}p(\vec{\boldsymbol{h}}_{n}|\tilde{\mathcal{I}}_{t})$, 
without loss of generality, we focus on the estimation 
problem for the first row of $\boldsymbol{H}$, denoted by 
$\vec{\boldsymbol{h}}_{1}$. The first row vector
$\vec{\boldsymbol{y}}_{\backslash t,1}\in
\mathbb{C}^{1\times (T_{\mathrm{c}}-1)}$ 
of (\ref{MIMO_channel_estimation}) is given by 
\begin{equation} \label{virtual_MIMO_n} 
\vec{\boldsymbol{y}}_{\backslash t,1} = \frac{1}{\sqrt{M}}
\vec{\boldsymbol{h}}_{1}\bar{\boldsymbol{X}}_{\backslash t} 
+ (\vec{\boldsymbol{0}},\vec{\boldsymbol{w}}_{(t,T_{\mathrm{c}}],1}) 
+ \vec{\boldsymbol{n}}_{\backslash t,1}, 
\end{equation} 
where $\vec{\boldsymbol{w}}_{(t,T_{\mathrm{c}}],1}
\in\mathbb{C}^{1\times (T_{\mathrm{c}}-t)}$ and 
$\vec{\boldsymbol{n}}_{\backslash t,1}
\in\mathbb{C}^{1\times (T_{\mathrm{c}}-1)}$ denote the first row vectors of 
the matrices $\boldsymbol{W}_{(t,T_{\mathrm{c}}]}$ and 
$\boldsymbol{N}_{\backslash t}$, respectively. 

The channel~(\ref{virtual_MIMO_n}) can be regarded as a MIMO channel 
with the channel matrix $\bar{\boldsymbol{X}}_{\backslash t}$ known to the 
receiver. The main difference between $\bar{\boldsymbol{X}}_{\backslash t}$ 
and zero-mean channel matrices considered in previous 
works~\cite{Tse99,Verdu99,Evans00,Tanaka02,Mueller04,Guo05} is that 
$\bar{\boldsymbol{X}}_{\backslash t}$ has the nonzero mean 
$\hat{\boldsymbol{X}}_{\backslash t}=
(\boldsymbol{O},\boldsymbol{\Theta}_{(T_{\mathrm{tr}},t)},
\boldsymbol{\Theta}_{(t,T_{\mathrm{c}}]})
\in\mathbb{C}^{M\times (T_{\mathrm{c}}-1)}$ 
conditioned on $\boldsymbol{\Theta}$. 
Let us decompose the channel matrix $\bar{\boldsymbol{X}}_{\backslash t}$ into 
the mean $\hat{\boldsymbol{X}}_{\backslash t}$ and the difference 
$\bar{\boldsymbol{X}}_{\backslash t}-\hat{\boldsymbol{X}}_{\backslash t}$. 
The problem would reduce to the zero-mean case if the two matrices were 
independent of each other, since the sum of two independent matrices 
with i.i.d.\ zero-mean entries is also a matrix with i.i.d.\ zero-mean 
entries. However, the two matrices are not independent 
while they are uncorrelated zero-mean matrices. Thus, we have to treat 
the influence of higher-order correlations carefully. 
  
The following proposition implies that the large-system results for each 
element of the error covariance matrix~(\ref{error_covariance})  
coincide with those for the case as if the two matrices 
$\hat{\boldsymbol{X}}_{\backslash t}$ and 
$\bar{\boldsymbol{X}}_{\backslash t}-\hat{\boldsymbol{X}}_{\backslash t}$ were  
mutually independent. In other words, higher-order correlations between 
the two matrices do not affect the results for each element of the 
error covariance matrix~(\ref{error_covariance}) in the large-system limit. 
Note that we do not claim the norm convergence 
$\|\boldsymbol{\Xi}_{t}-\xi^{2}(\tau)
\boldsymbol{I}_{M}\|\rightarrow0$. 

\begin{proposition} \label{proposition2} 
Suppose that Assumption~\ref{assumption1} and the RS assumption hold. Then, 
each diagonal element of the error covariance matrix~(\ref{error_covariance}) 
converges in probability to (\ref{xi}), defined 
by (\ref{fixed_point_tr}) and (\ref{fixed_point_c}), in the large-system 
limit. Furthermore, each off-diagonal element of 
the error covariance matrix~(\ref{error_covariance}) converges in probability 
to zero in the large-system limit.  
\end{proposition} 
\begin{IEEEproof} 
See Appendix~\ref{sec_deriv_proposition2}. 
\end{IEEEproof}

Proposition~\ref{proposition2} was rigorously proved without 
Assumption~\ref{assumption1} for the unbiased case $\theta_{m,t}=0$ in 
\cite{Evans00}. Since we cannot claim the norm convergence 
$\|\boldsymbol{\Xi}_{t}-\xi^{2}(\tau)
\boldsymbol{I}_{M}\|\rightarrow0$, a careful 
treatment of $\boldsymbol{\Xi}_{t}$ is required in the analysis of the 
equivalent channel~(\ref{equivalent_channel}).   

We remark that the convergence of each off-diagonal element to zero results 
from the fact that the MMSE estimate $\hat{\boldsymbol{h}}_{1}$ of 
$\vec{\boldsymbol{h}}_{1}$ and 
its error $\vec{\boldsymbol{h}}_{1}-\hat{\boldsymbol{h}}_{1}$ are uncorrelated 
with each other. In fact, we can show a stronger result for the off-diagonal 
elements without the replica method. 

\begin{lemma} \label{lemma1}
Suppose that Assumption~\ref{assumption1} holds. For a constant $A\geq0$, 
each off-diagonal element $(\boldsymbol{\Xi}_{t})_{\tilde{m},\tilde{m}'}$ 
($\tilde{m}\neq \tilde{m}'$) of (\ref{error_covariance}) satisfies 
\begin{equation}  
\limsup_{M\rightarrow\infty}M^{3/4}
|(\boldsymbol{\Xi}_{t})_{\tilde{m},\tilde{m}'}|=A
\quad \hbox{in probability,} 
\end{equation}
where the limit denotes the large-system limit. 
\end{lemma} 
\begin{IEEEproof}
We use the fact that the covariance matrices $\boldsymbol{\Xi}_{t}$ and 
$\boldsymbol{I}-\boldsymbol{\Xi}_{t}$ for 
$\vec{\boldsymbol{h}}_{1}-\hat{\boldsymbol{h}}_{1}$ and
$\hat{\boldsymbol{h}}_{1}$ are positive definite. 
Let $\{\lambda_{\tilde{m}}>0:\tilde{m}=1,\ldots,M\}$ denote the eigenvalues of 
$\boldsymbol{\Xi}_{t}$. The positive definiteness of 
$\boldsymbol{I}-\boldsymbol{\Xi}_{t}$ implies 
$1-\lambda_{\tilde{m}}>0$ for all $\tilde{m}$, or, 
$0<\lambda_{\tilde{m}}<1$ for all $\tilde{m}$. This observation implies that 
$\boldsymbol{I}_{M}-\boldsymbol{\Xi}_{t}^{k}$ is also positive definite 
for any $k\in\mathbb{N}$, or 
\begin{equation} \label{inequality} 
\limsup_{M\rightarrow\infty}\frac{1}{M}\mathrm{Tr}(\boldsymbol{\Xi}_{t}^{k}) 
< 1 
\end{equation} 
in the large-system limit. Note that Assumption~\ref{assumption1} implies 
that the left-hand side (LHS) of (\ref{inequality}) tends to the expected 
one $M^{-1}\mathrm{Tr}(\mathbb{E}[\boldsymbol{\Xi}_{t}^{k}])$. 

In order to prove Lemma~\ref{lemma1}, we evaluate  
$\mathrm{Tr}(\mathbb{E}[\boldsymbol{\Xi}_{t}^{4}])$. 
Let $\rho_{t}$ denote the elements of $\mathbb{E}[\boldsymbol{\Xi}_{t}]$ 
in the strictly upper triangular part. 
A direct calculation implies that the the leading term of 
$\mathrm{Tr}(\mathbb{E}[\boldsymbol{\Xi}_{t}^{4}])$ is given by 
$|\rho_{t}|^{2}M^{4}(2|\rho_{t}|^{2} - \Re[\rho_{t}^{2}])/3$ 
as $M\rightarrow\infty$. 
Applying this result and $\Re[\rho_{t}^{2}]\leq |\rho_{t}|^{2}$ 
to (\ref{inequality}), we have 
\begin{equation}
\limsup_{M\rightarrow\infty}M^{3}|\rho_{t}|^{4} < 3, 
\end{equation} 
which implies that Lemma~\ref{lemma1} holds. 
\end{IEEEproof}

Lemma~\ref{lemma1} implies $(\boldsymbol{\Xi}_{t})_{\tilde{m},\tilde{m}'}=
O(M^{-3/4})$ in the large-system limit. 
We believe that it is possible to prove 
$(\boldsymbol{\Xi}_{t})_{\tilde{m},\tilde{m}'}=O(M^{-1})$ by calculating 
$\mathrm{Tr}(\mathbb{E}[\boldsymbol{\Xi}_{t}^{k}])$  and 
taking $k\rightarrow\infty$ after the large-system limit. 
However, Lemma~\ref{lemma1} is sufficient for deriving 
Proposition~\ref{proposition1}. 

\subsection{Analysis of Detector} 
We focus on substage~$m$ within stage~$t$ and analyze the equivalent 
channel~(\ref{equivalent_channel}) in the large-system limit. 
It is shown that the equivalent channel reduces to a MIMO channel with 
perfect CSI at the receiver in the large-system limit. 
Let $\boldsymbol{\Xi}_{t}^{(\mathrm{c})}\in\mathbb{C}^{(M-m+1)\times(M-m+1)}$ 
denote the posterior covariance matrix 
of $(h_{1,m},\ldots,h_{1,M})^{\mathrm{T}}\in\mathbb{C}^{M-m+1}$ given 
$\tilde{\mathcal{I}}_{t}$, i.e., the bottom-right block of the 
error covariance matrix~(\ref{error_covariance}), 
\begin{equation}
\boldsymbol{\Xi}_{t} = 
\begin{pmatrix}
* & * \\ 
* & \boldsymbol{\Xi}_{t}^{(\mathrm{c})}
\end{pmatrix}. 
\end{equation}
The equivalent MIMO channel with perfect CSI at the receiver is defined as    
\begin{equation} \label{deterministic_MIMO} 
\underline{\boldsymbol{z}} = \alpha^{-1/2}
\sqrt{\boldsymbol{I}-\boldsymbol{\Xi}_{t}^{(\mathrm{c})}}
\boldsymbol{x}_{[m,M],t} + \underline{\boldsymbol{w}}, 
\end{equation}
with $\underline{\boldsymbol{w}}\sim\mathcal{CN}(\boldsymbol{0},\sigma^{2}
\boldsymbol{I}_{M-m+1})$. In (\ref{deterministic_MIMO}), 
the matrix $\sqrt{\boldsymbol{I}-\boldsymbol{\Xi}_{t}^{(\mathrm{c})}}$  
denotes a squared root of 
$\boldsymbol{I}-\boldsymbol{\Xi}_{t}^{(\mathrm{c})}$, i.e., 
$\boldsymbol{I}-\boldsymbol{\Xi}_{t}^{(\mathrm{c})}=
\sqrt{\boldsymbol{I}-\boldsymbol{\Xi}_{t}^{(\mathrm{c})}}^{\mathrm{H}}
\sqrt{\boldsymbol{I}-\boldsymbol{\Xi}_{t}^{(\mathrm{c})}}$. 

The equivalent channel between $x_{m,t}$ and the associated 
decoder for the MIMO channel~(\ref{deterministic_MIMO}) with perfect CSI at 
the receiver is given by 
\begin{IEEEeqnarray}{rl}  
& p(\tilde{x}_{m,t} | x_{m,t},\boldsymbol{\Xi}_{t}^{(\mathrm{c})},
\boldsymbol{\theta}_{[m,M],t}) \nonumber \\ 
=& \int p(x_{m,t}=\tilde{x}_{m,t} | 
\underline{\boldsymbol{z}},\boldsymbol{\Xi}_{t}^{(\mathrm{c})},
\boldsymbol{\theta}_{[m,M],t})
p(\underline{\boldsymbol{z}} | \boldsymbol{\Xi}_{t}^{(\mathrm{c})},
\boldsymbol{x}_{[m,M],t}) 
p(\boldsymbol{x}_{(m,M],t}
|\boldsymbol{\theta}_{(m,M],t})
d\boldsymbol{x}_{(m,M],t}d\underline{\boldsymbol{z}}, 
\label{equivalent_channel_deterministic}
\end{IEEEeqnarray}
where $p(x_{m,t} | \underline{\boldsymbol{z}},\boldsymbol{\Xi}_{t}^{(\mathrm{c})},
\boldsymbol{\theta}_{[m,M],t})$ represents the pdf of $x_{m,t}$ 
conditioned on $\underline{\boldsymbol{z}}$, $\boldsymbol{\Xi}_{t}^{(\mathrm{c})}$, and 
$\boldsymbol{\theta}_{[m,M],t}$. 

\begin{proposition} \label{proposition3} 
Suppose that Assumption~\ref{assumption2} and the RS assumption hold. Then, 
the equivalent channel~(\ref{equivalent_channel}) 
converges in law to the equivalent 
channel~(\ref{equivalent_channel_deterministic}) for the MIMO 
channel~(\ref{deterministic_MIMO}) with perfect CSI at the receiver in the 
large-system limit. In evaluating (\ref{equivalent_channel_deterministic}), 
the variance $\sigma^{2}$ of $\underline{\boldsymbol{w}}$ is given as the solution to the 
fixed-point equation, 
\begin{equation} \label{fixed_point_MIMO} 
\sigma^{2} = N_{0} + \lim_{M\rightarrow\infty}\frac{P}{M}
 \mathrm{Tr}(\boldsymbol{\Xi}_{t}) 
 + V(\sigma^{2}), 
\end{equation}  
with 
\begin{equation} \label{MMSE} 
V(\sigma^{2}) = \lim_{M\rightarrow\infty}\frac{1}{M}
\mathbb{E}\left[
 \left. 
  (\boldsymbol{x}_{[m,M],t} 
  - \langle \boldsymbol{x}_{[m,M],t} \rangle)^{\mathrm{H}} 
  (\boldsymbol{I} - \boldsymbol{\Xi}_{t}^{(\mathrm{c})})
  (\boldsymbol{x}_{[m,M],t} 
  - \langle \boldsymbol{x}_{[m,M],t} \rangle)
 \right| \boldsymbol{\Xi}_{t}^{(\mathrm{c})},
 \boldsymbol{\theta}_{[m,M],t} 
\right], 
\end{equation}
where $\langle \boldsymbol{x}_{[m,M],t} \rangle$ denotes the 
mean of $\boldsymbol{x}_{[m,M],t}$ with respect to 
the posterior pdf $p(\boldsymbol{x}_{[m,M],t} | \underline{\boldsymbol{z}},
\boldsymbol{\Xi}_{t}^{(\mathrm{c})},\boldsymbol{\theta}_{[m,M],t})$. 
If there are multiple solutions, one should choose the solution $\sigma^{2}$ 
minimizing the following quantity 
\begin{equation} \label{free_energy_MIMO} 
\lim_{M\rightarrow\infty}\frac{1}{M}
I(\boldsymbol{x}_{[m,M],t};\underline{\boldsymbol{z}}) 
+ \frac{1}{\alpha}\left[
 D_{2}(N_{0}\|\sigma^{2}) + \lim_{M\rightarrow\infty}
 \frac{\log_{2}\mathrm{e}}{\sigma^{2}M}\mathrm{Tr}(\boldsymbol{\Xi}_{t}) 
\right], 
\end{equation}
where $I(\boldsymbol{x}_{[m,M],t};\underline{\boldsymbol{z}})$ denotes the 
mutual information between $\boldsymbol{x}_{[m,M],t}$ and 
$\underline{\boldsymbol{z}}$ given realizations of $\boldsymbol{\Xi}_{t}$ and 
$\boldsymbol{\theta}_{[m,M],t}$. 
\end{proposition}
\begin{IEEEproof}
See Appendix~\ref{sec_deriv_proposition3}. 
\end{IEEEproof} 

We have implicitly assumed that the equivalent 
channel~(\ref{equivalent_channel_deterministic}) and the last two terms  
in (\ref{fixed_point_MIMO}) converge as $M\rightarrow\infty$.   
This assumption is justified below by using Proposition~\ref{proposition2} 
and Lemma~\ref{lemma1}. 

Proposition~\ref{proposition3} implies that the mutual information 
$I(x_{m,t};\tilde{x}_{m,t}|\tilde{\mathcal{I}}_{t},
\boldsymbol{x}_{[1,m),t},
\boldsymbol{\theta}_{[m,M],t})$ tends to 
the constrained capacity $I(x_{m,t};\tilde{x}_{m,t}|
\boldsymbol{\Xi}_{t}^{(\mathrm{c})},\boldsymbol{\theta}_{[m,M],t})$ 
of the MIMO channel~(\ref{deterministic_MIMO}) with perfect CSI at the 
receiver in the large-system limit. 
In order to complete the derivation of Proposition~\ref{proposition1}, 
we show that $I(x_{m,t};\tilde{x}_{m,t}|
\boldsymbol{\Xi}_{t}^{(\mathrm{c})},\boldsymbol{\theta}_{[m,M],t})$ 
tends to the integrand in (\ref{capacity_lower_bound}). 
A proof of this statement is given in Appendix~\ref{sec_deriv_proposition1}. 
One may expect that if the convergence of each off-diagonal element of the 
error covariance matrix~(\ref{error_covariance}) to zero is 
fast enough, the off-diagonal elements of the channel matrix 
$\sqrt{\boldsymbol{I}-\boldsymbol{\Xi}_{t}^{(\mathrm{c})}}$ for the MIMO 
channel~(\ref{deterministic_MIMO}) with perfect CSI at the receiver are  
negligible. Thus, the MIMO channel~(\ref{deterministic_MIMO}) with 
perfect CSI at the receiver is decoupled into the bank of the AWGN 
channels~(\ref{AWGN}). The proof presented in 
Appendix~\ref{sec_deriv_proposition1} implies that the convergence 
speed shown in Lemma~\ref{lemma1}, i.e., 
$(\boldsymbol{\Xi}_{t}^{(\mathrm{c})})_{\tilde{m},\tilde{m}'}=
O(M^{-3/4})$ for $\tilde{m}\neq\tilde{m}'$, is fast enough.  
In order to explain this argument intuitively, 
we apply the matched filter (MF) $\boldsymbol{r}=\alpha^{-1/2} 
\sqrt{\boldsymbol{I}-\boldsymbol{\Xi}_{t}^{(\mathrm{c})}}^{\mathrm{H}}
\underline{\boldsymbol{z}}$ 
for the received vector $\underline{\boldsymbol{z}}$ of the MIMO 
channel~(\ref{deterministic_MIMO}) with perfect CSI at the receiver,  
\begin{equation} \label{MF_output} 
\boldsymbol{r} = \frac{\boldsymbol{\xi}_{t,m}}{\alpha}x_{m,t}
+ \sum_{m'=m+1}^{M}\frac{\boldsymbol{\xi}_{t,m'}}{\alpha}x_{m',t} 
+ \boldsymbol{\eta}, 
\end{equation}
with $\boldsymbol{\eta}\sim\mathcal{CN}(\boldsymbol{0}, 
\sigma^{2}(\boldsymbol{I}-\boldsymbol{\Xi}_{t}^{(\mathrm{c})})/\alpha)$. 
In (\ref{MF_output}), the vector $\boldsymbol{\xi}_{t,m'}$ 
denotes the $(m'-m+1)$th column vector of 
$\boldsymbol{I}-\boldsymbol{\Xi}_{t}^{(\mathrm{c})}$ for $m'=m,\ldots,M$. 
Note that the MF output vector~(\ref{MF_output}) contains sufficient 
information for the estimation of $\boldsymbol{x}_{[m,M],t}$. 
The magnitude of the inter-stream interference, given by the second term of 
the RHS in (\ref{MF_output}), would be proportional to the magnitude of each 
interfering signal multiplied by $M-m$ 
if a {\em constructive} superposition of all interfering signals occurred. 
However, it does not occur due to the independence of data symbols 
with high probability. 
On average, the magnitude of the inter-stream interference is proportional to 
the magnitude of each interfering signal multiplied by $\sqrt{M-m}$. 
Since the magnitude of each interfering signal is $O(M^{-3/4})$, the 
magnitude of inter-stream interference is $O(M^{-1/4})$. Thus,  
the inter-stream interference is negligible in the large-system limit.  

We have so far presented the derivation of Proposition~\ref{proposition1}. 
Finally, we discuss the performance degradation caused by using the LMMSE 
channel estimator. Let us consider the estimation problem for the 
first row vector $\vec{\boldsymbol{h}}_{1}$ of $\boldsymbol{H}$ based on 
the first row vector of the received matrix $\boldsymbol{Y}_{\backslash t}$, 
instead of $\tilde{\boldsymbol{Y}}_{\backslash t}$. 
It is worth noticing the similarity between this problem 
and the detection problem of $x_{m,t}$ in stage~$t$. This similarity allows 
us to analyze the performance of the optimal channel 
estimator~(\ref{posterior_H}) in 
the large-system limit. 

\begin{proposition} \label{proposition4} 
Suppose that the error covariance matrix for the optimal channel 
estimator~(\ref{posterior_H}) is self-averaging in the large-system limit. 
Under the RS assumption, then, each diagonal element of the error covariance 
matrix converges in probability to the same value as that for the LMMSE 
channel estimator, defined in Proposition~\ref{proposition2}, 
in the large-system limit.   
\end{proposition}

The derivation of Proposition~\ref{proposition4} is omitted since it is 
straightforwardly derived by combining the methods for deriving 
Propositions~\ref{proposition2} and~\ref{proposition3}. 
Proposition~\ref{proposition4} allows us to expect that the gap between 
the achievable rate of the optimal SD receiver and its lower 
bound~(\ref{capacity_lower_bound}) may be quite small in the large-system 
limit, although we cannot immediately conclude that the lower 
bound~(\ref{capacity_lower_bound}) is tight in the large-system limit.     

\section{Derivation of Proposition~\ref{proposition2}}
\label{sec_deriv_proposition2} 
\subsection{Formulation} 
It is sufficient from Assumption~\ref{assumption1} to show that 
the averaged quantities $\bar{\xi}_{t}^{2}=M^{-1}\sum_{\tilde{m}=1}^{M}
\mathbb{E}[(\boldsymbol{\Xi}_{t})_{\tilde{m},\tilde{m}}]$ and 
$\bar{\rho}_{t} = (M-1)^{-1}\sum_{\tilde{m}=2}^{M}\mathbb{E}[
(\boldsymbol{\Xi}_{t})_{1,\tilde{m}}]$
converge to $\xi^{2}(\tau)$ and zero, respectively, in the large-system limit,  
in which $M$, $T_{\mathrm{c}}$, $T_{\mathrm{tr}}$, and $t$ tend to 
infinity while $\beta=M/T_{\mathrm{c}}$, 
$\tau_{0}=T_{\mathrm{tr}}/T_{\mathrm{c}}$, and 
$\tau=t/T_{\mathrm{c}}$ are kept constant. 

For notational convenience, hereinafter, we drop the subscript $1$ 
in (\ref{virtual_MIMO_n}) from all variables. 
For example, $\vec{\boldsymbol{y}}_{\backslash t,1}$ and 
$\vec{\boldsymbol{h}}_{1}$ are 
written as $\vec{\boldsymbol{y}}_{\backslash t}$ and $\vec{\boldsymbol{h}}$, 
respectively.      
Let $\vec{\boldsymbol{h}}^{(a)}=(h_{1}^{(a)},\ldots,h_{M}^{(a)})
\in\mathbb{C}^{1\times M}$ denote replicas of $\vec{\boldsymbol{h}}$ 
for $a\in\mathbb{N}$: 
$\{\vec{\boldsymbol{h}}^{(a)}:a\in\mathbb{N}\}$ are i.i.d.\ random vectors 
drawn from $p(\vec{\boldsymbol{h}})$. Furthermore, 
we write $\vec{\boldsymbol{h}}$ as 
$\vec{\boldsymbol{h}}^{(0)}=(h_{1}^{(0)},\ldots,h_{M}^{(0)})$. 
The replica analysis is based on the following lemma. 
\begin{lemma} \label{lemma_ave_MSE} 
Let us define a function $Z_{n}(\omega;f)$ as 
\begin{equation} \label{partition} 
Z_{n}(\omega;f) = \mathbb{E}\left[
 \int \mathrm{e}^{M\omega f}
 \left\{
  \int p(\vec{\boldsymbol{y}}_{\backslash t}|\vec{\boldsymbol{h}},
  \bar{\boldsymbol{X}}_{\backslash t}) 
  p(\vec{\boldsymbol{h}})d\vec{\boldsymbol{h}}
 \right\}^{n-2} 
 \prod_{a=0}^{2}\left\{
  p(\vec{\boldsymbol{y}}_{\backslash t}|
  \vec{\boldsymbol{h}}=\vec{\boldsymbol{h}}^{(a)},
  \bar{\boldsymbol{X}}_{\backslash t})p(\vec{\boldsymbol{h}}^{(a)})
  d\vec{\boldsymbol{h}}^{(a)} 
 \right\}d\vec{\boldsymbol{y}}_{\backslash t}
\right], 
\end{equation}
with a complex function $f$ of $\{\vec{\boldsymbol{h}}^{(a)}:a=0,1,2\}$. 
In (\ref{partition}), 
$p(\vec{\boldsymbol{y}}_{\backslash t}|\vec{\boldsymbol{h}},
\bar{\boldsymbol{X}}_{\backslash t})$ 
represents the virtual MIMO channel~(\ref{virtual_MIMO_n}).  
For $n\geq0$ and $\omega\in{R}$,  
\begin{equation} \label{ave_MSE_r} 
\bar{\xi}_{t}^{2} = \lim_{n\rightarrow+0}\lim_{\omega\rightarrow0} 
\frac{1}{M}\frac{\partial}{\partial \omega}\ln Z_{n}(\omega;f_{1}), 
\end{equation}
\begin{equation} \label{covariance} 
\bar{\rho}_{t} = \lim_{n\rightarrow+0}\lim_{\omega\rightarrow0} 
\frac{1}{M}\frac{\partial}{\partial \omega}\ln Z_{n}(\omega;f_{2}), 
\end{equation}
where the functions $f_{1}$ and $f_{2}$ are given by 
$f_{1}=M^{-1}\sum_{\tilde{m}=1}^{M}f_{\tilde{m},\tilde{m}}$ and 
$f_{2}=(M-1)^{-1}\sum_{\tilde{m}=2}^{M}f_{1,\tilde{m}}$, respectively, with  
\begin{equation}
f_{\tilde{m},\tilde{m}'}=(h_{\tilde{m}}^{(0)} - h_{\tilde{m}}^{(1)})
(h_{\tilde{m}'}^{(0)} - h_{\tilde{m}'}^{(2)})^{*}.
\end{equation}  
\end{lemma}
\begin{IEEEproof}
We only present the proof of (\ref{ave_MSE_r}) since the proof of 
(\ref{covariance}) is the same as that of (\ref{ave_MSE_r}).  
Let $\hat{\boldsymbol{h}}_{t}\in\mathbb{C}^{1\times M}$ denote the first 
row vector of the LMMSE estimate~(\ref{LMMSE_estimate}), i.e., 
the mean of $\vec{\boldsymbol{h}}$ with respect to 
$p(\vec{\boldsymbol{h}}|\tilde{\mathcal{I}}_{t})$. Then, we have 
\begin{equation}
\bar{\xi}_{t}^{2} 
= \frac{1}{M}\mathbb{E}\left[
 \int (\vec{\boldsymbol{h}} - \vec{\boldsymbol{h}}^{(2)})^{\mathrm{H}}
 (\vec{\boldsymbol{h}} - \vec{\boldsymbol{h}}^{(1)})
 \prod_{a=1}^{2}\left\{
  p(\vec{\boldsymbol{h}}=\vec{\boldsymbol{h}}^{(a)}|
  \tilde{\mathcal{I}}_{t})d\vec{\boldsymbol{h}}^{(a)}
 \right\}
\right], 
\end{equation}
where we have used the fact that the error covariance 
matrix~(\ref{error_covariance}) is the posterior covariance of 
$\vec{\boldsymbol{h}}$ given $\tilde{\mathcal{I}}_{t}$.  
The introduction of a non-negative real number $n$ gives  
\begin{equation} \label{ave_MSE_tmp}
\bar{\xi}_{t}^{2} = \lim_{n\rightarrow+0}\mathbb{E}\left[
 \int f_{1}
 \left\{
  \int p(\vec{\boldsymbol{y}}_{\backslash t}|\vec{\boldsymbol{h}},
  \bar{\boldsymbol{X}}_{\backslash t}) 
  p(\vec{\boldsymbol{h}})d\vec{\boldsymbol{h}}
 \right\}^{n-2} 
 \prod_{a=0}^{2}\left\{ 
  p(\vec{\boldsymbol{y}}_{\backslash t}|
  \vec{\boldsymbol{h}}=\vec{\boldsymbol{h}}^{(a)}, 
  \bar{\boldsymbol{X}}_{\backslash t}) 
  p(\vec{\boldsymbol{h}}^{(a)})d\vec{\boldsymbol{h}}^{(a)} 
 \right\}d\vec{\boldsymbol{y}}_{\backslash t}
\right]. 
\end{equation}
It is straightforward to confirm that (\ref{ave_MSE_r}) is equivalent to  
(\ref{ave_MSE_tmp}), since $Z_{n}(\omega;f_{1})=1$ as $n,\omega\rightarrow0$. 
\end{IEEEproof}

It is difficult to evaluate (\ref{partition}) for a real number $n$. 
The main trick of the replica method is that $n$ is regarded as a 
non-negative integer in evaluating (\ref{partition}). 
For $n=2,3,\ldots$, we have a simple expression of (\ref{partition}), 
\begin{equation} \label{partition_tmp1} 
Z_{n}(\omega;f) = \mathbb{E}\left[
 \mathrm{e}^{M\omega f} \int \prod_{a=0}^{n}p(\vec{\boldsymbol{y}}_{\backslash t}|
 \vec{\boldsymbol{h}}=\vec{\boldsymbol{h}}^{(a)},
 \bar{\boldsymbol{X}}_{\backslash t})d\vec{\boldsymbol{y}}_{\backslash t}
\right]. 
\end{equation} 
In order to use Lemma~\ref{lemma_ave_MSE}, we have to take the operations with 
respect to $\omega$ before the large-system limit. However, 
we need to take the operations after the large-system limit, since it is 
possible to get an analytical formula of (\ref{partition_tmp1}) only in the 
large-system limit, as shown in the next section. We circumvent this 
dilemma by assuming the commutativity of the large-system limit and the 
operations. 
\begin{assumption} \label{assumption_commutativity} 
For a non-negative integer $n$, 
\begin{equation}
\lim_{M\rightarrow\infty}\lim_{\omega\rightarrow0}
\frac{1}{M}\frac{\partial}{\partial \omega}\ln Z_{n}(\omega;f) 
= \lim_{\omega\rightarrow0}
\frac{\partial}{\partial \omega}\lim_{M\rightarrow\infty}
\frac{1}{M}\ln Z_{n}(\omega;f), 
\end{equation}
where $\lim_{M\rightarrow\infty}$ denotes the large-system limit. 
\end{assumption}

An analytical formula of (\ref{partition_tmp1}) obtained in the large-system 
limit is not generally defined for $n\geq0$. In order to {\em predict} the 
correct asymptotic formula of (\ref{partition}) in a neighborhood of $n=0$, 
we will assume a symmetric statistics with respect to replica indices, 
called the RS assumption. Assuming that the order of the large-system 
limit and the operations with respect to $n$ and $\omega$ in (\ref{ave_MSE_r}) 
and (\ref{covariance}) is commutative, we obtain analytical expressions of 
(\ref{ave_MSE_r}) and (\ref{covariance}) in the 
large-system limit. It is a challenging open problem to prove whether these 
assumptions are valid or whether the obtained result is correct. 

\subsection{Average over Non-Replicated Variables} 
In this section, we evaluate the expectations in (\ref{partition_tmp1}) with 
respect to the non-replicated variables 
$\boldsymbol{\Theta}_{(T_{\mathrm{tr}},t)}$ and 
$\bar{\boldsymbol{X}}_{\backslash t}
=(\boldsymbol{X}_{\mathcal{T}_{T_{\mathrm{tr}}}},
\boldsymbol{X}_{(T_{\mathrm{tr}},t)},
\boldsymbol{\Theta}_{(t,T_{\mathrm{c}}]})$. 
The matrix $\bar{\boldsymbol{X}}_{\backslash t}$ consists of three kinds of 
random vectors:  
$\{\boldsymbol{x}_{t'}\}$ for $t'=1,\ldots,T_{\mathrm{tr}}$ are the pilot 
symbol vectors, $\{\boldsymbol{x}_{t'}\}$ for $t'=T_{\mathrm{tr}}+1,\ldots,
t-1$ are the data symbol vectors decoded in the preceding stages, and 
$\{\boldsymbol{\theta}_{t'}\}$ for $t'=t+1,\ldots,T_{\mathrm{c}}$ 
are the bias vectors for the data symbol vectors unknown in the current stage. 
Since the elements of $\vec{\boldsymbol{y}}_{\backslash t}$, given by 
(\ref{virtual_MIMO_n}), are mutually independent conditioned on 
$\mathcal{H}=\{\vec{\boldsymbol{h}}^{(a)}:a=0,1,\ldots\}$ 
and $\bar{\boldsymbol{X}}_{\backslash t}$, 
(\ref{partition_tmp1}) yields 
\begin{equation}  
Z_{n}(\omega;f) = \mathbb{E}\left\{
 \mathrm{e}^{M\omega f}
 \left[
  e_{n}(v_{\mathrm{p}}^{(a)},N_{0},\mathcal{H})
 \right]^{T_{\mathrm{tr}}}
 \left[
  e_{n}(v_{\mathrm{d}}^{(a)},N_{0},\mathcal{H})
 \right]^{t-T_{\mathrm{tr}}-1}
 \prod_{t'=t+1}^{T_{\mathrm{c}}} \left[
  e_{n}(v_{\mathrm{c}}^{(a)},N_{0}+P-\sigma_{t'}^{2},\mathcal{H})
 \right]
\right\}, \label{partition_tmp2}
\end{equation}
with
\begin{equation}
e_{n}(v^{(a)},\sigma^{2},\mathcal{H}) = \mathbb{E}\left[
 \left. 
 \int \prod_{a=0}^{n}g(y;v^{(a)},\sigma^{2})dy
 \right| \mathcal{H} 
\right], 
\end{equation}
where $g(y;v^{(a)},\sigma^{2})$ denotes the pdf of a proper complex Gaussian 
random variable $y\in\mathbb{C}$ with mean $v^{(a)}$ and 
variance $\sigma^{2}$.  
In (\ref{partition_tmp2}), $\sigma_{t'}^{2}$ is given by 
$\sigma_{t'}^{2}= M^{-1}\sum_{\tilde{m}=1}^{M}|\theta_{\tilde{m},t'}|^{2}$. 
Furthermore, $v_{\mathrm{p}}^{(a)}\in\mathbb{C}$, 
$v_{\mathrm{d}}^{(a)}\in\mathbb{C}$, and $v_{\mathrm{c}}^{(a)}\in\mathbb{C}$ 
are given by 
\begin{equation}
v_{\mathrm{p}}^{(a)} 
= \frac{1}{\sqrt{M}}\sum_{\tilde{m}=1}^{M}h_{\tilde{m}}^{(a)}x_{\tilde{m},1}, 
\end{equation}
\begin{equation}
v_{\mathrm{d}}^{(a)} = \frac{1}{\sqrt{M}}\sum_{\tilde{m}=1}^{M}
h_{\tilde{m}}^{(a)}x_{\tilde{m},t-1},  
\end{equation}
\begin{equation}
v_{\mathrm{c}}^{(a)} = \frac{1}{\sqrt{M}}\sum_{\tilde{m}=1}^{M}
h_{\tilde{m}}^{(a)}\theta_{\tilde{m},T_{\mathrm{c}}},  
\end{equation} 
respectively. 

We first evaluate $e_{n}(v_{\mathrm{p}}^{(a)},N_{0},\mathcal{H})$ 
in the large-system limit, following \cite{Tanaka02,Guo05}. 
Calculating the Gaussian integration with respect to $y$, we obtain 
\begin{equation}
e_{n}(v_{\mathrm{p}}^{(a)},N_{0},\mathcal{H}) 
= \frac{
 \mathbb{E}\left[
  \left. 
   \mathrm{e}^{-N_{0}^{-1}\boldsymbol{v}_{\mathrm{p}}^{\mathrm{H}}\boldsymbol{A}
   \boldsymbol{v}_{\mathrm{p}}
   }
  \right| \mathcal{H} 
 \right]
}
{
 (\pi N_{0})^{n}(1+n)
}, 
\end{equation}
with $\boldsymbol{v}_{\mathrm{p}}=(v_{\mathrm{p}}^{(0)},\ldots, 
v_{\mathrm{p}}^{(n)})^{\mathrm{T}}$ and 
\begin{equation} \label{A} 
\boldsymbol{A} = \frac{1}{(1+n)}
\begin{pmatrix}
n & -\boldsymbol{1}_{n}^{\mathrm{T}} \\
-\boldsymbol{1}_{n} & (1+n)\boldsymbol{I}_{n}
- \boldsymbol{1}_{n}\boldsymbol{1}_{n}^{\mathrm{T}} 
\end{pmatrix}. 
\end{equation}
In $M\rightarrow\infty$, due to the central limit theorem, 
$\boldsymbol{v}_{\mathrm{p}}$ conditioned on $\mathcal{H}$ converges in 
distribution to a CSCG random vector with 
the covariance matrix $P\boldsymbol{Q}$, given by  
\begin{equation}
\boldsymbol{Q} = \frac{1}{M}\sum_{\tilde{m}=1}^{M} 
\boldsymbol{h}_{\tilde{m}}\boldsymbol{h}_{\tilde{m}}^{\mathrm{H}}, 
\end{equation}
with $\boldsymbol{h}_{\tilde{m}}=(h_{\tilde{m}}^{(0)},\ldots,
h_{\tilde{m}}^{(n)})^{\mathrm{T}}$. Thus, 
\begin{equation} \label{e_p} 
e_{n}(v_{\mathrm{p}}^{(a)},N_{0},\mathcal{H}) 
= \exp\left\{
 G\left(
  \frac{P}{N_{0}}\boldsymbol{Q}
 \right)
\right\} + O(M^{-1}), 
\end{equation}
in the large-system limit, in which the function $G(\boldsymbol{Q})$ is 
given by 
\begin{equation} \label{function_G} 
G(\boldsymbol{Q}) 
= -\ln\det(\boldsymbol{I}_{n+1} + \boldsymbol{A}\boldsymbol{Q}) 
- n\ln(\pi N_{0}) - \ln(1+n). 
\end{equation}

We next calculate $ e_{n}(v_{\mathrm{c}}^{(a)},N_{0}+P-\sigma_{t'}^{2},
\mathcal{H})$. Expanding it with respect to the difference 
$\sigma_{t'}^{2}-\sigma_{\theta}^{2}$, we have 
\begin{equation} \label{e_d_asym} 
e_{n}(v_{\mathrm{c}}^{(a)},N_{0}+P-\sigma_{t'}^{2},\mathcal{H})
= e_{n}(v_{\mathrm{c}}^{(a)},N_{0}+P-\sigma_{\theta}^{2},\mathcal{H})
+ O(M^{-1/2}), 
\end{equation} 
in the large-system limit, since the standard deviation of 
$\sigma_{t'}^{2}-\sigma_{\theta}^{2}$ is $O(1/\sqrt{M})$. 
In the same manner as in the derivation of (\ref{e_p}), we have 
\begin{equation} \label{e_c} 
e_{n}(v_{\mathrm{c}}^{(a)},N_{0}+P-\sigma_{t'}^{2},\mathcal{H}) 
= \exp\left\{
 G\left(
  \frac{\sigma_{\theta}^{2}}{N_{0}+P-\sigma_{\theta}^{2}}\boldsymbol{Q}
 \right)
\right\} + O(M^{-1/2}).  
\end{equation}

The quantity $e_{n}(v_{\mathrm{d}}^{(a)},N_{0},\mathcal{H})$ is 
different from the other two quantities since 
$\boldsymbol{v}_{\mathrm{d}}=(v_{\mathrm{d}}^{(0)},\ldots, 
v_{\mathrm{d}}^{(n)})^{\mathrm{T}}$ has the nonzero mean $\boldsymbol{v}_{\theta}
=(v_{\theta}^{(0)},\ldots,v_{\theta}^{(n)})^{\mathrm{T}}$, with 
\begin{equation}
v_{\theta}^{(a)} 
= \frac{1}{\sqrt{M}}\sum_{\tilde{m}=1}^{M}h_{\tilde{m}}^{(a)}
\theta_{\tilde{m},t-1}. 
\end{equation} 
The difference $\boldsymbol{v}_{\mathrm{d}}-\boldsymbol{v}_{\theta}$ 
conditioned on $\mathcal{H}$ and 
$\boldsymbol{\theta}_{t-1}$ converges in distribution to a CSCG random vector 
with the covariance matrix $\boldsymbol{Q}_{\mathrm{d}}=
P\boldsymbol{Q} - \frac{1}{M}\sum_{\tilde{m}=1}^{M}|\theta_{\tilde{m},t-1}|^{2}
\boldsymbol{h}_{\tilde{m}}\boldsymbol{h}_{\tilde{m}}^{\mathrm{H}}$ 
in the large-system limit. We first take the 
expectation with respect to $\boldsymbol{x}_{t-1}$ to obtain  
\begin{equation} \label{e_tmp1} 
e_{n}(v_{\mathrm{d}}^{(a)},N_{0},\mathcal{H}) 
= \mathbb{E}\left[
 \left. 
  \mathrm{e}^{
   G(N_{0}^{-1}\boldsymbol{Q}_{\mathrm{d}}) 
  }
  \mathrm{e}^{
   - \boldsymbol{v}_{\theta}^{\mathrm{H}}\boldsymbol{B}
   (\boldsymbol{Q}_{\mathrm{d}},N_{0}^{-1}\boldsymbol{A})
   \boldsymbol{v}_{\theta}
  }
 \right| \mathcal{H} 
\right] + O(M^{-1}), 
\end{equation}
with 
\begin{equation} \label{B}
\boldsymbol{B}(\boldsymbol{Q},\boldsymbol{A}) 
=\boldsymbol{Q}^{-1} - \boldsymbol{Q}^{-1} 
(\boldsymbol{A}+\boldsymbol{Q}^{-1})^{-1}\boldsymbol{Q}^{-1}.
\end{equation} 
In order to eliminate the dependence of $\boldsymbol{\theta}_{m-1}$ 
on $\boldsymbol{Q}_{\mathrm{d}}$, we use 
$\|\boldsymbol{Q}_{\mathrm{d}} - (P-\sigma_{\theta}^{2})\boldsymbol{Q}\|
=O(1/\sqrt{M})$ in the large-system limit. Expanding the 
exponent in (\ref{e_tmp1}) around $\boldsymbol{Q}_{\mathrm{d}}=
(P-\sigma_{\theta}^{2})\boldsymbol{Q}$, we obtain 
\begin{equation} \label{e_tmp2} 
e_{n}(v_{\mathrm{d}}^{(a)},N_{0},\mathcal{H}) 
= 
\exp\left\{ 
 G\left(
  \frac{P-\sigma_{\theta}^{2}}{N_{0}}\boldsymbol{Q}
 \right)
\right\} 
\mathbb{E}\left[
 \left. 
  \exp\left\{
   - N_{0}^{-1}\boldsymbol{v}_{\theta}^{\mathrm{H}}\boldsymbol{B}
   \left(
    \frac{P-\sigma_{\theta}^{2}}{N_{0}}\boldsymbol{Q},\boldsymbol{A}
   \right)\boldsymbol{v}_{\theta}
  \right\}
 \right| \mathcal{H} 
\right] + O(M^{-1/2}), 
\end{equation}
where we have used the identity 
$\boldsymbol{B}(\boldsymbol{Q},N_{0}^{-1}\boldsymbol{A}) 
=N_{0}^{-1}\boldsymbol{B}(N_{0}^{-1}\boldsymbol{Q},\boldsymbol{A})$. 
Applying the central limit theorem with respect to 
$\boldsymbol{v}_{\theta}$ to (\ref{e_tmp2}), after some calculation, 
we arrive at 
\begin{equation} \label{e_d} 
e_{n}(v_{\mathrm{d}}^{(a)},N_{0},\mathcal{H}) 
= \exp\left\{
 G\left(
  \frac{P}{N_{0}}\boldsymbol{Q}
 \right)
\right\} + O(M^{-1/2}). 
\end{equation}
It is interesting to compare (\ref{e_p}) for the pilot symbols and 
(\ref{e_d}) for the data symbols decoded in the preceding stages. 
These expressions imply that random biases do not contribute to the 
performance of channel estimation in the leading order. 

We substitute (\ref{e_p}), (\ref{e_c}), and (\ref{e_d}) into 
(\ref{partition_tmp2}) to obtain 
\begin{equation} \label{partition_tmp3} 
\frac{1}{M}\ln Z_{n}(\omega;f) 
= \frac{1}{M}\ln\mathbb{E}\left\{
 \mathrm{e}^{
  M\left[
   \omega f + \tilde{G}(\boldsymbol{Q}) 
  \right]
 }
\right\} + O(1/\sqrt{M}), 
\end{equation}
in the large-system limit, with 
\begin{equation}
\tilde{G}(\boldsymbol{Q}) 
= \frac{\tau}{\beta}G\left(
    \frac{P}{N_{0}}\boldsymbol{Q}
   \right)
   + \frac{1-\tau}{\beta}G\left(
    \frac{\sigma_{\theta}^{2}}{N_{0}+P-\sigma_{\theta}^{2}}\boldsymbol{Q}
   \right). 
\end{equation}
Differentiating (\ref{partition_tmp3}) with respect to $\omega$ 
and using Assumption~\ref{assumption_commutativity}, we have 
\begin{equation} \label{partition_tmp4} 
\lim_{M\rightarrow\infty}\lim_{\omega\rightarrow0}
\frac{1}{M}\frac{\partial}{\partial\omega}
\ln Z_{n}(\omega;f) = 
\lim_{M\rightarrow\infty}\frac{
 \mathbb{E}\left[
  f_{m_{1},m_{2}}\mathrm{e}^{
   M\tilde{G}(\boldsymbol{Q}) 
  } 
 \right]
}
{\mathbb{E}\left[
  \mathrm{e}^{
   M\tilde{G}(\boldsymbol{Q}) 
  } 
 \right]
},
\end{equation} 
with $f_{m_{1},m_{2}}=f_{1,1}$ for $f=f_{1}$ and 
$f_{m_{1},m_{2}}=f_{1,2}$ for $f=f_{2}$. 
Expression~(\ref{partition_tmp4}) implies that the problem of evaluating 
(\ref{ave_MSE_r}) and (\ref{covariance}) reduces to that of 
evaluating (\ref{partition_tmp4}) for $f=f_{m_{1},m_{2}}$.  

\subsection{Average over Replicated Variables} 
In this section, we take the expectation in (\ref{partition_tmp4}) 
with respect to the replicated variables $\mathcal{H}$, 
following \cite{Nakamura08}.  
For notational convenience, we define a set $\mathcal{M}=\{m_{1},m_{2}\}$ 
of integers. 
We first evaluate the conditional pdf $\mu(\boldsymbol{Q})$ of 
$\boldsymbol{Q}$ given $\mathcal{H}_{\mathcal{M}}=
\{\boldsymbol{h}_{\tilde{m}}:\tilde{m}\in\mathcal{M}\}$. 
\begin{equation} \label{measure} 
\mu(\boldsymbol{Q}) = \mathbb{E}\left[
 \left. 
  \delta\left(
   \boldsymbol{Q} - \frac{1}{M}\sum_{\tilde{m}=1}^{M}\boldsymbol{h}_{\tilde{m}}
   \boldsymbol{h}_{\tilde{m}}^{\mathrm{H}} 
  \right) 
 \right| \mathcal{H}_{\mathcal{M}} 
\right]. 
\end{equation}
It might be possible to obtain the analytical expression of the 
pdf~(\ref{measure}) since $\boldsymbol{Q}$ is a Wishart matrix. 
However, we derive an asymptotic expression in the large-system limit by 
using the inversion formula for the moment generating 
function $F(\tilde{\boldsymbol{Q}})$ of $\boldsymbol{Q}$ given by 
\begin{equation} \label{F} 
F(\tilde{\boldsymbol{Q}}) = \mathbb{E}\left[
 \left. 
  \mathrm{e}^{M\mathrm{Tr}(\boldsymbol{Q}\tilde{\boldsymbol{Q}})}  
 \right| \mathcal{H}_{\mathcal{M}} 
\right], 
\end{equation} 
where a positive definite $(n+1)\times(n+1)$ Hermitian matrix 
$\tilde{\boldsymbol{Q}}$ is given by 
\begin{equation} \label{Q_tilde} 
\tilde{\boldsymbol{Q}} 
= \begin{pmatrix}
\tilde{q}_{0,0} & \frac{1}{2}\tilde{q}_{0,1} & \cdots & 
\frac{1}{2}\tilde{q}_{0,n} \\ 
\frac{1}{2}\tilde{q}_{0,1}^{*} \ddots & \ddots & \vdots \\ 
\vdots & \ddots & \ddots & \frac{1}{2}\tilde{q}_{n-1,n} \\ 
\frac{1}{2}\tilde{q}_{0,n}^{*} & \cdots & \frac{1}{2}\tilde{q}_{n-1,n}^{*} 
& \tilde{q}_{n,n} 
\end{pmatrix}. 
\end{equation}
The inversion formula for moment generating functions implies 
\begin{equation} \label{measure_tmp1} 
\mu(\boldsymbol{Q}) = 
\left(
 \frac{M}{2\pi\mathrm{j}}
\right)^{(n+1)^{2}}\int \mathrm{e}^{-M\mathrm{Tr}
(\boldsymbol{Q}\tilde{\boldsymbol{Q}})}
F(\tilde{\boldsymbol{Q}})d\tilde{\boldsymbol{Q}},  
\end{equation}
with $d\tilde{\boldsymbol{Q}}=\prod_{a=0}^{n}d\tilde{q}_{a,a}
\prod_{a<a'}\{d\Re[\tilde{q}_{a,a'}]d\Im[\tilde{q}_{a,a'}]\}$. 
In (\ref{measure_tmp1}), the integrations with respect to 
$d\tilde{q}_{a,a}$, $d\Re[\tilde{q}_{a,a'}]$, and 
$d\Im[\tilde{q}_{a,a'}]$ are taken 
along the imaginary axes from $-\mathrm{j}\infty$ to $\mathrm{j}\infty$, 
respectively. Since $\{\boldsymbol{h}_{\tilde{m}}\}$ are i.i.d.\ for all 
$\tilde{m}$, the moment generating function~(\ref{F}) reduces to 
$\{F_{1}(\tilde{\boldsymbol{Q}})\}^{M-|\mathcal{M}|}
\prod_{\tilde{m}\in\mathcal{M}}\exp(\boldsymbol{h}_{\tilde{m}}^{\mathrm{H}}
\tilde{\boldsymbol{Q}}\boldsymbol{h}_{\tilde{m}})$, given by 
\begin{equation}
F_{1}(\tilde{\boldsymbol{Q}}) = \mathbb{E}\left[
 \mathrm{e}^{
  \boldsymbol{h}_{1}^{\mathrm{H}}\tilde{\boldsymbol{Q}}\boldsymbol{h}_{1} 
 }
\right]. 
\end{equation}   
Substituting this expression into (\ref{measure_tmp1}) gives 
\begin{equation} \label{measure_tmp2} 
\mu(\boldsymbol{Q}) = 
\left(
 \frac{M}{2\pi\mathrm{j}}
\right)^{(n+1)^{2}}\prod_{\tilde{m}\in\mathcal{M}}
\mathrm{e}^{\boldsymbol{h}_{\tilde{m}}^{\mathrm{H}}\tilde{\boldsymbol{Q}}
\boldsymbol{h}_{\tilde{m}}}
\int \mathrm{e}^{-MI(\boldsymbol{Q},\tilde{\boldsymbol{Q}})}
d\tilde{\boldsymbol{Q}}, 
\end{equation} 
with 
\begin{equation} \label{I} 
I(\boldsymbol{Q},\tilde{\boldsymbol{Q}}) = 
\mathrm{Tr}(\boldsymbol{Q}\tilde{\boldsymbol{Q}}) 
- \left(
 1 - \frac{|\mathcal{M}|}{M} 
\right)\ln F_{1}(\tilde{\boldsymbol{Q}}). 
\end{equation}
In order to obtain an analytical expression of (\ref{measure_tmp2}), 
we use the saddle-point method. 
Let us define $\tilde{\boldsymbol{q}}\in\mathbb{R}^{(n+1)^{2}}$ as 
$\tilde{\boldsymbol{q}}
=(\tilde{\boldsymbol{q}}_{0}^{\mathrm{T}},\ldots,
\tilde{\boldsymbol{q}}_{n}^{\mathrm{T}})^{\mathrm{T}}$, 
given by $\tilde{\boldsymbol{q}}_{a}=(\tilde{q}_{a,a},
\Re[\tilde{q}_{a,a+1}],\Im[\tilde{q}_{a,a+1}],\ldots, 
\Re[\tilde{q}_{a,n}],\Im[\tilde{q}_{a,n}])^{\mathrm{T}}
\in\mathbb{R}^{2(n-a)+1}$. 
Expanding (\ref{I}) with respect to $\tilde{\boldsymbol{Q}}$ around  
the saddle-point 
\begin{equation} \label{saddle_point1}
\tilde{\boldsymbol{Q}}_{\mathrm{s}} 
= \argsup_{\tilde{\boldsymbol{Q}}\in\mathcal{M}_{n+1}^{+}}
\lim_{M\rightarrow\infty}I(\boldsymbol{Q},\tilde{\boldsymbol{Q}}),  
\end{equation}
with $\mathcal{M}_{n+1}^{+}$ denoting the space of positive 
definite $(n+1)\times(n+1)$ Hermitian matrices, we have 
\begin{equation} \label{measure_tmp3} 
\mu(\boldsymbol{Q}) = 
\left(
 \frac{\sqrt{M}}{2\pi}
\right)^{(n+1)^{2}}
\prod_{\tilde{m}\in\mathcal{M}}
\mathrm{e}^{\boldsymbol{h}_{\tilde{m}}^{\mathrm{H}}\tilde{\boldsymbol{Q}}_{\mathrm{s}}
\boldsymbol{h}_{\tilde{m}}}
\mathrm{e}^{-MI(\boldsymbol{Q},\tilde{\boldsymbol{Q}}_{\mathrm{s}})}
\int_{\mathbb{R}^{(n+1)^{2}}}  
\exp\left\{
 \frac{1}{2}\tilde{\boldsymbol{q}}^{\mathrm{T}}\nabla_{\tilde{\boldsymbol{q}}}^{2}
I(\boldsymbol{Q},\tilde{\boldsymbol{Q}}_{\mathrm{s}}) 
 \tilde{\boldsymbol{q}}
\right\}\left[
 1+O(1/\sqrt{M})
\right]d\tilde{\boldsymbol{q}}, 
\end{equation}
where $\nabla_{\tilde{\boldsymbol{q}}}^{2}
I(\boldsymbol{Q},\tilde{\boldsymbol{Q}}_{\mathrm{s}})$ denotes the 
Hesse matrix of $I(\boldsymbol{Q},\tilde{\boldsymbol{Q}})$ with respect 
to $\tilde{\boldsymbol{q}}$. In the derivation of (\ref{measure_tmp3}), 
we have transformed the variable $\tilde{\boldsymbol{Q}}$ into 
$\tilde{\boldsymbol{Q}}'=\sqrt{M}(\tilde{\boldsymbol{Q}}
-\tilde{\boldsymbol{Q}}_{\mathrm{s}})/\mathrm{j}$ and then 
rewritten $\tilde{\boldsymbol{Q}}'$ as $\tilde{\boldsymbol{Q}}$. 
The Hesse matrix 
$\nabla_{\tilde{\boldsymbol{q}}}^{2}I(\boldsymbol{Q},
\tilde{\boldsymbol{Q}}_{\mathrm{s}})$ is 
negative definite since the cumulant generating function 
$\ln F_{1}(\tilde{\boldsymbol{Q}})$ is convex. Thus, we can perform the 
Gaussian integration in (\ref{measure_tmp3}) to obtain 
\begin{equation} \label{measure_tmp4} 
\mu(\boldsymbol{Q}) = 
\left(
 \sqrt{\frac{M}{2\pi}}
\right)^{(n+1)^{2}}
|\det\{\nabla_{\tilde{\boldsymbol{q}}}^{2}
I(\boldsymbol{Q},\tilde{\boldsymbol{Q}}_{\mathrm{s}})\}|^{-1}  
\prod_{\tilde{m}\in\mathcal{M}}
\mathrm{e}^{\boldsymbol{h}_{\tilde{m}}^{\mathrm{H}}\tilde{\boldsymbol{Q}}_{\mathrm{s}}
\boldsymbol{h}_{\tilde{m}}}
\mathrm{e}^{-MI(\boldsymbol{Q},\tilde{\boldsymbol{Q}}_{\mathrm{s}})}
\left[
 1+O(1/\sqrt{M})
\right].  
\end{equation}

We next calculate the numerator in (\ref{partition_tmp4})
 by using the pdf~(\ref{measure_tmp4}).   
Substituting (\ref{measure_tmp4}) into the quantity 
$\mathbb{E}[f_{m_{1},m_{2}}\mathrm{e}^{
M\tilde{G}(\boldsymbol{Q}) }]$ and then using the saddle-point 
method, we have 
\begin{equation} \label{numerator}  
\mathbb{E}\left[
 f_{m_{1},m_{2}}\mathrm{e}^{
  M\tilde{G}(\boldsymbol{Q}) 
 }
\right]
= C_{n}(\boldsymbol{Q},\boldsymbol{Q}_{\mathrm{s}}) 
\mathrm{e}^{
  -M\Phi(\boldsymbol{Q}_{\mathrm{s}}) 
 }
\mathbb{E}\left[
 f_{m_{1},m_{2}}\prod_{\tilde{m}\in\mathcal{M}}
 \mathrm{e}^{\boldsymbol{h}_{\tilde{m}}^{\mathrm{H}}\tilde{\boldsymbol{Q}}_{\mathrm{s}}
 \boldsymbol{h}_{\tilde{m}}}
\right]
\left[
 1+O(1/\sqrt{M})
\right], 
\end{equation} 
with $\Phi(\boldsymbol{Q})=I(\boldsymbol{Q},
\tilde{\boldsymbol{Q}}_{\mathrm{s}}) -\tilde{G}(\boldsymbol{Q})$ 
and $C_{n}(\boldsymbol{Q},\boldsymbol{Q}_{\mathrm{s}}) = 
|\det\{\nabla_{\tilde{\boldsymbol{q}}}^{2}
I(\boldsymbol{Q},\tilde{\boldsymbol{Q}}_{\mathrm{s}})\}|^{-1}  
\det\{\nabla^{2}\Phi(\boldsymbol{Q}_{\mathrm{s}})\}^{-1}$.   
In (\ref{numerator}), $\boldsymbol{Q}_{\mathrm{s}}$ denotes 
the saddle-point 
\begin{equation} \label{saddle_point2} 
\boldsymbol{Q}_{\mathrm{s}} 
= \arginf_{\boldsymbol{Q}\in\mathcal{M}_{n+1}^{+}}
\lim_{M\rightarrow\infty}\Phi(\boldsymbol{Q}). 
\end{equation}
Furthermore, $\nabla^{2}\Phi(\boldsymbol{Q}_{\mathrm{s}})$ 
represents the Hesse matrix of $\Phi(\boldsymbol{Q})$ 
at the saddle-point $\boldsymbol{Q}=\boldsymbol{Q}_{\mathrm{s}}$, and is 
assumed to be positive definite. 

Similarly, we can obtain an analytical expression of the denominator 
in (\ref{partition_tmp4}).   
Substituting the obtained expression and (\ref{numerator}) into 
(\ref{partition_tmp4}), we arrive at 
\begin{equation} \label{partition_tmp5} 
\lim_{M\rightarrow\infty}\lim_{\omega\rightarrow0}
\frac{1}{M}\frac{\partial}{\partial\omega}\ln Z_{n}(\omega;f) 
= \mathbb{E}\left[
 f_{m_{1},m_{2}}\prod_{\tilde{m}\in\mathcal{M}}\frac{ 
  \mathrm{e}^{\boldsymbol{h}_{\tilde{m}}^{\mathrm{H}}
  \tilde{\boldsymbol{Q}}_{\mathrm{s}}\boldsymbol{h}_{\tilde{m}}}}
 {
 \mathbb{E}[ 
  \mathrm{e}^{\boldsymbol{h}_{\tilde{m}}^{\mathrm{H}}\tilde{\boldsymbol{Q}}_{\mathrm{s}}
  \boldsymbol{h}_{\tilde{m}}} ] 
 }
\right], 
\end{equation} 
with $f_{m_{1},m_{2}}=f_{1,1}$ for $f=f_{1}$ and 
$f_{m_{1},m_{2}}=f_{1,2}$ for $f=f_{2}$.

The calculations of the stationarity conditions for (\ref{saddle_point1}) 
and (\ref{saddle_point2}) implies that 
$(\boldsymbol{Q}_{\mathrm{s}},\tilde{\boldsymbol{Q}}_{\mathrm{s}})$ is 
given as the solution to the coupled fixed-point equations 
\begin{equation} \label{fixed_point1_tmp1} 
\boldsymbol{Q} = \frac{
 \mathbb{E}\left[ 
  \boldsymbol{h}_{1}\boldsymbol{h}_{1}^{\mathrm{H}}
  \mathrm{e}^{
   \boldsymbol{h}_{1}^{\mathrm{H}}\tilde{\boldsymbol{Q}}\boldsymbol{h}_{1} 
  }
 \right]
}
{
 \mathbb{E}\left[ 
  \mathrm{e}^{
   \boldsymbol{h}_{1}^{\mathrm{H}}\tilde{\boldsymbol{Q}}\boldsymbol{h}_{1} 
  }
 \right]
}, 
\end{equation}
\begin{equation} \label{fixed_point2_tmp1} 
\tilde{\boldsymbol{Q}} = 
-\frac{\tau P}{\beta N_{0}}
\left(
 \boldsymbol{I}_{n+1}+\frac{P}{N_{0}}\boldsymbol{A}\boldsymbol{Q}
\right)^{-1}\boldsymbol{A}  
-\frac{(1-\tau)\sigma_{\theta}^{2}}{\beta(N_{0}+P-\sigma_{\theta}^{2})}
\left(
 \boldsymbol{I}_{n+1}+\frac{\sigma_{\theta}^{2}}{N_{0}+P-
 \sigma_{\theta}^{2}}\boldsymbol{A}\boldsymbol{Q}
\right)^{-1}\boldsymbol{A}.   
\end{equation}

\subsection{Replica Symmetry} 
The expression~(\ref{partition_tmp5}) is defined only for $n\in\mathbb{N}$, 
since $(n+1)$ is the dimension of $\boldsymbol{Q}$ and 
$\tilde{\boldsymbol{Q}}$. In order to obtain a formula of 
(\ref{partition_tmp5}) defined for $n\in\mathbb{R}$, we assume RS for the 
solution to the coupled fixed-point 
equations~(\ref{fixed_point1_tmp1}) and~(\ref{fixed_point2_tmp1}). 

\begin{assumption} \label{RS} 
The solution $(\boldsymbol{Q}_{\mathrm{s}},
\tilde{\boldsymbol{Q}}_{\mathrm{s}})$ is invariant under all permutations of 
replica indices: 
\begin{equation}
\boldsymbol{Q}_{\mathrm{s}} 
= \begin{pmatrix}
a & b\boldsymbol{1}_{n}^{\mathrm{T}} \\ 
b^{*}\boldsymbol{1}_{n} & (d-c)\boldsymbol{I}_{n} 
+ c\boldsymbol{1}_{n}\boldsymbol{1}_{n}^{\mathrm{T}} 
\end{pmatrix}, 
\end{equation}
\begin{equation}
\tilde{\boldsymbol{Q}}_{\mathrm{s}} 
= \begin{pmatrix}
\tilde{a} & \tilde{b}\boldsymbol{1}_{n}^{\mathrm{T}} \\ 
\tilde{b}^{*}\boldsymbol{1}_{n} & (\tilde{d}-\tilde{c})\boldsymbol{I}_{n} 
+ \tilde{c}\boldsymbol{1}_{n}\boldsymbol{1}_{n}^{\mathrm{T}} 
\end{pmatrix}. 
\end{equation}
\end{assumption} 


We first evaluate the fixed-point equation~(\ref{fixed_point2_tmp1}). 
Let us define $(\sigma_{\mathrm{tr}}^{(0)})^{2}$, 
$\sigma_{\mathrm{tr}}^{2}$, 
$(\sigma_{\mathrm{c}}^{(0)})^{2}$, and $\sigma_{\mathrm{c}}^{2}$ as 
\begin{IEEEeqnarray}{l} 
(\sigma_{\mathrm{tr}}^{(0)})^{2} = N_{0} + P(a-b-b^{*}+c),   
\nonumber \\ 
\sigma_{\mathrm{tr}}^{2} = N_{0} + P(d-c),   
\nonumber \\ 
(\sigma_{\mathrm{c}}^{(0)})^{2} = N_{0}+P-\sigma_{\theta}^{2}
+ \sigma_{\theta}^{2}(a-b-b^{*}+c),   
\nonumber \\ 
\sigma_{\mathrm{c}}^{2} = N_{0}+P-\sigma_{\theta}^{2}
+ \sigma_{\theta}^{2}(d-c), \label{fixed_point2_1}   
\end{IEEEeqnarray}
respectively. After some calculation for (\ref{fixed_point2_tmp1}), we obtain 
\begin{IEEEeqnarray}{l}
\tilde{a} = -\frac{\tau P}{\beta}\frac{n}{
 \sigma_{\mathrm{tr}}^{2} + n(\sigma_{\mathrm{tr}}^{(0)})^{2}
}
- \frac{(1-\tau)\sigma_{\theta}^{2}}{\beta}\frac{n}{
 \sigma_{\mathrm{c}}^{2} + n(\sigma_{\mathrm{c}}^{(0)})^{2}
}, \nonumber \\
\tilde{b} = \frac{\tau P}{\beta}\frac{1}{
 \sigma_{\mathrm{tr}}^{2} + n(\sigma_{\mathrm{tr}}^{(0)})^{2}
}
+ \frac{(1-\tau)\sigma_{\theta}^{2}}{\beta}\frac{1}{
 \sigma_{\mathrm{c}}^{2} + n(\sigma_{\mathrm{c}}^{(0)})^{2}
}, \nonumber \\
\tilde{c} = \frac{\tau P}{\beta}\frac{(\sigma_{\mathrm{tr}}^{(0)})^{2}}{
 (\sigma_{\mathrm{tr}}^{2} + n(\sigma_{\mathrm{tr}}^{(0)})^{2})
 \sigma_{\mathrm{tr}}^{2}
}
+ \frac{(1-\tau)\sigma_{\theta}^{2}}{\beta}\frac{
(\sigma_{\mathrm{c}}^{(0)})^{2}}{
 (\sigma_{\mathrm{c}}^{2} + n(\sigma_{\mathrm{c}}^{(0)})^{2})
 \sigma_{\mathrm{c}}^{2} 
}, \nonumber \\
\tilde{d} = \tilde{c} - \frac{\tau P}{\beta\sigma_{\mathrm{tr}}^{2}} 
- \frac{(1-\tau)\sigma_{\theta}^{2}}{\beta\sigma_{\mathrm{c}}^{2}}. 
\label{fixed_point2_tmp2} 
\end{IEEEeqnarray}

We next evaluate the fixed-point equation~(\ref{fixed_point1_tmp1}) by 
calculating $\mathrm{e}^{
 \boldsymbol{h}_{\tilde{m}}^{\mathrm{H}}\tilde{\boldsymbol{Q}}
\boldsymbol{h}_{\tilde{m}}}$ with (\ref{fixed_point2_tmp2}).  
\begin{equation} \label{quadratic} 
\mathrm{e}^{
 \boldsymbol{h}_{\tilde{m}}^{\mathrm{H}}\tilde{\boldsymbol{Q}}\boldsymbol{h}_{\tilde{m}}
} = 
\exp\left\{
 \frac{\tau P}{\beta}\left[
  \tilde{\sigma}_{\mathrm{tr}}^{2}\left|
   \sum_{a=0}^{n}\frac{h_{\tilde{m}}^{(a)}}{(\sigma_{\mathrm{tr}}^{(a)})^{2}}
  \right|^{2} 
  - \sum_{a=0}^{n}\frac{|h_{\tilde{m}}^{(a)}|^{2}}
  {(\sigma_{\mathrm{tr}}^{(a)})^{2}}
 \right]
 + \frac{(1-\tau)\sigma_{\theta}^{2}}{\beta}\left[
  \tilde{\sigma}_{\mathrm{c}}^{2}\left|
   \sum_{a=0}^{n}\frac{h_{\tilde{m}}^{(a)}}{(\sigma_{\mathrm{c}}^{(a)})^{2}}
  \right|^{2} 
  - \sum_{a=0}^{n}\frac{|h_{\tilde{m}}^{(a)}|^{2}}
  {(\sigma_{\mathrm{c}}^{(a)})^{2}}
 \right]
\right\}, 
\end{equation}
with $\tilde{\sigma}_{\mathrm{tr}}^{2}=(n\sigma_{\mathrm{tr}}^{-2}
+(\sigma_{\mathrm{tr}}^{(0)})^{-2})^{-1}$ and 
$\tilde{\sigma}_{\mathrm{c}}^{2}=
(n\sigma_{\mathrm{c}}^{-2}+(\sigma_{\mathrm{c}}^{(0)})^{-2})^{-1}$. 
In (\ref{quadratic}), $(\sigma_{\mathrm{tr}}^{(a)})^{2}$, 
and $(\sigma_{\mathrm{c}}^{(a)})^{2}$ are given by 
$(\sigma_{\mathrm{tr}}^{(a)})^{2}=\sigma_{\mathrm{tr}}^{2}$, 
and $(\sigma_{\mathrm{c}}^{(a)})^{2}=\sigma_{\mathrm{c}}^{2}$ 
for $a=1,\ldots,n$. 
In order to linearize the two quadratic forms in (\ref{quadratic}), 
we use the identity 
\begin{equation} \label{identity} 
\mathrm{e}^{\tilde{\sigma}^{2}|a|^{2}} 
= \int_{\mathbb{C}} \frac{1}{\pi\tilde{\sigma}^{2}} 
\mathrm{e}^{
 - \frac{|\underline{y}|^{2}}{\tilde{\sigma}^{2}} 
 + a^{*}\underline{y}+a\underline{y}^{*}
}d\underline{y}, 
\end{equation}
for $\underline{y}=\underline{y}_{\mathrm{tr}}\in\mathbb{C}$ or   
$\underline{y}=\underline{y}_{\mathrm{c}}\in\mathbb{C}$. 
Substituting (\ref{identity})   
with $(a,\tilde{\sigma}^{2})=
(\sqrt{\tau P/\beta}\sum_{a=0}^{n}h_{\tilde{m}}^{(a)}/
(\sigma_{\mathrm{tr}}^{(a)})^{2},\tilde{\sigma}_{\mathrm{tr}}^{2})$ or 
$(a,\tilde{\sigma}^{2})=
(\sqrt{(1-\tau)\sigma_{\theta}^{2}/\beta}\sum_{a=0}^{n}h_{\tilde{m}}^{(a)}/(
\sigma_{\mathrm{c}}^{(a)})^{2},
\tilde{\sigma}_{\mathrm{c}}^{2})$ into (\ref{quadratic}), we have  
\begin{equation} \label{quadratic_tmp1} 
\mathrm{e}^{
 \boldsymbol{h}_{\tilde{m}}^{\mathrm{H}}\tilde{\boldsymbol{Q}}\boldsymbol{h}_{\tilde{m}}
} = D_{n}\int\prod_{a=0}^{n}q(\underline{\boldsymbol{y}}|h_{\tilde{m}}^{(a)})
d\underline{\boldsymbol{y}}, 
\end{equation}
with $D_{n}=(\pi^{2}\sigma_{\mathrm{tr}}^{2}\sigma_{\mathrm{c}}^{2})^{n}
(1+n(\sigma_{\mathrm{tr}}^{(0)})^{2}/\sigma_{\mathrm{tr}}^{2})
(1+n(\sigma_{\mathrm{c}}^{(0)})^{2}/\sigma_{\mathrm{c}}^{2})$. 
In (\ref{quadratic_tmp1}), the function 
$q(\underline{\boldsymbol{y}}|h_{\tilde{m}}^{(a)})$ for 
$\underline{\boldsymbol{y}}=(\underline{y}_{\mathrm{tr}},
\underline{y}_{\mathrm{c}})^{\mathrm{T}}\in\mathbb{C}^{2}$ is defined as 
\begin{equation}
q(\underline{\boldsymbol{y}}|h_{\tilde{m}}^{(a)}) = 
q\left(
 \underline{y}_{\mathrm{tr}}\left| 
  \sqrt{\frac{\tau P}{\beta}}h_{\tilde{m}}^{(a)};\sigma_{\mathrm{tr}}^{(a)}
 \right. 
\right)
q\left(
 \underline{y}_{\mathrm{c}}\left| 
  \sqrt{\frac{(1-\tau)\sigma_{\theta}^{2}}{\beta}}h_{\tilde{m}}^{(a)};
  \sigma_{\mathrm{c}}^{(a)}
 \right. 
\right), 
\end{equation} 
with 
\begin{equation}
q(\underline{y}|h;\sigma) = \frac{1}{\pi\sigma^{2}}\mathrm{e}^{
 -\frac{|\underline{y}-h|^{2}}{\sigma^{2}} 
}. 
\end{equation}
Applying the expression~(\ref{quadratic_tmp1}) to (\ref{fixed_point1_tmp1}), 
we arrive at 
\begin{equation} \label{fixed_point1_1} 
a-b-b^{*}+c = \frac{
 \mathbb{E}\left[ 
  \int \left|
   h_{1}^{(0)} - \langle h_{1}^{(1)} \rangle
  \right|^{2} 
  q(\underline{\boldsymbol{y}}|h_{1}^{(0)})
  \left\{
   \mathbb{E}_{h_{1}^{(1)}}\left[
    q(\underline{\boldsymbol{y}}|h_{1}^{(1)})
   \right]
  \right\}^{n}
  d\underline{\boldsymbol{y}}
 \right]
}
{
 \mathbb{E}\left[ 
  \int q(\underline{\boldsymbol{y}}|h_{1}^{(0)})
  \left\{
   \mathbb{E}_{h_{1}^{(1)}}\left[
    q(\underline{\boldsymbol{y}}|h_{1}^{(1)})
   \right]
  \right\}^{n}
  d\underline{\boldsymbol{y}} 
 \right]
}, 
\end{equation} 
\begin{equation} \label{fixed_point1_2} 
d-c = \frac{
 \mathbb{E}\left[ 
  \int \left\langle
   \left|
    h_{1}^{(1)} - \langle h_{1}^{(1)} \rangle
   \right|^{2}
  \right\rangle  
  q(\underline{\boldsymbol{y}}|h_{1}^{(0)})
  \left\{
   \mathbb{E}_{h_{1}^{(1)}}\left[
    q(\underline{\boldsymbol{y}}|h_{1}^{(1)})
   \right]
  \right\}^{n}
  d\underline{\boldsymbol{y}}
 \right]
}
{
 \mathbb{E}\left[ 
  \int q(\underline{\boldsymbol{y}}|h_{1}^{(0)})
  \left\{
   \mathbb{E}_{h_{1}^{(1)}}\left[ 
    q(\underline{\boldsymbol{y}}|h_{1}^{(1)})
   \right]
  \right\}^{n}
  d\underline{\boldsymbol{y}} 
 \right]
}, 
\end{equation} 
with 
\begin{equation}
\langle h_{1}^{(1)} \rangle 
= \frac{
 \mathbb{E}_{h_{1}^{(1)}}\left[
  h_{1}^{(1)} q(\underline{\boldsymbol{y}}|h_{1}^{(1)})
 \right]
}
{
 \mathbb{E}_{h_{1}^{(1)}}\left[
  q(\underline{\boldsymbol{y}}|h_{1}^{(1)})
 \right]
}. 
\end{equation}

\subsection{Replica Continuity}
Equations~(\ref{fixed_point2_1}), (\ref{fixed_point1_1}), and 
(\ref{fixed_point1_2}) provide the coupled fixed-point equations of 
$(a-b-b^{*}+c,d-c)$ under the RS assumption, and 
are well defined for $n\in\mathbb{R}$.  
We regard $n$ as a real number and take the limit $n\rightarrow+0$ to obtain  
\begin{equation} \label{fixed_point_appen_1}
(\sigma_{\mathrm{tr}}^{(0)})^{2} = N_{0} + P\mathbb{E}\left[
 \left|
  h_{1}^{(0)} - \langle h_{1}^{(1)} \rangle
 \right|^{2}
\right],   
\end{equation}
\begin{equation} \label{fixed_point_appen_2}
\sigma_{\mathrm{tr}}^{2} = N_{0} + P\mathbb{E}\left[
 \left|
  h_{1}^{(1)} - \langle h_{1}^{(1)} \rangle
 \right|^{2}
\right],   
\end{equation}
\begin{equation} \label{fixed_point_appen_3}
(\sigma_{\mathrm{c}}^{(0)})^{2} = N_{0}+P-\sigma_{\theta}^{2}
+ \sigma_{\theta}^{2}\mathbb{E}\left[
 \left|
  h_{1}^{(0)} - \langle h_{1}^{(1)} \rangle
 \right|^{2}
\right],   
\end{equation} 
\begin{equation} \label{fixed_point_appen_4}
\sigma_{\mathrm{c}}^{2} = N_{0}+P-\sigma_{\theta}^{2}
+ \sigma_{\theta}^{2}\mathbb{E}\left[
 \left|
  h_{1}^{(1)} - \langle h_{1}^{(1)} \rangle
 \right|^{2}
\right],    
\end{equation}
where the expectations for $\underline{\boldsymbol{y}}$ are taken with 
respect to the measure 
$p(\underline{\boldsymbol{y}}|h_{1}^{(0)})d\underline{\boldsymbol{y}}$. 
Note that $\mathbb{E}[|h_{1}^{(1)} - \langle h_{1}^{(1)} \rangle|^{2}]$ 
and $\mathbb{E}[|h_{1}^{(1)} - \langle h_{1}^{(1)} \rangle|^{2}]$ 
depend on $(\sigma_{\mathrm{tr}}^{(0)})^{2}$, $\sigma_{\mathrm{tr}}^{2}$, 
$(\sigma_{\mathrm{c}}^{(0)})^{2}$, and $\sigma_{\mathrm{c}}^{2}$. 
Furthermore, the quantity~(\ref{partition_tmp5}) is given by 
\begin{equation} \label{partition_tmp7} 
\lim_{M\rightarrow\infty}\lim_{\omega\rightarrow0}
\frac{1}{M}\frac{\partial}{\partial\omega}\ln Z_{n}(\omega;f) 
= \left\{ 
\begin{array}{cc} 
\mathbb{E}[ |h_{1}^{(0)} - \langle h_{1}^{(1)} \rangle|^{2}] 
& \hbox{for $f=f_{1}$} \\ 
0 & \hbox{for $f=f_{2}$}. 
\end{array}
\right. 
\end{equation} 
Under the assumption of the commutativity between the large-system 
limit and the limit $n\rightarrow+0$, the substitution of 
(\ref{partition_tmp7}) into (\ref{ave_MSE_r}) or (\ref{covariance}) gives 
\begin{equation} \label{ave_MSE_r_tmp1}
\lim_{M\rightarrow\infty}\bar{\xi}_{t}^{2} = 
\mathbb{E}\left[
 \left|
  h_{1}^{(0)} - \langle h_{1}^{(1)} \rangle
 \right|^{2}
\right], 
\end{equation}
\begin{equation} \label{covariance_tmp1} 
\lim_{M\rightarrow\infty}\bar{\rho}_{t} = 0,  
\end{equation}
in the large-system limit. 
Note that we have implicitly assumed that the RHSs of 
(\ref{ave_MSE_r_tmp1}) and (\ref{covariance_tmp1}), obtained by the replica 
method, coincide with the correct ones. 

In order to complete the derivation of Proposition~\ref{proposition2}, 
we show that (\ref{ave_MSE_r_tmp1}) reduces to (\ref{xi}), defined by 
the fixed-point equations~(\ref{fixed_point_tr}) and (\ref{fixed_point_c}). 
Since $h_{1}^{(a)}\sim\mathcal{CN}(0,1)$, the 
quantities $\mathbb{E}[|h_{1}^{(0)} - \langle h_{1}^{(1)} \rangle|^{2}]$ and 
$\mathbb{E}[|h_{1}^{(1)} - \langle h_{1}^{(1)} \rangle|^{2}]$ reduce to 
\begin{equation} \label{MSE_appen} 
\mathbb{E}\left[
 \left|
  h_{1}^{(0)} - \langle h_{1}^{(1)} \rangle
 \right|^{2}
\right] = 
\xi^{4}\left(
 1 + \frac{(\sigma_{\mathrm{tr}}^{(0)})^{2}\tau P}
 {\sigma_{\mathrm{tr}}^{4}\beta}
 + \frac{(\sigma_{\mathrm{c}}^{(0)})^{2}(1-\tau)\sigma_{\theta}^{2}}
 {\sigma_{\mathrm{c}}^{4}\beta}
\right), 
\end{equation}
\begin{equation} \label{postulated_MSE_appen} 
\mathbb{E}\left[
 \left|
  h_{1}^{(1)} - \langle h_{1}^{(1)} \rangle
 \right|^{2}
\right] = \xi^{2}, 
\end{equation}
with 
\begin{equation}
\xi^{2} =  
\left(
 1 + \frac{\tau P}{\sigma_{\mathrm{tr}}^{2}\beta}
 + \frac{(1-\tau)\sigma_{\theta}^{2}}{\sigma_{\mathrm{c}}^{2}\beta}
\right)^{-1}. 
\end{equation}
Equations~(\ref{fixed_point_appen_2}), (\ref{fixed_point_appen_4}), 
and (\ref{postulated_MSE_appen}) provide a closed form for   
$(\sigma_{\mathrm{tr}}^{2},\sigma_{\mathrm{c}}^{2})$. Furthermore, 
(\ref{fixed_point_appen_1}), (\ref{fixed_point_appen_3}), 
and (\ref{MSE_appen}) for a given solution 
$(\sigma_{\mathrm{tr}}^{2},\sigma_{\mathrm{c}}^{2})$ form two independent 
linear equations with respect to  $(\sigma_{\mathrm{tr}}^{(0)})^{2}$ and 
$(\sigma_{\mathrm{c}}^{(0)})^{2}$, 
and have the unique solution $((\sigma_{\mathrm{tr}}^{(0)})^{2},
(\sigma_{\mathrm{c}}^{(0)})^{2})=(\sigma_{\mathrm{tr}}^{2},
\sigma_{\mathrm{c}}^{2})$. These observations indicate that the averaged 
MSE~(\ref{ave_MSE_r_tmp1}) is given as (\ref{xi}), defined by 
the fixed-point equations~(\ref{fixed_point_tr}) and (\ref{fixed_point_c}).

\section{Derivation of Proposition~\ref{proposition3}} 
\label{sec_deriv_proposition3} 
\subsection{Formulation}
Let $\mathbb{E}_{\tilde{\boldsymbol{Y}}_{\backslash t},
\boldsymbol{x}_{[1,m),t}}[\cdots]$ denote the expectation 
with respect to $\tilde{\boldsymbol{Y}}_{\backslash t}$ and 
$\boldsymbol{x}_{[1,m),t}$ given $\tilde{x}_{m,t}$, $x_{m,t}$, 
$\bar{\boldsymbol{X}}_{\backslash t}$ and $\boldsymbol{\theta}_{t}$. 
It is sufficient from Assumption~\ref{assumption2} to show that 
$\mathbb{E}_{\tilde{\boldsymbol{Y}}_{\backslash t},
\boldsymbol{x}_{[1,m),t}}[
p(\tilde{x}_{m,t}| x_{m,t}, \tilde{\mathcal{I}}_{t},
\boldsymbol{x}_{[1,m),t},
\boldsymbol{\theta}_{[m,M],t})]$, given by (\ref{equivalent_channel}),  
converges in law to the equivalent 
channel~(\ref{equivalent_channel_deterministic}) in the 
large-system limit. Substituting the posterior pdf~(\ref{posterior_x}) into 
(\ref{equivalent_channel}) and then introducing a non-negative number $n$, 
we have 
\begin{equation} \label{formulation} 
\mathbb{E}_{\tilde{\boldsymbol{Y}}_{\backslash t},
\boldsymbol{x}_{[1,m),t}}\left[
 p(\tilde{x}_{m,t}| x_{m,t}, \tilde{\mathcal{I}}_{t},
 \boldsymbol{x}_{[1,m),t},
 \boldsymbol{\theta}_{[m,M],t}) 
\right]
= \lim_{n\rightarrow+0}Z_{n}^{(\mathrm{d})}, 
\end{equation}
with 
\begin{IEEEeqnarray}{r}
Z_{n}^{(\mathrm{d})} 
= \mathbb{E}_{\tilde{\boldsymbol{Y}}_{\backslash t},\boldsymbol{x}_{[1,m),t}}
\left[
 \int 
 \left\{
  \int p(\boldsymbol{y}_{t} | \tilde{\boldsymbol{x}}_{t},
  \tilde{\mathcal{I}}_{t})p(\tilde{\boldsymbol{x}}_{[m,M],t}|
  \boldsymbol{\theta}_{[m,M],t})
  d\tilde{\boldsymbol{x}}_{[m,M],t}
 \right\}^{n-1}
 p(\boldsymbol{y}_{t} | \tilde{\boldsymbol{x}}_{t},\tilde{\mathcal{I}}_{t}) 
\right. \nonumber \\
\left.
 \times 
 p(\tilde{\boldsymbol{x}}_{[m,M],t} | 
 \boldsymbol{\theta}_{[m,M],t})
 d\tilde{\boldsymbol{x}}_{(m,M],t} 
 p(\boldsymbol{y}_{t} | \boldsymbol{x}_{t},\tilde{\mathcal{I}}_{t}) 
 p(\boldsymbol{x}_{(m,M],t} |
 \boldsymbol{\theta}_{(m,M],t})
 d\boldsymbol{x}_{(m,M],t}d\boldsymbol{y}_{t}
\right], \label{partition_d} 
\end{IEEEeqnarray}
where we have introduced 
$\tilde{\boldsymbol{x}}_{t}=((\boldsymbol{x}_{[1,m),t})^{\mathrm{T}},
(\tilde{\boldsymbol{x}}_{[m,M],t})^{\mathrm{T}})^{\mathrm{T}}$, in which 
$\tilde{\boldsymbol{x}}_{[m,M],t}=(\tilde{x}_{m,t},\ldots,
\tilde{x}_{M,t})^{\mathrm{T}}$ has the same statistical 
properties as $\boldsymbol{x}_{[m,M],t}$. Furthermore,  
$\tilde{\boldsymbol{x}}_{(m,M],t}$ is given by 
$\tilde{\boldsymbol{x}}_{(m,M],t}
=(\tilde{x}_{m+1,t},\ldots,\tilde{x}_{M,t})^{\mathrm{T}}$.  
Note that (\ref{partition_d}) is a quantity of $O(1)$, while 
(\ref{partition}) is exponential in $M$. 
Thus, we have to evaluate (\ref{partition_d}) up to $O(1)$ in the 
large-system limit. 

Let us regard $n$ in (\ref{partition_d}) as a positive integer. For 
$n=2,3,\ldots$, (\ref{partition_d}) reduces to a special expression,   
\begin{IEEEeqnarray}{r}  
Z_{n}^{(\mathrm{d})} 
= \mathbb{E}_{\tilde{\boldsymbol{Y}}_{\backslash t},\boldsymbol{x}_{[1,m),t}}
\left[
 \int \prod_{a=0}^{n}
 p(\boldsymbol{y}_{t} | \boldsymbol{x}_{t}^{(a)}, \tilde{\mathcal{I}}_{t})
 \prod_{a=2}^{n}\left\{
  p(\boldsymbol{x}_{[m,M],t}^{(a)} | \boldsymbol{\theta}_{[m,M],t})
  d\boldsymbol{x}_{[m,M],t}^{(a)}  
 \right\}
\right. \nonumber \\ 
\left. 
 \times p(\boldsymbol{x}_{(m,M],t}^{(0)} | 
 \boldsymbol{\theta}_{(m,M],t})
 d\boldsymbol{x}_{(m,M],t}^{(0)}
 p(\boldsymbol{x}_{[m,M],t}^{(1)} | \boldsymbol{\theta}_{[m,M],t})
 d\boldsymbol{x}_{(m,M],t}^{(1)}d\boldsymbol{y}_{t}  
\right].  \label{partition_d_tmp1}
\end{IEEEeqnarray}
In (\ref{partition_d_tmp1}), 
$\boldsymbol{x}_{t}^{(a)}=((\boldsymbol{x}_{[1,m),t})^{\mathrm{T}},
(\boldsymbol{x}_{[m,M],t}^{(a)})^{\mathrm{T}})^{\mathrm{T}}\in\mathbb{C}^{M}$ denotes replicas of 
$\tilde{\boldsymbol{x}}_{t}$ for $a=2,\ldots,n$, in which 
$\{\boldsymbol{x}_{[m,M],t}^{(a)}=
(x_{m,t}^{(a)},\ldots,x_{M,t}^{(a)})^{\mathrm{T}}\}$ conditioned on 
$\boldsymbol{\theta}_{[m,M],t}$ are independent random 
vectors drawn from $p(\boldsymbol{x}_{[m,M],t}|
\boldsymbol{\theta}_{[m,M],t})$. 
For notational convenience, we have written 
$\boldsymbol{x}_{[m,M],t}$ and 
$\tilde{\boldsymbol{x}}_{[m,M],t}$ as $\boldsymbol{x}_{[m,M],t}^{(a)}
=(x_{m,t}^{(a)},\ldots,x_{M,t}^{(a)})^{\mathrm{T}}$ for $a=0$ and $a=1$, 
respectively. 
The vector $\boldsymbol{x}_{(m,M],t}^{(a)}$ 
is defined as $\boldsymbol{x}_{(m,M],t}^{(a)}
=(x_{m+1,t}^{(a)},\ldots,x_{M,t}^{(a)})^{\mathrm{T}}$.  
Furthermore, we have written 
$\boldsymbol{x}_{t}$ and $\tilde{\boldsymbol{x}}_{t}$ as 
$\boldsymbol{x}_{t}^{(a)}=((\boldsymbol{x}_{[1,m),t})^{\mathrm{T}},
(\boldsymbol{x}_{[m,M],t}^{(a)})^{\mathrm{T}})^{\mathrm{T}}$ 
for $a=0$ and $a=1$, respectively.

\subsection{Average over Non-Replicated Variables} 
The goal of this section is to evaluate the expectation in 
(\ref{partition_d_tmp1}) with respect to the non-replicated  
variables $\tilde{\boldsymbol{Y}}_{\backslash t}$. 
We first calculate the integration in (\ref{partition_d_tmp1}) with respect 
to $\boldsymbol{y}_{t}$. The substitution of (\ref{conditional_channel}) into 
(\ref{partition_d_tmp1}) gives 
\begin{IEEEeqnarray}{rl}
Z_{n}^{(\mathrm{d})} 
= \mathbb{E}_{\tilde{\boldsymbol{Y}}_{\backslash t},\boldsymbol{x}_{[1,m),t}}
&\left[
 \int \prod_{a=0}^{n}\left\{
  p(\boldsymbol{y}_{t} | \boldsymbol{H}^{(a)},\boldsymbol{x}_{t}^{(a)})
  p(\boldsymbol{H}^{(a)} | \tilde{\mathcal{I}}_{t})d\boldsymbol{H}^{(a)} 
 \right\}
\prod_{a=2}^{n}\left\{
   p(\boldsymbol{x}_{[m,M],t}^{(a)} | \boldsymbol{\theta}_{[m,M],t})
   d\boldsymbol{x}_{[m,M],t}^{(a)}
  \right\}
\right. \nonumber \\
&\left. 
 \times
 p(\boldsymbol{x}_{(m,M],t}^{(0)} | 
 \boldsymbol{\theta}_{(m,M],t})
 d\boldsymbol{x}_{(m,M],t}^{(0)}
 p(\boldsymbol{x}_{[m,M],t}^{(1)} | \boldsymbol{\theta}_{[m,M],t})
 d\boldsymbol{x}_{(m,M],t}^{(1)}d\boldsymbol{y}_{t}
\right],  \label{partition_d_tmp2} 
\end{IEEEeqnarray}
where $\boldsymbol{H}^{(a)}=((\vec{\boldsymbol{h}}_{1}^{(a)})^{\mathrm{T}},\ldots,
(\vec{\boldsymbol{h}}_{N}^{(a)})^{\mathrm{T}})^{\mathrm{T}}
\in\mathbb{C}^{N\times M}$ denotes 
replicas of $\boldsymbol{H}$ for $a=0,\ldots,n$: 
$\{\boldsymbol{H}^{(a)}\}$ 
conditioned on $\tilde{\mathcal{I}}_{t}$ are mutually independent 
random matrices drawn from $p(\boldsymbol{H}|\tilde{\mathcal{I}}_{t})$. 
This expression is useful since the covariance matrix of 
$\boldsymbol{y}_{t}$ for 
$p(\boldsymbol{y}_{t}|\boldsymbol{H}^{(a)}, \boldsymbol{x}_{t}^{(a)})$ 
does not depend on $\boldsymbol{x}_{t}^{(a)}$, while the covariance matrix of 
$\boldsymbol{y}_{t}$ for 
$p(\boldsymbol{y}_{t}|\boldsymbol{x}_{t}^{(a)},\tilde{\mathcal{I}}_{t})$ 
depends on $\boldsymbol{x}_{t}^{(a)}$. 
Using the fact that the row vectors $\{\vec{\boldsymbol{h}}_{n'}^{(a)}\}$ 
of $\boldsymbol{H}^{(a)}$ are mutually independent, we obtain 
\begin{equation} \label{partition_d_tmp3} 
Z_{n}^{(\mathrm{d})} 
= p(x_{m,t}^{(1)}|\theta_{m,t}^{(1)})\mathbb{E}\left\{
 \left. 
  \prod_{n'=1}^{N}\mathbb{E}_{\{\vec{\boldsymbol{h}}_{n'}^{(a)}\}}\left[
   \int \prod_{a=0}^{n}\left\{
    \frac{1}{\pi N_{0}}\mathrm{e}^{
     -\frac{1}{N_{0}}|y-v_{n',m,t}^{(a)}|^{2}
    }
   \right\}dy
  \right]
 \right| x_{m,t}^{(1)},x_{m,t}^{(0)},\bar{\boldsymbol{X}}_{\backslash t}, 
 \boldsymbol{\theta}_{t} 
\right\}, 
\end{equation}
with 
\begin{equation}
v_{n',m,t}^{(a)}=\frac{1}{\sqrt{M}}\left[
 \sum_{m'=1}^{m-1}\Delta h_{n',m',t}^{(a)}x_{m',t}
 + \sum_{m'=m}^{M}h_{n',m'}^{(a)}x_{m',t}^{(a)}
\right], 
\end{equation}
where $h_{n',m'}^{(a)}$ and $\Delta h_{n',m',t}^{(a)}$ denote the 
$(n',m')$th element of $\boldsymbol{H}^{(a)}$  
and the LMMSE estimation error $\Delta h_{n',m',t}^{(a)}=h_{n',m'}^{(a)}-
\hat{h}_{n',m',t}$, respectively, with $\hat{h}_{n',m',t}$ denoting the 
$(n',m')$th element of the LMMSE estimate~(\ref{LMMSE_estimate}). 
In (\ref{partition_d_tmp3}), the expectation 
$\mathbb{E}_{\{\vec{\boldsymbol{h}}_{n'}^{(a)}\}}[\cdots]$ is taken with 
respect to the measure 
$p(\vec{\boldsymbol{h}}_{n'}^{(a)}|\tilde{\mathcal{I}}_{t})
d\vec{\boldsymbol{h}}_{n'}^{(a)}$.  
In the derivation of (\ref{partition_d_tmp3}), we have eliminated the 
bias $b=M^{-1/2}\sum_{m'=1}^{m-1}\hat{h}_{n',m',t}x_{m',t}$ known to the 
receiver by transforming 
$(\boldsymbol{y}_{t})_{n'}$ into $y=(\boldsymbol{y}_{t})_{n'}-b$. 
Performing the Gaussian integration with respect to $y$, we have 
\begin{equation} \label{partition_d_tmp4} 
Z_{n}^{(\mathrm{d})} 
= p(x_{m,t}^{(1)}|\theta_{m,t}^{(1)})\mathbb{E}\left[
 \left. 
  \prod_{n'=1}^{N}\frac{
   \mathbb{E}_{\{\vec{\boldsymbol{h}}_{n'}^{(a)}\}}\left[
    \mathrm{e}^{
     -N_{0}^{-1}\boldsymbol{v}_{n',m,t}^{\mathrm{H}}
     \boldsymbol{A}\boldsymbol{v}_{n',m,t}
    }
   \right]
  }
  {
   (\pi N_{0})^{n}(1+n)
  } 
 \right| x_{m,t}^{(1)},x_{m,t}^{(0)},\bar{\boldsymbol{X}}_{\backslash t},
 \boldsymbol{\theta}_{t}
\right], 
\end{equation}
with $\boldsymbol{v}_{n',m,t}=(v_{n',m,t}^{(0)},\ldots,
v_{n',m,t}^{(n)})^{\mathrm{T}}$. In (\ref{partition_d_tmp4}), 
the matrix $\boldsymbol{A}$ is given by (\ref{A}). 

We next calculate the expectation in (\ref{partition_d_tmp4}) with respect 
to $\{\vec{\boldsymbol{h}}_{n'}^{(a)}\}$. Since $\{h_{n',m'}^{(a)}\}$ 
conditioned on $\tilde{\mathcal{I}}_{t}$ are proper complex Gaussian 
random vectors\footnote{
We could not immediately conclude the Gaussianity of $\boldsymbol{v}_{n',m,t}$ 
if the optimal channel estimator~(\ref{posterior_H}) were used. 
}, the random vector $\boldsymbol{v}_{n',m,t}$ conditioned on 
$\mathcal{X}_{t}=\{\boldsymbol{x}_{t}^{(a)}:\hbox{for all $a$}\}$ and 
$\tilde{\mathcal{I}}_{t}$ is also a proper complex 
Gaussian random variable with mean 
\begin{equation}
\boldsymbol{u}_{n',m,t} = \frac{1}{\sqrt{M}}
\sum_{m'=m}^{M}\hat{h}_{n',m',t}\boldsymbol{x}_{m',t}, 
\end{equation}
and with the covariance matrix 
$\boldsymbol{D} = M^{-1}\mathrm{diag}\{
 (\boldsymbol{x}_{t}^{(0)})^{\mathrm{H}}\boldsymbol{\Xi}_{t}
 \boldsymbol{x}_{t}^{(0)},
 \ldots,(\boldsymbol{x}_{t}^{(n)})^{\mathrm{T}}\boldsymbol{\Xi}_{t}
 \boldsymbol{x}_{t}^{(n)}
\}$, in which $\boldsymbol{x}_{m',t}\in\mathbb{C}^{n+1}$ is given by 
$\boldsymbol{x}_{m',t}=(x_{m',t}^{(0)},\ldots,x_{m',t}^{(n)})^{\mathrm{T}}$.  
In the same manner as in the derivation of (\ref{e_tmp1}), 
we take the expectation with respect to $\boldsymbol{v}_{n',m,t}$ conditioned 
on $\mathcal{X}_{t}$ and $\tilde{\mathcal{I}}_{t}$ to obtain 
\begin{equation} \label{partition_d_tmp5} 
Z_{n}^{(\mathrm{d})} 
= p(x_{m,t}^{(1)}|\theta_{m,t}^{(1)})\mathbb{E}\left[
 \left. 
  \prod_{n'=1}^{N}\mathrm{e}^{
   G(N_{0}^{-1}\boldsymbol{D})    
   -\boldsymbol{u}_{n',m,t}^{\mathrm{H}}\boldsymbol{B}(\boldsymbol{D},N_{0}^{-1}
   \boldsymbol{A})\boldsymbol{u}_{n',m,t}
  }
 \right| x_{m,t}^{(1)},x_{m,t}^{(0)},\bar{\boldsymbol{X}}_{\backslash t},
 \boldsymbol{\theta}_{t}
\right], 
\end{equation}
where $G(\boldsymbol{Q})$ and $\boldsymbol{B}(\boldsymbol{D},N_{0}^{-1}
\boldsymbol{A})$ are given by (\ref{function_G}) and (\ref{B}), respectively. 

Finally, we evaluate the expectation in (\ref{partition_d_tmp5}) with respect 
to $\tilde{\boldsymbol{Y}}_{\backslash t}$. 
Expression~(\ref{LMMSE_estimate}) implies that the LMMSE estimates 
$\{(\hat{h}_{n',m,t},\ldots,\hat{h}_{n',M,t}):\hbox{for all $n'$}\}$ 
conditioned on $\bar{\boldsymbol{X}}_{\backslash t}$ 
are mutually independent CSCG random vectors 
with the covariance matrix 
$\boldsymbol{I}-\boldsymbol{\Xi}_{t}^{(\mathrm{c})}$. Thus, 
the vectors $\{\boldsymbol{u}_{n',m,t}\}$ conditioned on 
$\bar{\boldsymbol{X}}_{\backslash t}$ and $\mathcal{X}_{t}$ are also 
mutually independent CSCG random vectors 
with covariance 
$\mathbb{E}[
(\boldsymbol{u}_{n',m,t})_{a}(\boldsymbol{u}_{n',m,t})_{a'}^{*} 
| \bar{\boldsymbol{X}}_{\backslash t},\mathcal{X}_{t}]
=M^{-1}(\boldsymbol{x}_{[m,M],t}^{(a')})^{\mathrm{H}}
(\boldsymbol{I}-\boldsymbol{\Xi}_{t}^{(\mathrm{c})})
\boldsymbol{x}_{[m,M],t}^{(a)}$ for all $n'$. Taking the expectation 
with respect to $\tilde{\boldsymbol{Y}}_{\backslash t}$, 
after some calculation, we have 
\begin{equation} \label{partition_d_tmp6} 
Z_{n}^{(\mathrm{d})} 
= p(x_{m,t}^{(1)}|\theta_{m,t}^{(1)})\mathbb{E}\left[
 \left. 
  \mathrm{e}^{
   NG(N_{0}^{-1}\boldsymbol{Q}_{\mathrm{d}})
  }
 \right| x_{m,t}^{(1)},x_{m,t}^{(0)},\boldsymbol{\Xi}_{t},
 \boldsymbol{\theta}_{t} 
\right], 
\end{equation} 
where the $(n+1)\times(n+1)$ Hermitian matrix $\boldsymbol{Q}_{\mathrm{d}}$ 
is given by  
\begin{equation}
(\boldsymbol{Q}_{\mathrm{d}})_{a,a'} = 
M^{-1}\delta_{a,a'}(\boldsymbol{x}_{t}^{(a)})^{\mathrm{H}}\boldsymbol{\Xi}_{t}
\boldsymbol{x}_{t}^{(a)}  
+ M^{-1}(\boldsymbol{x}_{[m,M],t}^{(a')})^{\mathrm{H}}
(\boldsymbol{I}-\boldsymbol{\Xi}_{t}^{(\mathrm{c})})
\boldsymbol{x}_{[m,M],t}^{(a)}.  
\end{equation}

\subsection{Average over Replicated Variables} 
In order to evaluate the conditional expectation in (\ref{partition_d_tmp6}) 
with respect to the replicated variables $\boldsymbol{Q}_{\mathrm{d}}$, 
we evaluate the pdf of $\boldsymbol{Q}_{\mathrm{d}}$ conditioned on 
$x_{m,t}^{(1)}$, $x_{m,t}^{(0)}$, $\boldsymbol{\Xi}_{t}$, and 
$\boldsymbol{\theta}_{t}$. 
Let us define the function $\tilde{I}_{\mathrm{d}}(\boldsymbol{Q}_{\mathrm{d}},
\tilde{\boldsymbol{Q}}_{\mathrm{d}})$ as  
\begin{equation} \label{function_I_inf}
\tilde{I}_{\mathrm{d}}(\boldsymbol{Q}_{\mathrm{d}},
\tilde{\boldsymbol{Q}}_{\mathrm{d}}) = 
\mathrm{Tr}(\boldsymbol{Q}_{\mathrm{d}}\tilde{\boldsymbol{Q}}_{\mathrm{d}}) 
- \lim_{M\rightarrow\infty}\frac{1}{M}
 \ln\tilde{F}_{\mathrm{d}}(\tilde{\boldsymbol{Q}}_{\mathrm{d}}), 
\end{equation} 
with 
\begin{equation}
\tilde{F}_{\mathrm{d}}(\tilde{\boldsymbol{Q}}_{\mathrm{d}}) = 
\mathbb{E}\left[
 \left. 
  \mathrm{e}^{M\mathrm{Tr}(\boldsymbol{Q}_{\mathrm{d}}
  \tilde{\boldsymbol{Q}}_{\mathrm{d}})}  
 \right| \boldsymbol{\Xi}_{t}, \boldsymbol{\theta}_{t}  
\right], 
\end{equation}
where a positive definite $(n+1)\times(n+1)$ Hermitian matrix 
$\tilde{\boldsymbol{Q}}_{\mathrm{d}}$ is defined in the same manner as in 
(\ref{Q_tilde}). In (\ref{function_I_inf}), we have implicitly assumed 
that the limit in the RHS of (\ref{function_I_inf}) exists. 
Furthermore, we define the saddle-point 
$\tilde{\boldsymbol{Q}}_{\mathrm{d}}^{(\mathrm{s})}$ as 
\begin{equation} \label{fixed_point_d1} 
\tilde{\boldsymbol{Q}}_{\mathrm{d}}^{(\mathrm{s})} 
=\argsup_{\tilde{\boldsymbol{Q}}_{\mathrm{d}}\in\mathcal{M}_{n+1}^{+}}
\tilde{I}_{\mathrm{d}}(\boldsymbol{Q}_{\mathrm{d}},
\tilde{\boldsymbol{Q}}_{\mathrm{d}}). 
\end{equation}
We represent the pdf $\mu(\boldsymbol{Q}_{\mathrm{d}})$ of 
$\boldsymbol{Q}_{\mathrm{d}}$ conditioned on 
$x_{m,t}^{(1)}$, $x_{m,t}^{(0)}$, $\boldsymbol{\Xi}_{t}$, and 
$\boldsymbol{\theta}_{t}$ by using the inversion formula for the moment 
generating function of $\boldsymbol{Q}_{\mathrm{d}}$, given by 
\begin{equation} \label{moment_generating_c} 
F_{\mathrm{d}}(\tilde{\boldsymbol{Q}}_{\mathrm{d}}) = 
\mathbb{E}\left[
 \left. 
  \mathrm{e}^{M\mathrm{Tr}(\boldsymbol{Q}_{\mathrm{d}}
  \tilde{\boldsymbol{Q}}_{\mathrm{d}})}  
 \right| x_{m,t}^{(1)},x_{m,t}^{(0)},\boldsymbol{\Xi}_{t}, 
 \boldsymbol{\theta}_{t}  
\right]. 
\end{equation} 
Using the saddle-point method in the same manner as in the derivation of 
(\ref{measure_tmp4}) gives  
\begin{equation} \label{measure_c} 
\mu(\boldsymbol{Q}_{\mathrm{d}}) = 
\left(
 \frac{M}{2\pi}
\right)^{(n+1)^{2}}
|\det\{\nabla_{\tilde{\boldsymbol{Q}}_{\mathrm{d}}}^{2}
I_{\mathrm{d}}(\boldsymbol{Q}_{\mathrm{d}},
\tilde{\boldsymbol{Q}}_{\mathrm{d}}^{(\mathrm{s})})\}|^{-1}
\mathrm{e}^{-MI_{\mathrm{d}}(\boldsymbol{Q}_{\mathrm{d}},
\tilde{\boldsymbol{Q}}_{\mathrm{d}}^{(\mathrm{s})})+O(M^{-1})}
[1+O(M^{-1/2})],   
\end{equation}
in the large-system limit. In (\ref{measure_c}), 
the function $I_{\mathrm{d}}(\boldsymbol{Q}_{\mathrm{d}},
\tilde{\boldsymbol{Q}}_{\mathrm{d}})$ is given by 
\begin{equation} \label{function_I_c} 
I_{\mathrm{d}}(\boldsymbol{Q}_{\mathrm{d}},
\tilde{\boldsymbol{Q}}_{\mathrm{d}}) = 
\mathrm{Tr}(\boldsymbol{Q}_{\mathrm{d}}\tilde{\boldsymbol{Q}}_{\mathrm{d}}) 
- \frac{1}{M}\ln F_{\mathrm{d}}(\tilde{\boldsymbol{Q}}_{\mathrm{d}}). 
\end{equation}
Furthermore, $\nabla_{\tilde{\boldsymbol{Q}}_{\mathrm{d}}}^{2}
I_{\mathrm{d}}(\boldsymbol{Q},\tilde{\boldsymbol{Q}}_{\mathrm{d}})$ 
denotes the Hesse matrix of (\ref{function_I_c}) with respect to 
$\tilde{\boldsymbol{Q}}_{\mathrm{d}}$. 
 
The factor $O(M^{-1})$ in the exponent in (\ref{measure_c}) is due to a small 
deviation of the saddle-point~(\ref{fixed_point_d1}). The removal or addition 
of one transmit antenna results in a small change of 
$M\mathrm{Tr}(\boldsymbol{Q}_{\mathrm{d}}
\tilde{\boldsymbol{Q}}_{\mathrm{d}})$, more precisely, in a change of $O(1)$. 
This observation implies that 
$I_{\mathrm{d}}(\boldsymbol{Q}_{\mathrm{d}},
\tilde{\boldsymbol{Q}}_{\mathrm{d}}) 
= \tilde{I}_{\mathrm{d}}(\boldsymbol{Q}_{\mathrm{d}},
\tilde{\boldsymbol{Q}}_{\mathrm{d}})+ O(M^{-1})$ in the large-system limit. 
Differentiating both sides with respect to 
$\tilde{\boldsymbol{Q}}_{\mathrm{d}}$ at the 
saddle-point~(\ref{fixed_point_d1}), we find that 
the gradient $\nabla_{\tilde{\boldsymbol{Q}}_{\mathrm{d}}}
I_{\mathrm{d}}(\boldsymbol{Q}_{\mathrm{d}},
\tilde{\boldsymbol{Q}}_{\mathrm{d}}^{(\mathrm{s})})$ of 
(\ref{function_I_c}) with respect to $\tilde{\boldsymbol{Q}}_{\mathrm{d}}$ 
at the saddle-point is $O(M^{-1})$, which explains the factor 
$O(M^{-1})$ in the exponent in (\ref{measure_c}) since a deviation of 
the saddle-point results in a deviation of the exponent which is proportional 
to $M\|\nabla_{\tilde{\boldsymbol{Q}}_{\mathrm{d}}}
I_{\mathrm{d}}(\boldsymbol{Q}_{\mathrm{d}},
\tilde{\boldsymbol{Q}}_{\mathrm{d}}^{(\mathrm{s})})\|^{2}$. 

We repeat the same argument to evaluate (\ref{partition_d_tmp6}). 
Applying (\ref{measure_c}) to (\ref{partition_d_tmp6}) and using the 
saddle-point method, we arrive at 
\begin{equation} \label{partition_d_tmp7} 
Z_{n}^{(\mathrm{d})}  
= p(x_{m,t}^{(1)}|\theta_{m,t}^{(1)})
C_{n}^{(\mathrm{d})}(\boldsymbol{Q}_{\mathrm{d}}^{(\mathrm{s})},
\tilde{\boldsymbol{Q}}_{\mathrm{d}}^{(\mathrm{s})}) 
\mathrm{e}^{-M\Phi_{\mathrm{d}}(\boldsymbol{Q}_{\mathrm{d}}^{(\mathrm{s})})} 
[ 1 + O(M^{-1/2}) ]. 
\end{equation}
In (\ref{partition_d_tmp7}), the function 
$\Phi_{\mathrm{d}}(\boldsymbol{Q}_{\mathrm{d}})$ is defined as 
\begin{equation} \label{Phi_c}
\Phi_{\mathrm{d}}(\boldsymbol{Q}_{\mathrm{d}}) = 
I_{\mathrm{d}}(\boldsymbol{Q}_{\mathrm{d}},
\tilde{\boldsymbol{Q}}_{\mathrm{d}}^{(\mathrm{s})}) 
- \alpha^{-1}G(N_{0}^{-1}\boldsymbol{Q}_{\mathrm{d}}). 
\end{equation} 
The saddle-point $\boldsymbol{Q}_{\mathrm{d}}^{(\mathrm{s})}$ 
is given by 
\begin{equation} \label{fixed_point_d2} 
\boldsymbol{Q}_{\mathrm{d}}^{(\mathrm{s})} 
= \arginf_{\boldsymbol{Q}_{\mathrm{d}}\in\mathcal{M}_{n+1}^{+}}
\tilde{\Phi}_{\mathrm{d}}(\boldsymbol{Q}_{\mathrm{d}}) , 
\end{equation}
with  
\begin{equation} \label{Phi_c_inf} 
\tilde{\Phi}_{\mathrm{d}}(\boldsymbol{Q}_{\mathrm{d}}) = 
\tilde{I}_{\mathrm{d}}(\boldsymbol{Q}_{\mathrm{d}},
\tilde{\boldsymbol{Q}}_{\mathrm{d}}^{(\mathrm{s})}) 
- \alpha^{-1}G(N_{0}^{-1}\boldsymbol{Q}_{\mathrm{d}}). 
\end{equation} 
Furthermore, $C_{n}^{(\mathrm{d})}
(\boldsymbol{Q}_{\mathrm{d}},\tilde{\boldsymbol{Q}}_{\mathrm{d}})$ 
is defined as 
\begin{equation}
C_{n}^{(\mathrm{d})}
(\boldsymbol{Q}_{\mathrm{d}},\tilde{\boldsymbol{Q}}_{\mathrm{d}}) 
= |\det\{\nabla_{\tilde{\boldsymbol{Q}}_{\mathrm{d}}}^{2}
I_{\mathrm{d}}(\boldsymbol{Q}_{\mathrm{d}},
\tilde{\boldsymbol{Q}}_{\mathrm{d}})\}|^{-1}  
\det\{\alpha^{-1}\nabla_{\boldsymbol{Q}_{\mathrm{d}}}^{2}
G(N_{0}^{-1}\boldsymbol{Q}_{\mathrm{d}})\}^{-1},  
\end{equation}
where $\nabla_{\boldsymbol{Q}_{\mathrm{d}}}^{2}
G(N_{0}^{-1}\boldsymbol{Q}_{\mathrm{d}})$ denotes the Hesse matrix of 
$G(N_{0}^{-1}\boldsymbol{Q}_{\mathrm{d}})$ with respect to 
$\boldsymbol{Q}_{\mathrm{d}}$. 
Note that we have assumed the positive definiteness of the 
Hesse matrix $\nabla_{\boldsymbol{Q}_{\mathrm{d}}}^{2}
G(N_{0}^{-1}\boldsymbol{Q}_{\mathrm{d}})$ at the saddle-point 
$\boldsymbol{Q}_{\mathrm{d}}=\boldsymbol{Q}_{\mathrm{d}}^{(\mathrm{s})}$. 

The calculation of the stationarity conditions for (\ref{fixed_point_d1}) 
and (\ref{fixed_point_d2}) implies that 
$(\boldsymbol{Q}_{\mathrm{d}}^{(\mathrm{s})},
\tilde{\boldsymbol{Q}}_{\mathrm{d}}^{(\mathrm{s})})$ is given as the 
solution to the coupled fixed-point equations 
\begin{equation} \label{fixed_point_d1_tmp1} 
\boldsymbol{Q}_{\mathrm{d}} = \lim_{M\rightarrow\infty}
\frac{
 \mathbb{E}\left[
  \left. 
   \boldsymbol{Q}_{\mathrm{d}}\mathrm{e}^{
    M\mathrm{Tr}(\boldsymbol{Q}_{\mathrm{d}}
    \tilde{\boldsymbol{Q}}_{\mathrm{d}})
   }  
  \right| \boldsymbol{\Xi}_{t}, \boldsymbol{\theta}_{t}
 \right]
}
{
 \mathbb{E}\left[
  \left. 
   \mathrm{e}^{
    M\mathrm{Tr}(\boldsymbol{Q}_{\mathrm{d}}
    \tilde{\boldsymbol{Q}}_{\mathrm{d}})
   }  
  \right| \boldsymbol{\Xi}_{t}, \boldsymbol{\theta}_{t} 
 \right]
}, 
\end{equation} 
\begin{equation} \label{fixed_point_d2_tmp1} 
\tilde{\boldsymbol{Q}}_{\mathrm{d}} = 
-\frac{\alpha^{-1}}{N_{0}}\left(
 \boldsymbol{I}_{n+1} + \frac{\boldsymbol{A}}{N_{0}}
 \boldsymbol{Q}_{\mathrm{d}}
\right)^{-1}\boldsymbol{A}. 
\end{equation}

\subsection{Evaluation of Fixed-Point Equations} 
In order to evaluate the coupled fixed-point 
equations~(\ref{fixed_point_d1_tmp1}) and 
(\ref{fixed_point_d2_tmp1}), we assume RS. The assumption of RS is 
consistent with Assumption~\ref{assumption2}, i.e., the assumption of 
the self-averaging property for the 
equivalent channel~(\ref{equivalent_channel})~\cite{Pastur91}. 
  
\begin{assumption} \label{RS_d} 
The solution $(\boldsymbol{Q}_{\mathrm{d}}^{(\mathrm{s})},
\tilde{\boldsymbol{Q}}_{\mathrm{d}}^{(\mathrm{s})})$ is invariant under all 
permutations of replica indices: 
\begin{equation}
\boldsymbol{Q}_{\mathrm{d}}^{(\mathrm{s})} 
= \begin{pmatrix}
a_{\mathrm{d}} & b_{\mathrm{d}}\boldsymbol{1}_{n}^{\mathrm{T}} \\ 
b_{\mathrm{d}}^{*}\boldsymbol{1}_{n} & 
(d_{\mathrm{d}}-c_{\mathrm{d}})\boldsymbol{I}_{n} 
+ c_{\mathrm{d}}\boldsymbol{1}_{n}\boldsymbol{1}_{n}^{\mathrm{T}} 
\end{pmatrix}, 
\end{equation}
\begin{equation}
\tilde{\boldsymbol{Q}}_{\mathrm{d}}^{(\mathrm{s})} 
= \begin{pmatrix}
\tilde{a}_{\mathrm{d}} & \tilde{b}_{\mathrm{d}}\boldsymbol{1}_{n}^{\mathrm{T}} \\ 
\tilde{b}_{\mathrm{d}}^{*}\boldsymbol{1}_{n} & 
(\tilde{d}_{\mathrm{d}}-\tilde{c}_{\mathrm{d}})\boldsymbol{I}_{n} 
+ \tilde{c}_{\mathrm{d}}\boldsymbol{1}_{n}\boldsymbol{1}_{n}^{\mathrm{T}}  
\end{pmatrix}. 
\end{equation}
\end{assumption} 

We first evaluate the fixed-point equation~(\ref{fixed_point_d2_tmp1}). 
Let us define $\sigma_{0}^{2}$ and  $\sigma^{2}$ as 
\begin{equation} \label{sigma_0}
\sigma_{0}^{2} = N_{0} + (a_{\mathrm{d}}-b_{\mathrm{d}}
-b_{\mathrm{d}}^{*}+c_{\mathrm{d}}),   
\end{equation}
\begin{equation} \label{sigma}
\sigma^{2} = N_{0} + (d_{\mathrm{d}}-c_{\mathrm{d}}),   
\end{equation}
respectively. After some calculation, we obtain 
\begin{equation} \label{fixed_point_d2_tmp2} 
\tilde{a}_{\mathrm{d}} = -\frac{\alpha^{-1}n}{\sigma^{2} + n\sigma_{0}^{2}},\; 
\tilde{b}_{\mathrm{d}} = \frac{\alpha^{-1}}{\sigma^{2} + n\sigma_{0}^{2}},\; 
\tilde{c}_{\mathrm{d}} = \frac{\alpha^{-1}\sigma_{0}^{2}}{
 (\sigma^{2} + n\sigma_{0}^{2})\sigma^{2}},\; 
\tilde{d}_{\mathrm{d}} = \tilde{c}_{\mathrm{d}} 
- \frac{\alpha^{-1}}{\sigma^{2}}. 
\end{equation}

We next calculate the quantity $\exp\{M\mathrm{Tr}(
\boldsymbol{Q}_{\mathrm{d}}^{(\mathrm{s})}
\tilde{\boldsymbol{Q}}_{\mathrm{d}}^{(\mathrm{s})})\}$. 
Let $\sqrt{\boldsymbol{I}-\boldsymbol{\Xi}_{t}^{(\mathrm{c})}}$ denote 
a square root of $\boldsymbol{I}-\boldsymbol{\Xi}_{t}^{(\mathrm{c})}$, 
i.e., $\boldsymbol{I}-\boldsymbol{\Xi}_{t}^{(\mathrm{c})}
=\sqrt{\boldsymbol{I}-\boldsymbol{\Xi}_{t}^{(\mathrm{c})}}^{\mathrm{H}}
\sqrt{\boldsymbol{I}-\boldsymbol{\Xi}_{t}^{(\mathrm{c})}}$. 
Substituting (\ref{fixed_point_d2_tmp2}) into that quantity gives  
\begin{IEEEeqnarray}{r}
\mathrm{e}^{M\mathrm{Tr}(
\boldsymbol{Q}_{\mathrm{d}}^{(\mathrm{s})}
\tilde{\boldsymbol{Q}}_{\mathrm{d}}^{(\mathrm{s})})} 
= \exp\left\{
 \tilde{\sigma}_{0}^{2}\left\|
  \sum_{a=0}^{n}\frac{
  \sqrt{\boldsymbol{I}-\boldsymbol{\Xi}_{t}^{(\mathrm{c})}}
  \boldsymbol{x}_{[m,M],t}^{(a)}
  }{\sqrt{\alpha}\sigma_{a}^{2}}
 \right\|^{2} 
 - \sum_{a=0}^{n}\frac{
  \|
   \sqrt{\boldsymbol{I}-\boldsymbol{\Xi}_{t}^{(\mathrm{c})}}
   \boldsymbol{x}_{[m,M],t}^{(a)}
  \|^{2} 
 }{\alpha\sigma_{a}^{2}}
\right. \nonumber \\
\left. 
 +\tilde{a}_{\mathrm{d}}(\boldsymbol{x}_{t}^{(0)})^{\mathrm{H}} 
 \boldsymbol{\Xi}_{t}\boldsymbol{x}_{t}^{(0)} 
 +\tilde{d}_{\mathrm{d}}\sum_{a=1}^{n}(\boldsymbol{x}_{t}^{(a)})^{\mathrm{H}} 
 \boldsymbol{\Xi}_{t}\boldsymbol{x}_{t}^{(a)} 
\right\}, \label{quadratic_d} 
\end{IEEEeqnarray}  
with $\tilde{\sigma}_{0}^{2}=(n\sigma^{-2}+\sigma_{0}^{-2})^{-1}$ and 
$\sigma_{a}^{2}=\sigma^{2}$ for $a=1,\ldots,n$. 
In order to linearize the quadratic form in (\ref{quadratic_d}), 
we use the identity 
\begin{equation} 
\mathrm{e}^{\tilde{\sigma}_{0}^{2}\|\boldsymbol{a}\|^{2}} 
= \int_{\mathbb{C}^{M-m+1}} \frac{1}{(\pi\tilde{\sigma}_{0}^{2})^{M-m+1}} 
\mathrm{e}^{
 - \frac{\|\underline{\boldsymbol{z}}\|^{2}}{\tilde{\sigma}_{0}^{2}} 
 + \boldsymbol{a}^{\mathrm{H}}\underline{\boldsymbol{z}}
 +\underline{\boldsymbol{z}}^{\mathrm{H}}\boldsymbol{a}
}d\underline{\boldsymbol{z}}, 
\end{equation}
with $\boldsymbol{a}=\sum_{a=0}^{n}
\sqrt{\boldsymbol{I}-\boldsymbol{\Xi}_{t}^{(\mathrm{c})}}
\boldsymbol{x}_{[m,M],t}^{(a)}/(\sqrt{\alpha}\sigma_{a}^{2})$ to obtain 
\begin{equation} \label{quadratic_d_tmp1} 
\mathrm{e}^{M\mathrm{Tr}(
\boldsymbol{Q}_{\mathrm{d}}^{(\mathrm{s})}
\tilde{\boldsymbol{Q}}_{\mathrm{d}}^{(\mathrm{s})})} 
= D_{n}^{(\mathrm{d})}\int \prod_{a=0}^{n}\left\{
 \tilde{q}_{a}(\underline{\boldsymbol{z}} | \boldsymbol{x}_{t}^{(a)},
 \boldsymbol{\Xi}_{t})
\right\}d\underline{\boldsymbol{z}}, 
\end{equation}
with $D_{n}^{(\mathrm{d})}=(\pi\sigma^{2})^{n(M-m+1)}
(1+n\sigma_{0}^{2}/\sigma^{2})^{M-m+1}$. 
In (\ref{quadratic_d_tmp1}), the functions 
$\tilde{q}_{0}(\underline{\boldsymbol{z}}|\boldsymbol{x}_{t}^{(0)},
\boldsymbol{\Xi}_{t})$ 
and  
$\tilde{q}_{a}(\underline{\boldsymbol{z}}|\boldsymbol{x}_{t}^{(a)},
\boldsymbol{\Xi}_{t})$ for $a=1,\ldots,n$ are given by  
\begin{equation}
\tilde{q}_{0}(\underline{\boldsymbol{z}}|\boldsymbol{x}_{t}^{(0)},
\boldsymbol{\Xi}_{t}) 
= q_{0}(\underline{\boldsymbol{z}}|\boldsymbol{x}_{[m,M],t}^{(0)},
\boldsymbol{\Xi}_{t}^{(\mathrm{c})})
\mathrm{e}^{
 \tilde{a}_{\mathrm{d}}(\boldsymbol{x}_{t}^{(0)})^{\mathrm{H}} 
 \boldsymbol{\Xi}_{t}\boldsymbol{x}_{t}^{(0)}
},
\end{equation}
\begin{equation}
\tilde{q}_{a}(\underline{\boldsymbol{z}}|\boldsymbol{x}_{t}^{(a)},
\boldsymbol{\Xi}_{t}) 
= q_{a}(\underline{\boldsymbol{z}}|\boldsymbol{x}_{[m,M],t}^{(a)},
\boldsymbol{\Xi}_{t}^{(\mathrm{c})})
\mathrm{e}^{
 \tilde{d}_{\mathrm{d}}(\boldsymbol{x}_{t}^{(a)})^{\mathrm{H}} 
 \boldsymbol{\Xi}_{t}\boldsymbol{x}_{t}^{(a)}
},
\end{equation}
respectively, where 
$q_{a}(\underline{\boldsymbol{z}} | \boldsymbol{x}_{[m,M],t}^{(a)},
\boldsymbol{\Xi}_{t}^{(\mathrm{c})})$ represents the pdf of a  
proper complex Gaussian random vector 
$\underline{\boldsymbol{z}}\in\mathbb{C}^{M-m+1}$ 
with mean $\sqrt{\boldsymbol{I}-\boldsymbol{\Xi}_{t}^{(\mathrm{c})}}
\boldsymbol{x}_{[m,M],t}^{(a)}/\sqrt{\alpha}$ and covariance 
$\sigma_{a}^{2}\boldsymbol{I}$, i.e.,  
\begin{equation}
q_{a}(\underline{\boldsymbol{z}} | \boldsymbol{x}_{[m,M],t}^{(a)},
\boldsymbol{\Xi}_{t}^{(\mathrm{c})})
= \frac{1}{(\pi\sigma_{a}^{2})^{M-m+1}}\exp\left\{
 -\frac{
 \| 
  \underline{\boldsymbol{z}} 
  - \sqrt{\boldsymbol{I}-\boldsymbol{\Xi}_{t}^{(\mathrm{c})}}
  \boldsymbol{x}_{[m,M],t}^{(a)}/\sqrt{\alpha}
 \|^{2}
 }{\sigma_{a}^{2}}
\right\}. 
\end{equation}
 
Finally, we evaluate the fixed-point equation~(\ref{fixed_point_d1_tmp1}). 
Substitution of (\ref{quadratic_d_tmp1}) into (\ref{fixed_point_d1_tmp1}) 
gives expressions of 
$a_{\mathrm{d}}-b_{\mathrm{d}}-b_{\mathrm{d}}^{*}+c_{\mathrm{d}}$ 
and $d_{\mathrm{d}}-c_{\mathrm{d}}$ well-defined for $n\in\mathbb{R}$. 
Taking $n\rightarrow+0$, we have 
\begin{IEEEeqnarray}{rl}
\lim_{n\rightarrow+0}
(a_{\mathrm{d}}-b_{\mathrm{d}}-b_{\mathrm{d}}^{*}+c_{\mathrm{d}})
&= \lim_{M\rightarrow\infty}\frac{1}{M}\mathbb{E}\left\{ 
 \int  q_{0}(\underline{\boldsymbol{z}} | \boldsymbol{x}_{[m,M],t}^{(0)},
 \boldsymbol{\Xi}_{t}^{(\mathrm{c})})
\right. \nonumber \\ 
&\left. 
 \left. 
  \times\left( 
   (\boldsymbol{x}_{t}^{(0)})^{\mathrm{H}}\boldsymbol{\Xi}_{t}
   \boldsymbol{x}_{t}^{(0)}   
   + \left\|
    \sqrt{\boldsymbol{I} - \boldsymbol{\Xi}_{t}^{(\mathrm{c})}}
    (\boldsymbol{x}_{[m,M],t}^{(0)} - \langle \boldsymbol{x}_{[m,M],t}^{(1)} 
    \rangle)
   \right\|^{2}
  \right)d\underline{\boldsymbol{z}}
 \right| \boldsymbol{\Xi}_{t}, \boldsymbol{\theta}_{t} 
\right\}, 
\end{IEEEeqnarray} 
\begin{IEEEeqnarray}{rl}
\lim_{n\rightarrow+0}(d_{\mathrm{d}}-c_{\mathrm{d}})
=& \lim_{M\rightarrow\infty}\frac{1}{M}\mathbb{E}\left\{
 \int q_{0}(\underline{\boldsymbol{z}} | \boldsymbol{x}_{[m,M],t}^{(0)},
 \boldsymbol{\Xi}_{t}^{(\mathrm{c})})
\right. \nonumber \\
&\left. 
 \left. 
  \times\left\langle 
   (\boldsymbol{x}_{t}^{(1)})^{\mathrm{H}}\boldsymbol{\Xi}_{t}
   \boldsymbol{x}_{t}^{(1)} + \left\|
    \sqrt{\boldsymbol{I} - \boldsymbol{\Xi}_{t}^{(\mathrm{c})}}
    (\boldsymbol{x}_{[m,M],t}^{(1)} - \langle \boldsymbol{x}_{[m,M],t}^{(1)} 
    \rangle)
   \right\|^{2}
  \right\rangle d\underline{\boldsymbol{z}}
 \right| \boldsymbol{\Xi}_{t}, \boldsymbol{\theta}_{t} 
\right\},
\end{IEEEeqnarray} 
with 
\begin{equation}
\langle \boldsymbol{x}_{t}^{(1)} \rangle = 
\frac{ 
 \int\boldsymbol{x}_{t}^{(1)} 
 \tilde{q}_{1}(\underline{\boldsymbol{z}} | \boldsymbol{x}_{t}^{(1)},
 \boldsymbol{\Xi}_{t}) 
 p(\boldsymbol{x}_{t}^{(1)}|\boldsymbol{\theta}_{t})d\boldsymbol{x}_{t}^{(1)} 
}
{
 \int \tilde{q}_{1}(\underline{\boldsymbol{z}} | \boldsymbol{x}_{t}^{(1)},
 \boldsymbol{\Xi}_{t}) 
 p(\boldsymbol{x}_{t}^{(1)}|\boldsymbol{\theta}_{t})d\boldsymbol{x}_{t}^{(1)} 
}. 
\end{equation}
Substituting these expressions into (\ref{sigma_0}) or (\ref{sigma}), we 
have the coupled fixed-point equations 
\begin{equation} \label{fixed_point1_tmp} 
\sigma_{0}^{2} = N_{0} + \lim_{M\rightarrow\infty}\frac{1}{M}\left\{
 P\mathrm{Tr}(\boldsymbol{\Xi}_{t})
 + \mathbb{E}\left[ 
  \left. 
   (\boldsymbol{x}_{[m,M],t}^{(0)} - \langle \boldsymbol{x}_{[m,M],t}^{(1)} 
   \rangle)^{\mathrm{H}}(\boldsymbol{I} - \boldsymbol{\Xi}_{t}^{(\mathrm{c})})
   (\boldsymbol{x}_{[m,M],t}^{(0)} - \langle \boldsymbol{x}_{[m,M],t}^{(1)} 
   \rangle) 
  \right| \boldsymbol{\Xi}_{t}^{(\mathrm{c})},\boldsymbol{\theta}_{t} 
 \right]
\right\},  
\end{equation}
\begin{equation} \label{fixed_point2_tmp} 
\sigma^{2} = N_{0} + \lim_{M\rightarrow\infty}\frac{1}{M}\left\{
 P\mathrm{Tr}(\boldsymbol{\Xi}_{t})
 + \mathbb{E}\left[ 
  \left. 
   (\boldsymbol{x}_{[m,M],t}^{(1)} - \langle \boldsymbol{x}_{[m,M],t}^{(1)} 
   \rangle)^{\mathrm{H}}(\boldsymbol{I} - \boldsymbol{\Xi}_{t}^{(\mathrm{c})})
   (\boldsymbol{x}_{[m,M],t}^{(1)} - \langle \boldsymbol{x}_{[m,M],t}^{(1)} 
   \rangle)
  \right| \boldsymbol{\Xi}_{t}^{(\mathrm{c})},\boldsymbol{\theta}_{t} 
 \right]
\right\}, 
\end{equation}
where the average over $\underline{\boldsymbol{z}}$ is taken with respect to 
the measure $q_{0}(\underline{\boldsymbol{z}} | \boldsymbol{x}_{[m,M],t}^{(0)},
\boldsymbol{\Xi}_{t}^{(\mathrm{c})})d\underline{\boldsymbol{z}}$.

The coupled fixed-point equations~(\ref{fixed_point1_tmp}) and 
(\ref{fixed_point2_tmp}) have the solution $\sigma_{0}^{2}=\sigma^{2}$.  
Nishimori's result~\cite{Nishimori01} implies that 
$\sigma_{0}^{2}=\sigma^{2}$ is the correct solution. Assuming 
$\sigma_{0}^{2}=\sigma^{2}$, we have the single fixed-point equation 
\begin{equation} \label{fixed_point_d} 
\sigma^{2} = N_{0} 
+ \lim_{M\rightarrow\infty}\frac{P}{M}\mathrm{Tr}(\boldsymbol{\Xi}_{t}) 
+ V(\sigma^{2}), 
\end{equation}
with 
\begin{equation}
V(\sigma^{2}) = \lim_{M\rightarrow\infty}\frac{1}{M}\mathbb{E}\left[
 \left. 
  (\boldsymbol{x}_{[m,M],t}^{(1)} - \langle \boldsymbol{x}_{[m,M],t}^{(1)} 
  \rangle)^{\mathrm{H}}(\boldsymbol{I} - \boldsymbol{\Xi}_{t}^{(\mathrm{c})})
  (\boldsymbol{x}_{[m,M],t}^{(1)} - \langle \boldsymbol{x}_{[m,M],t}^{(1)} 
  \rangle)
 \right| \boldsymbol{\Xi}_{t}^{(\mathrm{c})}, \boldsymbol{\theta}_{[m,M],t} 
\right]. 
\end{equation} 
In (\ref{fixed_point_d}), the average over $\underline{\boldsymbol{z}}$ is 
taken with respect to the measure 
$q_{0}(\underline{\boldsymbol{z}} | \boldsymbol{x}_{[m,M],t}^{(0)},
\boldsymbol{\Xi}_{t}^{(\mathrm{c})})d\underline{\boldsymbol{z}}$ with 
$\sigma_{0}^{2}=\sigma^{2}$. 
Furthermore, $\langle \boldsymbol{x}_{[m,M],t}^{(1)} \rangle$ denotes 
the expectation of $\boldsymbol{x}_{[m,M],t}^{(1)}$ with respect to the 
posterior measure 
$q_{1}(\boldsymbol{x}_{[m,M],t}^{(1)}|\underline{\boldsymbol{z}},
\boldsymbol{\Xi}_{t}^{(\mathrm{c})})d\boldsymbol{x}_{[m,M],t}^{(1)}$, given by 
\begin{equation} \label{posterior_rs} 
q_{1}(\boldsymbol{x}_{[m,M],t}^{(1)}|\underline{\boldsymbol{z}},
\boldsymbol{\Xi}_{t}^{(\mathrm{c})},\boldsymbol{\theta}_{[m,M],t}) 
= \frac{
 q_{1}(\underline{\boldsymbol{z}} | \boldsymbol{x}_{[m,M],t}^{(1)},
 \boldsymbol{\Xi}_{t}^{(\mathrm{c})})
 p(\boldsymbol{x}_{[m,M],t}^{(1)}|\boldsymbol{\theta}_{m,t}^{(c)})
}
{
 \int q_{1}(\underline{\boldsymbol{z}} | \boldsymbol{x}_{[m,M],t}^{(1)},
 \boldsymbol{\Xi}_{t}^{(\mathrm{c})})
 p(\boldsymbol{x}_{[m,M],t}^{(1)}|\boldsymbol{\theta}_{[m,M],t})
 d\boldsymbol{x}_{[m,M],t}^{(1)}
}. 
\end{equation}
Note that the fixed-point equation~(\ref{fixed_point_d}) is equivalent to 
(\ref{fixed_point_MIMO}). 

\subsection{Replica Continuity} 
We evaluate (\ref{partition_d_tmp7}) under the RS assumption 
(Assumption~\ref{RS_d}). 
The function $G(N_{0}^{-1}\boldsymbol{Q}_{\mathrm{d}}^{(\mathrm{s})})$, 
given by (\ref{function_G}), reduces to~\cite{Takeuchi101} 
\begin{equation} \label{function_G_tmp} 
G(N_{0}^{-1}\boldsymbol{Q}_{\mathrm{d}}^{(\mathrm{s})}) 
= - (n-1)\ln\sigma^{2} - \ln(\sigma^{2}+n\sigma_{0}^{2}) - n\ln\pi, 
\end{equation}
which is well defined for $n\in\mathbb{R}$ and tends to zero as  
$n\rightarrow+0$. 

Applying (\ref{sigma_0}), (\ref{sigma}), and (\ref{fixed_point_d2_tmp2}) to 
(\ref{function_I_c}), we obtain  
\begin{equation} \label{function_I_c_tmp} 
I_{\mathrm{d}}(\boldsymbol{Q}_{\mathrm{d}}^{(\mathrm{s})},
\tilde{\boldsymbol{Q}}_{\mathrm{d}}^{(\mathrm{s})}) 
= -\frac{n}{\alpha}\left[
 1 - \frac{N_{0}}{\sigma^{2}+n\sigma_{0}^{2}} - \frac{N_{0}}{\sigma^{2}}
 + \frac{N_{0}\sigma_{0}^{2}}{\sigma^{2}(\sigma^{2}+n\sigma_{0}^{2})}
\right] 
- \frac{1}{M}\ln F_{\mathrm{d}}
(\tilde{\boldsymbol{Q}}_{\mathrm{d}}^{(\mathrm{s})}). 
\end{equation}
Substituting (\ref{quadratic_d_tmp1}) into 
the moment generating function~(\ref{moment_generating_c}), 
we have an expression of (\ref{function_I_c_tmp}) well defined for 
$n\in\mathbb{R}$.  
Taking $n\rightarrow+0$, under the assumption of $\sigma_{0}^{2}=\sigma^{2}$, 
we have   
\begin{IEEEeqnarray}{rl}  
&\lim_{n\rightarrow+0}
I_{\mathrm{d}}(\boldsymbol{Q}_{\mathrm{d}}^{(\mathrm{s})},
\tilde{\boldsymbol{Q}}_{\mathrm{d}}^{(\mathrm{s})})  
\nonumber \\ 
=& -\frac{1}{M}\ln
\int q_{1}(x_{m,t}^{(1)}|\underline{\boldsymbol{z}},
\boldsymbol{\Xi}_{t}^{(\mathrm{c})}, 
\boldsymbol{\theta}_{[m,M],t})
q_{0}(\underline{\boldsymbol{z}}|\boldsymbol{x}_{[m,M],t}^{(0)}, 
\boldsymbol{\Xi}_{t}^{(\mathrm{c})})
p(\boldsymbol{x}_{(m,M],t}^{(0)}|
\boldsymbol{\theta}_{(m,M],t})
d\boldsymbol{x}_{(m,M],t}^{(0)}
d\underline{\boldsymbol{z}}, \label{moment_generating_tmp}
\end{IEEEeqnarray}  
with the marginal $q_{1}(x_{m,t}^{(1)}|\underline{\boldsymbol{z}},
\boldsymbol{\Xi}_{t}^{(\mathrm{c})},\boldsymbol{\theta}_{[m,M],t}) 
= \int q_{1}(\boldsymbol{x}_{[m,M],t}^{(1)}|\underline{\boldsymbol{z}},
\boldsymbol{\Xi}_{t}^{(\mathrm{c})},\boldsymbol{\theta}_{[m,M],t})
d\boldsymbol{x}_{(m,M],t}^{(1)}$ of (\ref{posterior_rs}). Substituting  
(\ref{function_G_tmp}) and (\ref{moment_generating_tmp}) into (\ref{Phi_c}) 
and assuming that the obtained expression is correct as  
$n\rightarrow+0$, from (\ref{formulation}), we arrive at 
\begin{IEEEeqnarray}{rl}
\mathbb{E}_{\tilde{\boldsymbol{Y}}_{\backslash t},\boldsymbol{x}_{[1,m),t}}
&\left[
 p(\tilde{x}_{m,t}| x_{m,t}, \tilde{\mathcal{I}}_{t},
 \boldsymbol{x}_{[1,m),t},
 \boldsymbol{\theta}_{[m,M],t})
\right] = \lim_{n\rightarrow+0}C_{n}^{(\mathrm{d})}
(\boldsymbol{Q}_{\mathrm{d}}^{(\mathrm{s})},
\tilde{\boldsymbol{Q}}_{\mathrm{d}}^{(\mathrm{s})})
\nonumber \\ 
&\times
\int q_{1}(x_{m,t}^{(1)}|\underline{\boldsymbol{z}},
\boldsymbol{\Xi}_{t}^{(\mathrm{c})}, 
\boldsymbol{\theta}_{[m,M],t})
q_{0}(\underline{\boldsymbol{z}}|\boldsymbol{x}_{[m,M],t}^{(0)}, 
\boldsymbol{\Xi}_{t}^{(\mathrm{c})})
p(\boldsymbol{x}_{(m,M],t}^{(0)}|
\boldsymbol{\theta}_{(m,M],t})
d\boldsymbol{x}_{(m,M],t}^{(0)}
d\underline{\boldsymbol{z}},  \label{result}
\end{IEEEeqnarray} 
where we have assumed that the large-system limit and the limit 
$n\rightarrow+0$ are commutative. Due to the normalization of pdfs, 
the quantity $C_{n}^{(\mathrm{d})}(\boldsymbol{Q}_{\mathrm{d}}^{(\mathrm{s})},
\tilde{\boldsymbol{Q}}_{\mathrm{d}}^{(\mathrm{s})})$ should tend to $1$ 
as $n\rightarrow+0$. This observation implies that the RHS of 
(\ref{result}) is equal to the equivalent 
channel~(\ref{equivalent_channel_deterministic}) between $x_{m,t}$ and 
the associated decoder for the MIMO channel~(\ref{deterministic_MIMO}) 
with perfect CSI at the receiver. 

\subsection{Multiple Solutions} 
The fixed-point equation~(\ref{fixed_point_d}) may have multiple solutions. 
In that case, one has to choose the solution minimizing the 
quantity~(\ref{Phi_c_inf}). 
Due to $\lim_{n\rightarrow+0}\tilde{\Phi}_{\mathrm{d}}
(\boldsymbol{Q}_{\mathrm{d}}^{(\mathrm{s})})=0$, 
the quantity $\tilde{\Phi}_{\mathrm{d}}(\boldsymbol{Q}_{\mathrm{d}})$ is 
given by $\tilde{\Phi}_{\mathrm{d}}
(\boldsymbol{Q}_{\mathrm{d}}^{(\mathrm{s})})=
nF + O(n^{2})$ as $n\rightarrow+0$, with the so-called free 
energy $F=\lim_{n\rightarrow+0}\frac{\partial }{\partial n}
\tilde{\Phi}_{\mathrm{d}}
(\boldsymbol{Q}_{\mathrm{d}}^{(\mathrm{s})})$. 
Thus, one should choose the solution minimizing the free energy $F$.

In order to calculate the free energy $F$, in the same manner as in the 
derivation of (\ref{function_I_c_tmp}), 
we evaluate (\ref{function_I_inf}) as 
\begin{IEEEeqnarray}{r}  
\tilde{I}_{\mathrm{d}}(\boldsymbol{Q}_{\mathrm{d}}^{(\mathrm{s})},
\tilde{\boldsymbol{Q}}_{\mathrm{d}}^{(\mathrm{s})}) 
= -\frac{n}{\alpha}\left[
 1 - \frac{N_{0}}{\sigma^{2}+n\sigma_{0}^{2}} - \frac{N_{0}}{\sigma^{2}}
 + \frac{N_{0}\sigma_{0}^{2}}{\sigma^{2}(\sigma^{2}+n\sigma_{0}^{2})}
\right]
- \lim_{M\rightarrow\infty}\frac{1}{M}
\ln D_{n}^{(\mathrm{d})} \nonumber \\ 
- \lim_{M\rightarrow\infty}\frac{1}{M}\ln 
\int \mathbb{E}_{\boldsymbol{x}_{t}^{(0)}}
[\tilde{q}_{0}(\underline{\boldsymbol{z}} | \boldsymbol{x}_{t}^{(0)}, 
\boldsymbol{\Xi}_{t})]
\left\{ 
 \mathbb{E}_{\boldsymbol{x}_{t}^{(1)}}
 [\tilde{q}_{1}(\underline{\boldsymbol{z}} | \boldsymbol{x}_{t}^{(1)}, 
 \boldsymbol{\Xi}_{t})]
\right\}^{n}d\underline{\boldsymbol{z}}. \label{function_I_inf_tmp}
\end{IEEEeqnarray}
We differentiate (\ref{function_G_tmp}) and (\ref{function_I_inf_tmp}) 
with respect to $n$ at $n=0$ to obtain  
\begin{equation} \label{free_energy_rs}
F = \lim_{M\rightarrow\infty}\frac{1}{M}
I(\boldsymbol{x}_{[m,M],t};\underline{\boldsymbol{z}}) 
+\frac{1}{\alpha}\left[
 D_{\mathrm{e}}(N_{0}\|\sigma^{2}) 
 + \lim_{M\rightarrow\infty}\frac{1}{M\sigma^{2}}
 \mathrm{Tr}(\boldsymbol{\Xi}_{t}) 
 + \ln(\pi\mathrm{e}N_{0}) 
\right],    
\end{equation} 
with 
\begin{IEEEeqnarray}{rl}
I(\boldsymbol{x}_{[m,M],t};\underline{\boldsymbol{z}}) 
=& \int q_{0}(\underline{\boldsymbol{z}}|\boldsymbol{x}_{[m,M],t}^{(0)},
\boldsymbol{\Xi}_{t}^{(\mathrm{c})})
p(\boldsymbol{x}_{[m,M],t}^{(0)}|\boldsymbol{\theta}_{[m,M],t})
\nonumber \\  
&\times\ln\frac{
 q_{1}(\underline{\boldsymbol{z}}|\boldsymbol{x}_{[m,M],t}^{(0)},
 \boldsymbol{\Xi}_{t}^{(\mathrm{c})})
}
{
 \int q_{1}(\underline{\boldsymbol{z}}|\boldsymbol{x}_{[m,M],t}^{(1)},
 \boldsymbol{\Xi}_{t}^{(\mathrm{c})})
 p(\boldsymbol{x}_{[m,M],t}^{(1)}|\boldsymbol{\theta}_{[m,M],t}) 
 d\boldsymbol{x}_{[m,M],t}^{(1)} 
}d\boldsymbol{x}_{[m,M],t}^{(0)}d\underline{\boldsymbol{z}}.  
\end{IEEEeqnarray}
Minimizing (\ref{free_energy_rs}) is equivalent to minimizing 
(\ref{free_energy_MIMO}). 

\section{Reduction of Proposition~\ref{proposition3} to Proposition~\ref{proposition1}} \label{sec_deriv_proposition1} 
Let us prove that the fixed-point equation~(\ref{fixed_point_MIMO}) 
coincides with the fixed-point equation~(\ref{fixed_point}). 
We first show that the last term in (\ref{fixed_point}) is  a lower  
bound on the last term in (\ref{fixed_point_MIMO}), by considering the 
MIMO channel~(\ref{deterministic_MIMO}) with additional side information. 
Let a genie inform the receiver about the correct values of the data 
symbols $\boldsymbol{x}_{(m,M],t}$. 
The MSE~(\ref{MMSE}) for the genie-aided receiver should provide a lower 
bound on the original one. In order to eliminate the inter-stream 
interference from the MF output vector~(\ref{MF_output}), the genie-aided 
receiver calculates $\boldsymbol{r}_{\mathrm{s}} = \boldsymbol{r} - 
\sum_{m'=m+1}^{M}\boldsymbol{\xi}_{t,m'}x_{m',t}/\alpha$, given by  
\begin{equation} \label{free_channel} 
\boldsymbol{r}_{\mathrm{s}} =  
\frac{1}{\alpha}\boldsymbol{\xi}_{t,m}x_{m,t} + \boldsymbol{\eta}. 
\end{equation}  
The performance of the interference-free channel~(\ref{free_channel}), 
such as the MSE and the constrained capacity, is determined by the SNR 
\begin{equation} \label{SNR}
\overline{\mathrm{snr}} = \frac{P\|\boldsymbol{\xi}_{t,m}\|^{4}}
{\alpha\sigma^{2}\boldsymbol{\xi}_{t,m}^{\mathrm{H}}(\boldsymbol{I} 
-\boldsymbol{\Xi}_{t}^{(\mathrm{c})})\boldsymbol{\xi}_{t,m}}. 
\end{equation} 
Proposition~\ref{proposition2} and Lemma~\ref{lemma1} imply that 
the numerator and denominator in (\ref{SNR}) are given by 
$P\|\boldsymbol{\xi}_{t,m}\|^{4}=P(1-\xi^{2}(\tau))^{4} + O(M^{-1/4})$ 
and $\alpha\sigma^{2}\boldsymbol{\xi}_{t,m}^{\mathrm{H}}(\boldsymbol{I} 
-\boldsymbol{\Xi}_{t}^{(\mathrm{c})})\boldsymbol{\xi}_{t,m} 
=\alpha\sigma^{2}(1-\xi^{2}(\tau))^{3} + O(M^{-1/4})$ in the large-system 
limit, respectively. Thus, the SNR~(\ref{SNR}) converges in probability 
to $\overline{\mathrm{snr}}=(1-\xi^{2}(\tau))P/(\alpha\sigma^{2})$ in the 
large-system limit, which coincides with the SNR for the AWGN 
channel~(\ref{AWGN}) with $\sigma^{2}(\tau,\mu)=\sigma^{2}$. 
This expression implies that the last term~(\ref{MMSE}) 
in the fixed-point equation~(\ref{fixed_point_MIMO}) is bounded from 
below by $(1-\mu)(1-\xi^{2}(\tau))\mathbb{E}[
\mathrm{MSE}(\sigma^{2},\theta_{m,t})]$ 
in the large-system limit. 

We next prove that the last term in (\ref{fixed_point}) is an upper  
bound on the last term in (\ref{fixed_point_MIMO}). 
Let us consider a suboptimal receiver, which estimates $x_{m,t}$ only from the 
first element $r_{m}$ of the MF output vector~(\ref{MF_output}), given by 
\begin{equation} \label{MF_output1} 
r_{m} = \frac{1-(\boldsymbol{\Xi}_{t}^{(\mathrm{c})})_{m,m}}{\alpha}x_{m,t} 
+ \sum_{m'=m+1}^{M}\frac{(\boldsymbol{\Xi}_{t}^{(\mathrm{c})})_{m,m'}}
{\alpha}x_{m',t} + \eta_{m}, 
\end{equation}
with $\eta_{m}$ denoting the first element of $\boldsymbol{\eta}$. 
In order to evaluate an upper bound on the MSE~(\ref{MMSE}) for this suboptimal 
receiver, we replace the inter-stream interference in (\ref{MF_output1}) by 
the AWGN with the same variance. The MSE~(\ref{MMSE}) for the obtained channel 
provides an upper bound on the original one, and is determined 
by the SNR 
\begin{equation} \label{SNR_lower}
\underline{\mathrm{snr}} 
= \frac{P(1-(\boldsymbol{\Xi}_{t}^{(\mathrm{c})})_{m,m})^{2}}
{P\sum_{m'=m+1}^{M}|(\boldsymbol{\Xi}_{t}^{(\mathrm{c})})_{m,m'}|^{2}  
 + \alpha\sigma^{2}(1-(\boldsymbol{\Xi}_{t}^{(\mathrm{c})})_{m,m})}, 
\end{equation} 
which converges in probability to 
$\underline{\mathrm{snr}}=(1-\xi^{2}(\tau))P/(\alpha\sigma^{2})$ in the 
large-system limit, due to Proposition~\ref{proposition2} and 
Lemma~\ref{lemma1}. This result implies that the last term~(\ref{MMSE}) in the 
fixed-point equation~(\ref{fixed_point_MIMO}) is bounded from above by 
$(1-\mu)(1-\xi^{2}(\tau))\mathbb{E}[
\mathrm{MSE}(\sigma^{2},\theta_{m,t})]$ in the large-system limit. 
Combining the two bounds, we find that the fixed-point 
equation~(\ref{fixed_point_MIMO}) is equal to the fixed-point 
equation~(\ref{fixed_point}). 

The argument described above implies that the inter-stream interference is 
negligible in the large-system limit. 
It is straightforward to confirm that the mutual information 
$I(x_{m,t};\tilde{x}_{m,t}|\boldsymbol{\Xi}_{t}^{(\mathrm{c})},
\boldsymbol{\theta}_{[m,M],t})$ converges to the constrained 
capacity of the AWGN channel~(\ref{AWGN}), i.e., 
the integrand in (\ref{capacity_lower_bound}), by repeating the same 
argument. Similarly, it is straightforward to find that 
(\ref{free_energy_MIMO}) is equal to (\ref{free_energy}). 
Combining these results and the argument described in 
Appendix~\ref{sec_sketch}, we find that Proposition~\ref{proposition1} holds.

\ifCLASSOPTIONcaptionsoff
  \newpage
\fi



\bibliographystyle{IEEEtran}
\bibliography{IEEEabrv,kt-it2010_2}

\begin{thebibliography}{10}
\providecommand{\url}[1]{#1}
\csname url@samestyle\endcsname
\providecommand{\newblock}{\relax}
\providecommand{\bibinfo}[2]{#2}
\providecommand{\BIBentrySTDinterwordspacing}{\spaceskip=0pt\relax}
\providecommand{\BIBentryALTinterwordstretchfactor}{4}
\providecommand{\BIBentryALTinterwordspacing}{\spaceskip=\fontdimen2\font plus
\BIBentryALTinterwordstretchfactor\fontdimen3\font minus
  \fontdimen4\font\relax}
\providecommand{\BIBforeignlanguage}[2]{{%
\expandafter\ifx\csname l@#1\endcsname\relax
\typeout{** WARNING: IEEEtran.bst: No hyphenation pattern has been}%
\typeout{** loaded for the language `#1'. Using the pattern for}%
\typeout{** the default language instead.}%
\else
\language=\csname l@#1\endcsname
\fi
#2}}
\providecommand{\BIBdecl}{\relax}
\BIBdecl

\bibitem{Foschini98}
G.~J. Foschini and M.~J. Gans, ``On limits of wireless communications in a
  fading environment when using multiple antennas,'' \emph{Wireless Pers.
  Commun.}, vol.~6, pp. 311--335, 1998.

\bibitem{Telatar99}
E.~Telatar, ``Capacity of multi-antenna {Gaussian} channels,'' \emph{Euro.
  Trans. Telecommun.}, vol.~10, no.~6, pp. 585--595, Nov.--Dec. 1999.

\bibitem{Tulino05}
A.~M. Tulino, A.~Lozano, and S.~Verd\'{u}, ``Impact of antenna correlation on
  the capacity of multiantenna channels,'' \emph{{IEEE} Trans. Inf. Theory},
  vol.~51, no.~7, pp. 2491--2509, Jul. 2005.

\bibitem{Marzetta99}
T.~L. Marzetta and B.~M. Hochwald, ``Capacity of a mobile multiple-antenna
  communication link in {Rayleigh} flat fading,'' \emph{{IEEE} Trans. Inf.
  Theory}, vol.~45, no.~1, pp. 139--157, Jan. 1999.

\bibitem{Lapidoth03}
A.~Lapidoth and S.~M. Moser, ``Capacity bounds via duality with applications to
  multiple-antenna systems on flat-fading channels,'' \emph{{IEEE} Trans. Inf.
  Theory}, vol.~49, no.~10, pp. 2426--2467, Oct. 2003.

\bibitem{Moser09}
S.~M. Moser, ``The fading number of multiple-input multiple-output fading
  channels with memory,'' \emph{{IEEE} Trans. Inf. Theory}, vol.~55, no.~6, pp.
  2716--2755, Jun. 2009.

\bibitem{Hochwald00}
B.~M. Hochwald and T.~L. Marzetta, ``Unitary space-time modulation for
  multiple-antenna communications in rayleigh flat fading,'' \emph{{IEEE}
  Trans. Inf. Theory}, vol.~46, no.~2, pp. 543--564, Mar. 2000.

\bibitem{Hassibi02}
B.~Hassibi and T.~L. Marzetta, ``Multiple-antennas and isotropically random
  unitary inputs: The received signal density in closed form,'' \emph{{IEEE}
  Trans. Inf. Theory}, vol.~48, no.~6, pp. 1473--1484, Jun. 2002.

\bibitem{Moustakas06}
A.~L. Moustakas, S.~H. Simon, and T.~L. Marzetta, ``Capacity of differential
  versus nondifferential unitary space-time modulation for {MIMO} channels,''
  \emph{{IEEE} Trans. Inf. Theory}, vol.~52, no.~8, pp. 3622--3634, Aug. 2006.

\bibitem{Zheng02}
L.~Zheng and D.~N.~C. Tse, ``Communication on the {Grassmann} manifold: A
  geometric approach to the noncoherent multiple-antenna channel,''
  \emph{{IEEE} Trans. Inf. Theory}, vol.~48, no.~2, pp. 359--383, Feb. 2002.

\bibitem{Yang13}
W.~Yang, G.~Durisi, and E.~Riegler, ``On the capacity of large-{MIMO}
  block-fading channels,'' \emph{{IEEE} J. Sel. Areas Commun.}, vol.~31, no.~2,
  pp. 117--132, Feb. 2013.

\bibitem{Verdu02}
S.~Verd\'u, ``Spectral efficiency in the wideband regime,'' \emph{{IEEE} Trans.
  Inf. Theory}, vol.~48, no.~6, pp. 1319--1343, Jun. 2002.

\bibitem{Zheng07}
L.~Zheng, D.~N.~C. Tse, and M.~M\'edard, ``Channel coherence in the low-{SNR}
  regime,'' \emph{{IEEE} Trans. Inf. Theory}, vol.~53, no.~3, pp. 976--997,
  Mar. 2007.

\bibitem{Ray07}
S.~Ray, M.~M\'edard, and L.~Zheng, ``On noncoherent {MIMO} channels in the
  wideband regime: Capacity and reliability,'' \emph{{IEEE} Trans. Inf.
  Theory}, vol.~53, no.~6, pp. 1983--2009, Jun. 2007.

\bibitem{Hassibi03}
B.~Hassibi and B.~M. Hochwald, ``How much training is needed in
  multiple-antenna wireless link?'' \emph{{IEEE} Trans. Inf. Theory}, vol.~49,
  no.~4, pp. 951--963, Apr. 2003.

\bibitem{Medard00}
M.~M\'edard, ``The effect upon channel capacity in wireless communications of
  perfect and imperfect knowledge of the channel,'' \emph{{IEEE} Trans. Inf.
  Theory}, vol.~46, no.~3, pp. 933--946, May 2000.

\bibitem{Takeuchi101}
K.~Takeuchi, M.~Vehkaper\"a, T.~Tanaka, and R.~R. M\"uller, ``Large-system
  analysis of joint channel and data estimation for {MIMO DS-CDMA} systems,''
  \emph{{IEEE} Trans. Inf. Theory}, vol.~58, no.~3, pp. 1385--1412, Mar. 2012.

\bibitem{Li07}
T.~Li and O.~M. Collins, ``A successive decoding strategy for channels with
  memory,'' \emph{{IEEE} Trans. Inf. Theory}, vol.~53, no.~2, pp. 628--646,
  Feb. 2007.

\bibitem{Padmanabhan08}
K.~Padmanabhan, S.~Venkatraman, and O.~M. Collins, ``Tight upper and lower
  bounds on the constrained capacity of non-coherent multi-antenna channels,''
  in \emph{Proc. 2008 IEEE Int. Symp. Inf. Theory}, Toronto, Canada, Jul. 2008,
  pp. 2588--2592.

\bibitem{Takeuchi092}
K.~Takeuchi, R.~R. M\"uller, M.~Vehkaper\"a, and T.~Tanaka, ``Practical
  signaling with vanishing pilot-energy for large noncoherent block-fading
  {MIMO} channels,'' in \emph{Proc. 2009 IEEE Int. Symp. Inf. Theory}, Seoul,
  Korea, Jun. 2009, pp. 759--763.

\bibitem{Takeuchi111}
K.~Takeuchi, R.~R. M\"uller, and M.~Vehkaper\"a, ``Bias-based training for
  iterative channel estimation and data decoding in fast fading channels,''
  \emph{IEICE Trans. Commun.}, vol. E94-B, no.~7, pp. 2161--2165, Jul. 2011.

\bibitem{Takeuchi112}
------, ``A construction of turbo-like codes for iterative channel estimation
  based on probabilistic bias,'' in \emph{Proc. IEEE Global Commun. Conf.
  (GLOBECOME 2011)}, Houston, Texas, USA, Dec. 2011.

\bibitem{Takeuchi12}
------, ``Iterative {LMMSE} channel estimation and decoding based on
  probabilistic bias,'' \emph{{\rm submitted to} IEEE Trans. Commun.}, 2012.

\bibitem{Tse99}
D.~N.~C. Tse and S.~V. Hanly, ``Linear multiuser receivers: effective
  interference, effective bandwidth and user capacity,'' \emph{{IEEE} Trans.
  Inf. Theory}, vol.~45, no.~2, pp. 641--657, Mar. 1999.

\bibitem{Verdu99}
S.~Verd\'u and S.~{Shamai (Shitz)}, ``Spectral efficiency of {CDMA} with random
  spreading,'' \emph{{IEEE} Trans. Inf. Theory}, vol.~45, no.~2, pp. 622--640,
  Mar. 1999.

\bibitem{Evans00}
J.~Evans and D.~N.~C. Tse, ``Large system performance of linear multiuser
  receivers in multipath fading channels,'' \emph{{IEEE} Trans. Inf. Theory},
  vol.~46, no.~6, pp. 2059--2078, Sep. 2000.

\bibitem{Shamai01}
S.~{Shamai (Shitz)} and S.~Verd\'u, ``The impact of frequency-flat fading on
  the spectral efficiency of {CDMA},'' \emph{{IEEE} Trans. Inf. Theory},
  vol.~47, no.~4, pp. 1302--1327, May 2001.

\bibitem{Tanaka02}
T.~Tanaka, ``A statistical-mechanics approach to large-system analysis of
  {CDMA} multiuser detectors,'' \emph{{IEEE} Trans. Inf. Theory}, vol.~48,
  no.~11, pp. 2888--2910, Nov. 2002.

\bibitem{Moustakas03}
A.~L. Moustakas, S.~H. Simon, and A.~M. Sengupta, ``{MIMO} capacity through
  correlated channels in the presence of correlated interferers and noise: A
  (not so) large ${N}$ analysis,'' \emph{{IEEE} Trans. Inf. Theory}, vol.~49,
  no.~10, pp. 2545--2561, Oct. 2003.

\bibitem{Mueller04}
R.~R. M\"{u}ller and W.~H. Gerstacker, ``On the capacity loss due to separation
  of detection and decoding,'' \emph{{IEEE} Trans. Inf. Theory}, vol.~50,
  no.~8, pp. 1769--1778, Aug. 2004.

\bibitem{Guo05}
D.~Guo and S.~Verd\'u, ``Randomly spread {CDMA}: Asymptotics via statistical
  physics,'' \emph{{IEEE} Trans. Inf. Theory}, vol.~51, no.~6, pp. 1983--2010,
  Jun. 2005.

\bibitem{Takeda06}
K.~Takeda, S.~Uda, and Y.~Kabashima, ``Analysis of {CDMA} systems that are
  characterized by eigenvalue spectrum,'' \emph{Europhys. Lett.}, vol.~76,
  no.~6, pp. 1193--1199, 2006.

\bibitem{Wen07}
C.~K. Wen and K.~K. Wong, ``Asymptotic analysis of spatially correlated {MIMO}
  multiple-access channels with arbitrary signaling inputs for joint and
  separate decoding,'' \emph{{IEEE} Trans. Inf. Theory}, vol.~53, no.~1, pp.
  252--268, Jan. 2007.

\bibitem{Takeuchi082}
K.~Takeuchi, T.~Tanaka, and T.~Yano, ``Asymptotic analysis of general multiuser
  detectors in {MIMO DS-CDMA} channels,'' \emph{{IEEE} J. Sel. Areas Commun.},
  vol.~26, no.~3, pp. 486--496, Apr. 2008.

\bibitem{Sherrington75}
D.~Sherrington and S.~Kirkpatrick, ``Solvable model of a spin-glass,''
  \emph{Phys. Rev. Lett.}, vol.~35, no.~26, pp. 1792--1796, Dec. 1975.

\bibitem{Nishimori01}
H.~Nishimori, \emph{Statistical Physics of Spin Glasses and Information
  Processing}.\hskip 1em plus 0.5em minus 0.4em\relax New York: Oxford
  University Press, 2001.

\bibitem{Mezard87}
M\'ezard, G.~Parisi, and M.~A. Virasoro, \emph{Spin Glass Theory and
  Beyond}.\hskip 1em plus 0.5em minus 0.4em\relax Singapore: World Scientific,
  1987.

\bibitem{Fischer91}
K.~H. Fischer and J.~A. Hertz, \emph{Spin Glasses}.\hskip 1em plus 0.5em minus
  0.4em\relax Cambridge, UK: Cambridge University Press, 1991.

\bibitem{Guerra032}
F.~Guerra, ``Broken replica symmetry bounds in the mean field spin glass
  model,'' \emph{Commun. Math. Phys.}, vol. 233, pp. 1--12, 2003.

\bibitem{Talagrand06}
M.~Talagrand, ``The {Parisi} formula,'' \emph{Annals of Mathematics}, vol. 163,
  pp. 221--263, 2006.

\bibitem{Neeser93}
F.~D. Neeser and J.~L. Massey, ``Proper complex random processes with
  applications to information theory,'' \emph{{IEEE} Trans. Inf. Theory},
  vol.~39, no.~4, pp. 1293--1302, Jul. 1993.

\bibitem{Forney92}
G.~D. {Forney, Jr.}, ``Trellis shaping,'' \emph{{IEEE} Trans. Inf. Theory},
  vol.~38, no.~2, pp. 281--300, Mar. 1992.

\bibitem{Takeuchi102}
K.~Takeuchi, R.~R. M\"uller, M.~Vehkaper\"a, and T.~Tanaka, ``An achievable
  rate of large block-fading {MIMO} systems with no {CSI} via successive
  decoding,'' in \emph{Proc. 2010 Int. Symp. Inf. Theory and its Appl \& Int.
  Symp. Spread Spectrum Tech. and Appl.}, Taichung, Taiwan, Oct. 2010, pp.
  519--524.

\bibitem{Cover06}
T.~M. Cover and J.~A. Thomas, \emph{Elements of Information Theory},
  2nd~ed.\hskip 1em plus 0.5em minus 0.4em\relax New Jersey: Wiley, 2006.

\bibitem{Ledoux01}
M.~Ledoux, \emph{The Concentration of Measure Phenomenon}.\hskip 1em plus 0.5em
  minus 0.4em\relax Providence, RI: American Mathmatical Society, 2001.

\bibitem{Guerra02}
F.~Guerra and F.~L. Toninelli, ``The thermodynamic limit in mean field spin
  glass models,'' \emph{Commun. Math. Phys.}, vol. 230, pp. 71--79, 2002.

\bibitem{Guerra031}
------, ``The infinite volume limit in generalized mean field disordered
  models,'' \emph{Markov. Proc. Rel. Fields}, vol.~9, pp. 195--207, 2003.

\bibitem{Cottatellucci05}
L.~Cottatellucci and R.~R. M\"uller, ``A systematic approach to multistage
  detectors in multipath fading channels,'' \emph{{IEEE} Trans. Inf. Theory},
  vol.~51, no.~9, pp. 3146--3158, Sep. 2005.

\bibitem{Nishimori021}
H.~Nishimori, ``Comment on ``statistical mechanics of {CDMA} multiuser
  demodulation" by {Tanaka},'' \emph{Europhys. Lett.}, vol.~57, no.~2, pp.
  302--303, Jan. 2002.

\bibitem{Verdu98}
S.~Verd\'u, \emph{Multiuser Detection}.\hskip 1em plus 0.5em minus 0.4em\relax
  New York: Cambridge University Press, 1998.

\bibitem{Korada11}
S.~B. Korada and A.~Montanari, ``Applications of the {Lindeberg} principle in
  communicaitons and statistical learning,'' \emph{{IEEE} Trans. Inf. Theory},
  vol.~57, no.~4, pp. 2440--2450, Apr. 2011.

\bibitem{Pastur91}
L.~A. Pastur and M.~V. Shcherbina, ``Absence of self-averaging of the order
  parameter in the {Sherrington-Kirkpatrick} model,'' \emph{J. Stat. Phys.},
  vol.~62, no. 1/2, pp. 1--19, 1991.

\bibitem{Tse05}
D.~N.~C. Tse and P.~Viswanath, \emph{Fundamentals of Wireless
  Communication}.\hskip 1em plus 0.5em minus 0.4em\relax Cambridge, UK:
  Cambridge University Press, 2005.

\bibitem{Vehkaperae093}
M.~Vehkaper\"a, K.~Takeuchi, R.~R. M\"uller, and T.~Tanaka, ``How much training
  is needed for iterative multiuser detection and decoding?'' in \emph{Proc.
  IEEE Global Communications Conference (GLOBECOM 2009)}, Honolulu, USA, Nov.
  2009.

\bibitem{Kitagawa10}
K.~Kitagawa and T.~Tanaka, ``Optimization of sequences in {CDMA} systems: A
  statistical-mechanics approach,'' \emph{Computer Networks}, vol.~54, pp.
  917--924, 2010.

\bibitem{Gursoy09}
M.~C. Gursoy, ``On the capacity and energy efficiency of training-based
  transmissions over fading channels,'' \emph{{IEEE} Trans. Inf. Theory},
  vol.~55, no.~10, pp. 4543--4567, Oct. 2009.

\bibitem{Nakamura08}
K.~Nakamura and T.~Tanaka, ``Microscopic analysis for decoupling principle of
  linear vector channel,'' in \emph{Proc. 2008 IEEE Int. Symp. Inf. Theory},
  Toronto, Canada, Jul. 2008, pp. 519--523.

\end{thebibliography}
\end{document}